\pdfoutput=1
\documentclass{article}
\usepackage[utf8]{inputenc}
\usepackage{csquotes}
\usepackage[english]{babel}
\usepackage[x11names]{xcolor}
\usepackage{tabularx}
\usepackage{amsmath}
\usepackage{amsfonts}
\usepackage{amsbsy}
\usepackage{amssymb}
\usepackage{amsthm}
\usepackage{amscd}
\usepackage{enumitem}
\usepackage{float}
\usepackage{hyperref}
\usepackage{siunitx}
\usepackage{comment}
\usepackage{placeins} \usepackage{multirow}
\usepackage[font=footnotesize]{caption}
\usepackage{subcaption}
\usepackage[numbers]{natbib}
\usepackage{graphicx}

\newcommand{\R}{\mathbb R}
\newcommand{\cb}{\boldsymbol{c}}
\newcommand{\chat}{{\widehat c}}
\newcommand{\cbhat}{{\widehat \cb}}
\newcommand{\kb}{\boldsymbol{k}}
\newcommand{\yb}{\boldsymbol{y}}
\newcommand{\stilde}{\widetilde s}
\newcommand{\ytilde}{\widetilde y}
\newcommand{\Kb}{\boldsymbol{K}}
\newcommand{\thetamax}{\theta_{\mathrm{max}}}

\renewcommand{\theta}{\vartheta}
\renewcommand{\phi}{\varphi}
\newcommand{\1}{\mathbf 1}
\newcommand{\fhat}{\widehat f}
\newcommand{\FRSWhat}{\widehat F_{\textrm{RSW}}}
\newcommand{\ncand}{N_{\textrm{cand}}}
\newcommand{\ninit}{N_{\textrm{init}}}
\newcommand{\nseq}{N_{\textrm{seq}}}
\newcommand{\ntotal}{N_{\textrm{total}}}

\newcommand{\xcand}{{\widetilde \theta}}
\newcommand{\Dc}{\mathcal D}
\newcommand{\Dctilde}{\widetilde{\Dc}}
\newcommand{\FRSW}{F_{\textrm{RSW}}}
\newcommand{\nuhat}{{\widehat \nu}}

\DeclareMathOperator*{\argmax}{arg\,max}

\definecolor{forestgreen}{RGB}{0, 148, 0}

\makeatletter
\def\@fnsymbol#1{\ensuremath{\ifcase#1\or *\or **\else\@ctrerr\fi}}
\makeatother

\title{Efficient prediction of grain boundary energies from atomistic simulations via  sequential design}

\author{Martin Kroll\thanks{Ruhr-Universität Bochum, Fakultät für Mathematik, D-44780 Bochum. Contact E-mail: Martin.Kroll-k9x@rub.de / Holger.Dette@rub.de}
\and Timo Schmalofski\thanks{Ruhr-Universität Bochum,   Interdisciplinary Centre for Advanced Materials Simulation (ICAMS), D-44780 Bochum. Contact E-mail: Timo.Schmalofski@rub.de / Rebecca.Janisch@rub.de}
\and Holger Dette\footnotemark[1]
\and Rebecca Janisch\footnotemark[2]} 
\date{\today}

\begin{document}

\maketitle

\begin{abstract}
Data based materials science is the new promise to accelerate materials design. Especially in computational materials science, data generation can easily be automatized. Usually, the focus is on processing and evaluating the data to derive rules or to discover new materials, while less attention is being paid on the strategy to generate the data. In this work, we show that by a sequential design of experiment scheme, the process of generating and learning from the data can be combined to discover the relevant sections of the parameter space. Our example is the energy of grain boundaries as a function of their geometric degrees of freedom, calculated via atomistic simulations. The sampling of this grain boundary energy space, or even subspaces of it, represents a challenge due to the presence of deep cusps of the energy, which are located at irregular intervals of the geometric parameters. Existing approaches to sample grain boundary energy subspaces therefore either need a huge amount of datapoints or a~priori knowledge of the positions of these cusps. We combine statistical methods with atomistic simulations and a sequential sampling technique and compare this strategy to a regular sampling technique. We thereby demonstrate that this sequential design is able to sample a subspace with a minimal amount of points while finding unknown cusps automatically.

\end{abstract} 

\section{Introduction}

Grain boundaries (GBs) play an important role in materials science but they are still among the least understood crystal defects. Nano-structured materials, i.e.~materials with a high density of GBs, dominate the development of new technologies for energy harvesting and storage \citep{banerjee2020interfaces}, for data processing, but also of solutions for structural applications \cite{li2020metallic, semenova2021advanced}, where the presence of interfaces leads to strengthening of the material \cite{Han2020AlloyDS,cordero2016six}. The stability of such nanostructures can be tuned by the chemical composition, or rather the distribution of chemical elements in the microstructure and their segregation to grain boundaries. This process is influenced by the GB energy, respectively, the anisotropy of the grain boundary energy. Similarly, the mobility of GBs, and with this the grain growth during heat treatment or deformation, depends on the change of grain boundary energy with grain boundary geometry (in this case mostly the plane inclination). Finally, the resistance of interfaces to decohesion and fracture depends on the difference between the interface energy and the energy of the created surfaces. For precise predictions about any of these processes, one needs accurate data of the grain boundary energy as a function of the geometric degrees of freedom. Thus, great efforts have been made to provide this data by systematic, high-throughput atomistic calculations \cite{Yang2019,ratanaphan2015grain,tschopp2015symmetric,Tschopp2012,kim2011identification,olmsted2009survey}.

In this paper we modify the regular atomistic high-throughput approach by combining a statistical assessment of the parameter space via a sequential design of experiments techniques with an interpolation method to establish a quick and reliable algorithm for accurate prediction of grain boundary energies based on atomistic simulations. By a more dense sampling in  the regions of interest  the new algorithm substantially reduces the number of necessary simulation experiments compared to a regular high-throughput scheme, without loss in accuracy. We will now recap the basics of the mathematical description of GBs, review some recent work to sample the resulting parameter space, and explain our own contribution in more detail.

A grain boundary is fully described by defining the 
five macroscopic degrees of freedom (DOF): three angles for the misorientation and the plane normal vector $\vec{n}$ (with $|\vec{n}|=1$), and the three microscopic ones, the latter being the three components of a translation vector $\vec{t}$, which separates and shifts the grains relative to each other, without changing the misorientation. While the macroscopic DOF can be prescribed, the microscopic DOFs adjust themselves to minimize the energy of the system.

To determine the GB energy as a function of any of the macroscopic DOFs, one needs to vary one of them while keeping the others fixed, i.e., one needs to probe a one dimensional subspace of the five dimensional parameter space. For example, an important property for the understanding of microstructures is
the angle-dependent GB energy for fixed misorientation axes. To obtain it, one has to prescribe the GB plane. The vector $\vec{n}$ can be parallel to the misorientation axis, in which case the GB is a pure twist GB, or perpendicular, which results in a pure tilt GB, or it can have a mixed character. If the GB is a tilt grain boundary and the plane represents a mirror plane between the two grains, the GB is called a symmetrical tilt grain boundary (STGB).  In this case, for each given rotation axis, the misorientation angle is the only DOF left. The energy as a function of misorientation angle for STGBs shows pronounced minima, so-called cusps. They occur at special misorientations,
for which the rotation of the two grains leads to a periodic superlattice of coincidence sites across both grains. Such coincidence-site lattice (CSL) based GBs are characterized by their $\Sigma$ value, where $\Sigma=V_{\mathrm{CSL}}/V_0$, the ratio of the volume of a periodic unit of the CSL lattice to the periodic unit of the crystal lattice \citep{sutton1995interfaces}. This paper will exemplarily focus on such 1-dimensional STGB subspaces with different fixed rotation axes.

The so-called Read-Shockley model of GB energies \cite{read1950dislocation} (see Section \ref{sec42}) with the empirical extension of Wolf \cite{wolf1989read} (called Read-Shockley-Wolf (RSW) model in the following) provides a good description of the one-dimensional energy subspaces. Olmsted et al.~\cite{olmsted2009survey} calculated the energies of several tilt, twist and mixed grain boundaries, from atomistic simulations, starting with one-dimensional subspaces and moving further in the parameter space to expand these sets. Bulatov et al.~\cite{bulatov2014grain} then used the RSW model to fit the energy in the one-dimensional subspaces, and a simple interpolation was applied between them. 
In this approach, only several hundreds
of datapoints are required for a satisfiying fit. However, the overall strategy requires a priori knowledge of the topography of the energy space, and the model still represents a rather crude approximation of the (up to now rarely explored) higher dimensional parameter spaces.

As alternative strategy, a high-throughput simulation method was employed by Kim et al.~\cite{kim2011identification}
to generate a database of more than 16.000 grain boundary energies in bcc iron, for fixed steps in misorientation and inclination angle. While such a regular sampling approach is quick, since it can easily be parallelised, it misses the special $\Sigma$ grain boundaries. In \cite{kim2011identification} these structures were added to the list of sampled grain boundaries manually. 
Restrepo et al.~\cite{restrepo2014artificial} used this data
to train and test an artificial neural network (ANN) for the prediction of grain boundary energies. 25\% of the data, including low-energy, special grain boundaries, was used for the training of the network, 75\% for testing. This study showed, that a trained ANN is able to predict unknown grain boundary energies with an error of less then $4\%$. Still, one needs thousands of grain boundaries for the training and additional knowledge about the position of the cusps.

To arrive at a more effective method for sampling the grain boundary parameter space, we replace in this paper the regular sampling technique by a sequential design of experiments approach, and combine it with an 
alternative interpolation method. The combination of these two recipes leads to an algorithm, 
which can determine the energy cusps 
with a minimal number of sequential steps and high accuracy.

In  Section \ref{subs:series_vs_kriging}, we first compare two interpolation methods for exploring 
one-dimensional sections of the energy landscape which can be parametrized by one single parameter, usually an angle.
The first one is an expansion of the grain boundary energy in a truncated Fourier series (introduced in Section~\ref{subs:Fourier}), inspired by the use of spherical harmonics in materials science \cite{Hoyt2001,Haxhimali2006,Banadaki2016}.
The second interpolation method is the so-called \emph{Kriging}, which is a well-established method in geostatistical applications \citep{stein1999interpolation} and  has recently attracted interest in the materials science literature \citep{noack2020autonomous}. This method  predicts the value of a function at a given point by computing a weighted average of the already determined values of the function in the neighborhood of the point. It is somewhat related to a regression analysis, but Kriging exactly interpolates through the observations already made. 
Our choice of Kriging interpolator is, among other parameters, defined by a smoothness parameter $\nu$ (see Section \ref{subs:Kriging} for more details). The smaller the value of $\nu$, the rougher is the resulting interpolation, in the extreme case close to a piecewise linear function. A typical Kriging estimator for $\nu = 2.5$ is displayed in Figure~\eqref{fig:Kriging}. In practical applications, $\nu$ can be fixed in advance or estimated from the given data. In our study, we have considered both scenarios with respect to the quality of the fit.

An important advantage of  Kriging consists in the fact that it comes along with a natural measure of uncertainty quantification for predictions at potential locations for future observations.
We will use this measure  to develop a sequential design strategy, which advises the experimenter 
to put new  observations  into areas with large 
uncertainty. 
The method is explained in detail in Section~\ref{subs:sequential} and  computes a criterion function on a set of potential future sampling locations.
It measures the uncertainty concerning the target energy function at any given location. The next sampling point is then selected as a maximizer of the criterion function. Altogether, this leads to a procedure where the next design point depends on both the already chosen sampling locations and the corresponding energies.

In the first part of the paper, we use the RSW model \citep{read1950dislocation,wolf1989read} as a quick method to generate grain boundary energies as benchmark data. The mathematical details of the model are given in Section \ref{sec42}.
The numerical results obtained for the RSW data 
in Section \ref{subs:series_vs_kriging} suggest a better performance of Kriging interpolation compared to series estimation. The Kriging method is then further refined by means of a sequential sampling technique. 
The key idea of this sampling technique, which is described in detail in Section \ref{subs:sequential},
is to make observations at regions of interest like cusps more likely. As a consequence the sequentially sampled data gives a better overall picture of the grain boundary energy landscape.
Our results in Section~\ref{sec22} demonstrate that substantial improvements can be obtained by sequential sampling. 

In Section \ref{sec23} we investigate the performance of the 
Kriging interpolator and the sequential design for atomistic simulations. Grain boundary energies for a reference data set as well as during the sequential experiments are determined via molecular statics with a semi-spherical grain method and an embedded atom-method potential
(see  Section  \ref{sec:Atomistic_simulations} for technical details). Using the atomistic data, details of the Kriging method such as  the choice of a smoothness parameter are optimized. 

Our study demonstrates that, in contrast to the regular sampling techniques, the sequential sampling technique proposed in this paper is able to sample a subspace with a small number of observations and to identify at the same time unknown cusps.

\section{Results}
\subsection{Kriging versus series estimation: a comparison using RSW model data}
\label{subs:series_vs_kriging}
We begin 
with a comparison of the two interpolation  procedures using  data predicted by the RSW model.
Note that (in contrast to the atomistic simulations approach presented below) evaluations in the RSW are not computationally expensive because an explicit representation for the energy function is available.
Therefore this model  provides an efficient tool for testing the potential applicability of the proposed
interpolation procedures.

\begin{figure}
     \centering
     \begin{subfigure}[b]{0.49\textwidth}
         \centering
         \caption{}
         \includegraphics[width=\textwidth]{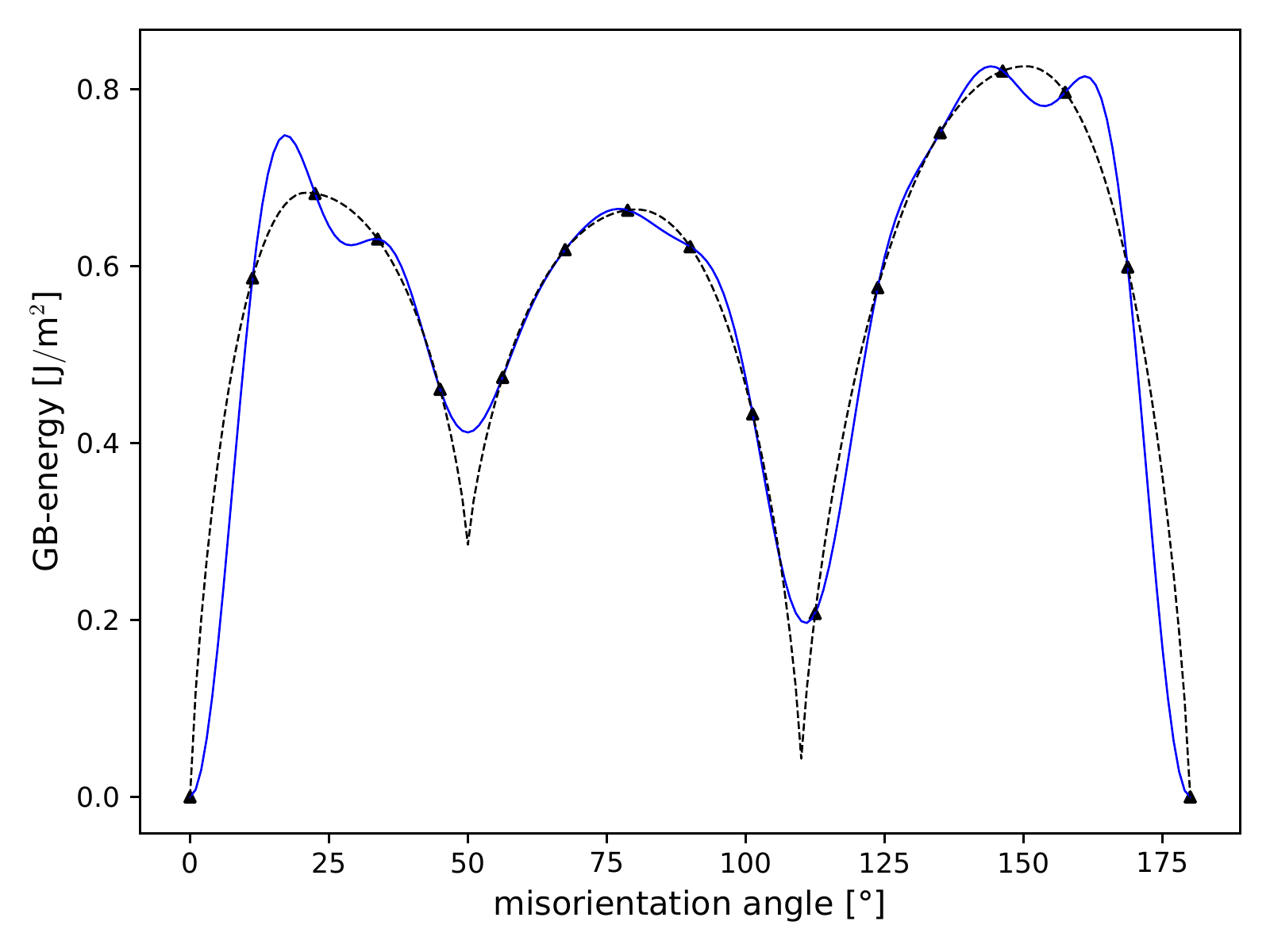}
         \label{fig:trig}
     \end{subfigure}
     \begin{subfigure}[b]{0.49\textwidth}
         \centering
         \caption{}
         \includegraphics[width=\textwidth]{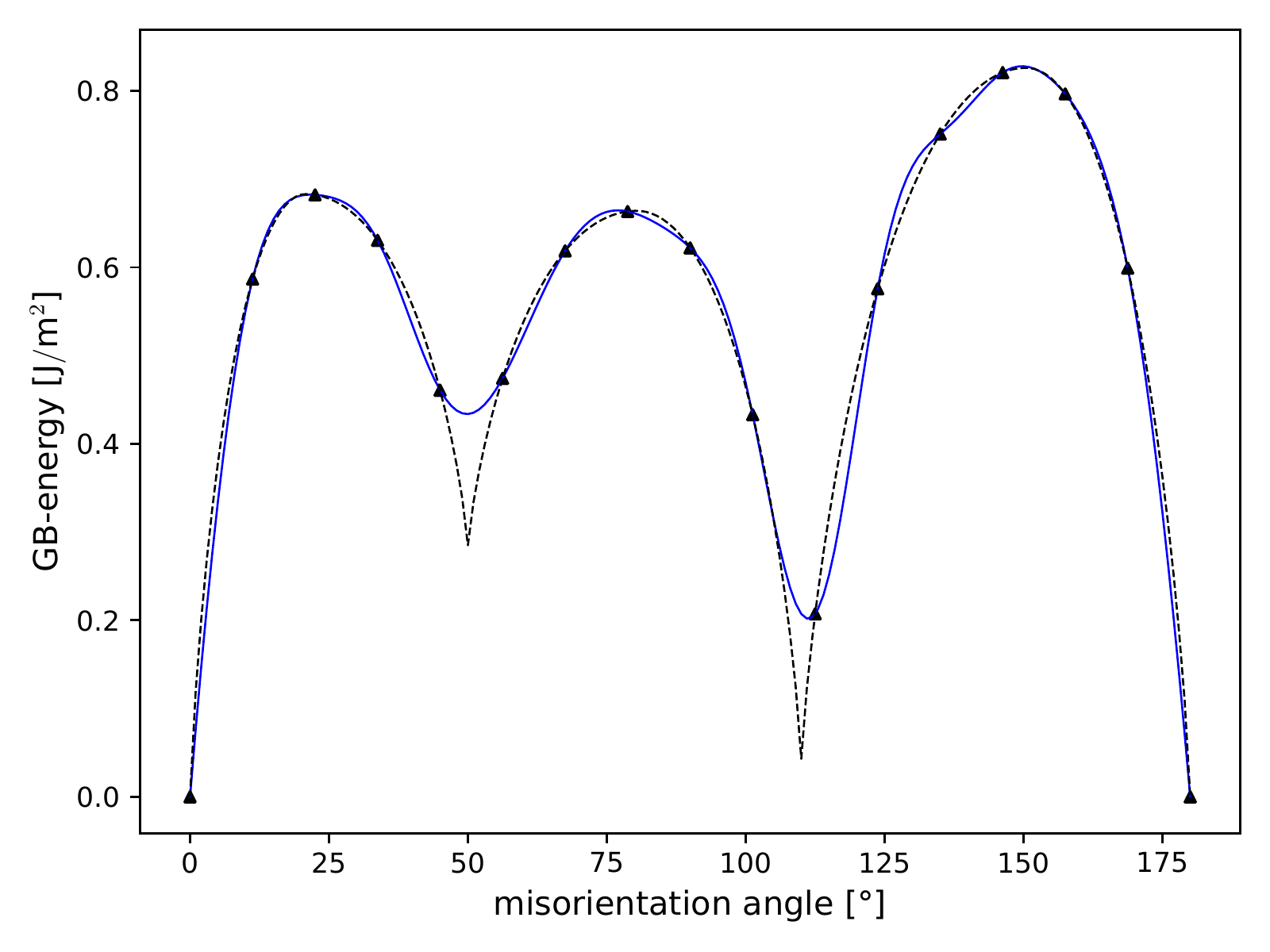}
         \label{fig:Kriging}
     \end{subfigure}
    \caption{Trigonometric (left plot) and Kriging interpolator (right plot, based on the Mat\'ern kernel with $\nu = 2.5$) for $N=17$ observations in the RSW model. In both plots the interpolator is the blue solid line, and the true energy function the black dashed line. Note that indeed both approaches interpolate exactly through the observed data points indicated as $\blacktriangle$.
}
    \label{fig:plots:RSW}
\end{figure}

To be precise, we chose the energy function of the [110] symmetrical tilt grain boundaries (where [110] is the rotation axis) in a cubic metal, because it contains several energy cusps and maxima. It is obtained by specifying the energy function, denoted by  $F_{\mathrm{RSW}}$, to contain 7 logarithmic segments positioned at the misorientation angles and energy offset values 
(see Section \ref{sec43} for details). This function is displayed by the dashed line 
in Figure \ref{fig:plots:RSW}. We  now assume that the function is only observed at a sample of $N = 9,17,33$, and $65$ equally spaced  points, say $(\theta_1, \FRSW(\theta_1) ), \ldots ,(\theta_N,\FRSW(\theta_N))$, which are marked by black triangles in Figure \ref{fig:plots:RSW} for the case $N=17$. We compare the prediction properties of two interpolation methods, which are described in detail in Section \ref{s:comparison}: trigonometric series and Kriging. The predictions of  both procedures, which interpolate exactly through the observed data points, are indicated by the blue solid lines  where the left panel corresponds to series and the right panel to Kriging interpolation.

As the performance measure for a generic interpolator $\FRSWhat$ we use throughout the paper the maximum absolute error \begin{equation} \label{h1}
    \max_{\theta \in [0,\thetamax]} \lvert \FRSWhat (\theta)  - \FRSW(\theta) \rvert 
\end{equation}
between the true function $\FRSW$ and
its interpolator  $\FRSWhat$ calculated from the sample
$$(\theta_1, \FRSW(\theta_1) ), \ldots ,(\theta_N,\FRSW(\theta_N)).$$
Here, $\thetamax=180\si{\degree}$ is the maximum misorientation angle of the considered grain boundary space, taking into account the symmetries of the cubic crystal system.

Corresponding results for the different sample sizes are shown in  Table~\ref{table:RSW} for the trigonometric series 
and the Kriging interpolator, using different fixed smoothness parameters $\nu$, as well as a data adaptive choice, which is denoted by $\nuhat$ 
and obtained by maximum likelihood estimation (see Section \ref{subs:Kriging} for details).
We observe that in all cases but one, the Kriging interpolator outperforms the trigonometric interpolator, independently of the choice of the parameter $\nu$, or whether $\widehat{\nu}$ was determined with the data-adaptive maximum likelihood estimator. The exception is $N=33$ and $\nu =0.5$, which results in a maximum error of $0.1424$, which is larger than the one for trigonometric interpolation, $0.1245$ (note that for all other smoothing parameters Kriging outperforms trigonometric interpolation).
We have considered numerous other tilt and twist grain boundaries in the RSW model, and observed a similar superiority of the Kriging approach. The results are displayed in Tables~\ref{tab:RSW1}--\ref{tab:RSW6} in Appendix~\ref{app:subs:comparison}.
From these numerical experiments we conclude that Kriging generally outperforms the series interpolator, and that for data which can be represented within the RSW model,
Kriging interpolation with a fixed  smoothing parameter 
$\nu=2.5$ is a very good choice. 
In the next step, we introduce a sequential design strategy aiming at a further improvement of the Kriging interpolator.

\begin{table}
\begin{center}
 \begin{tabular}{c|c|c|c|c|c}
 $N$ & trig. series & $\nu=0.5$ & $\nu=1.5$ & $\nu=2.5$ & $\nuhat$ via MLE\\
 \hline
 9 & 0.3636 & 0.2570 & 0.1924 & \bfseries 0.1784 & 0.2569 \footnotesize($\nuhat$ = 0.5062)\\
 17 & 0.2117 & 0.1751 & 0.1357 & \bfseries 0.1317 & 0.1557 \footnotesize($\nuhat$ = 0.6572)\\
 33 & 0.1245 & 0.1424 & 0.1079 & \bfseries 0.1040 & 0.1090 \footnotesize($\nuhat$ = 1.3726)\\
 65 & 0.0688 & 0.0361 & 0.0335 & 0.0352 & \textbf{0.0322} \footnotesize($\nuhat$ = 1.2561)
\end{tabular}
\caption{Maximum absolute error \eqref{h1} for trigonometric series and Kriging interpolation. Different values for the parameter $\nu$ including a data-adaptive choice $\nuhat$ by means of maximum likelihood estimation (MLE) were considered for the Kriging interpolator.}\label{table:RSW}
\end{center}
\end{table}

\subsection{Sequential designs  in the RSW model}
\label{sec22} 
In a second series of experiments we test a sequential design strategy for the Kriging interpolator. Roughly speaking, this  procedure has two steps. A  certain number of experiments  (initial design) is conducted to calculate a first Kriging interpolator.
The remaining experiments are carried out according to a sequential design, reflecting the previous output of the interpolator and updating it, such that more interesting regions of the grain boundary energy space are better explored. The details of this sequential procedure are described in Section \ref{sec44}.

We assume that we have an overall budget of $\ntotal=  9,17,33,65$ observations. 
As  initial designs  we consider regular equidistant grids on the  interval $[0,\thetamax]$,
where the number of points of the initial design 
is $\ninit =  9,17,33$, and $65$ (note that by this choice the initial designs are nested, which makes them directly comparable).
We then conducted $\ntotal - \ninit$ further simulation experiments, where  the experimental conditions are chosen according to  the sequential design algorithm described in Section \ref{subs:sequential}.
For the smoothness parameter $\nu$ of the Kriging interpolator, we considered here only the value $\nu = 2.5$ which performed very well in the experiments with fixed equidistant start design (see~Table~\ref{table:RSW}).

\begin{figure}
     \centering
     \begin{subfigure}[t]{0.49\textwidth}
         \centering
         \caption{}\label{fig:RSW:dynamics}
         \includegraphics[width=\textwidth]{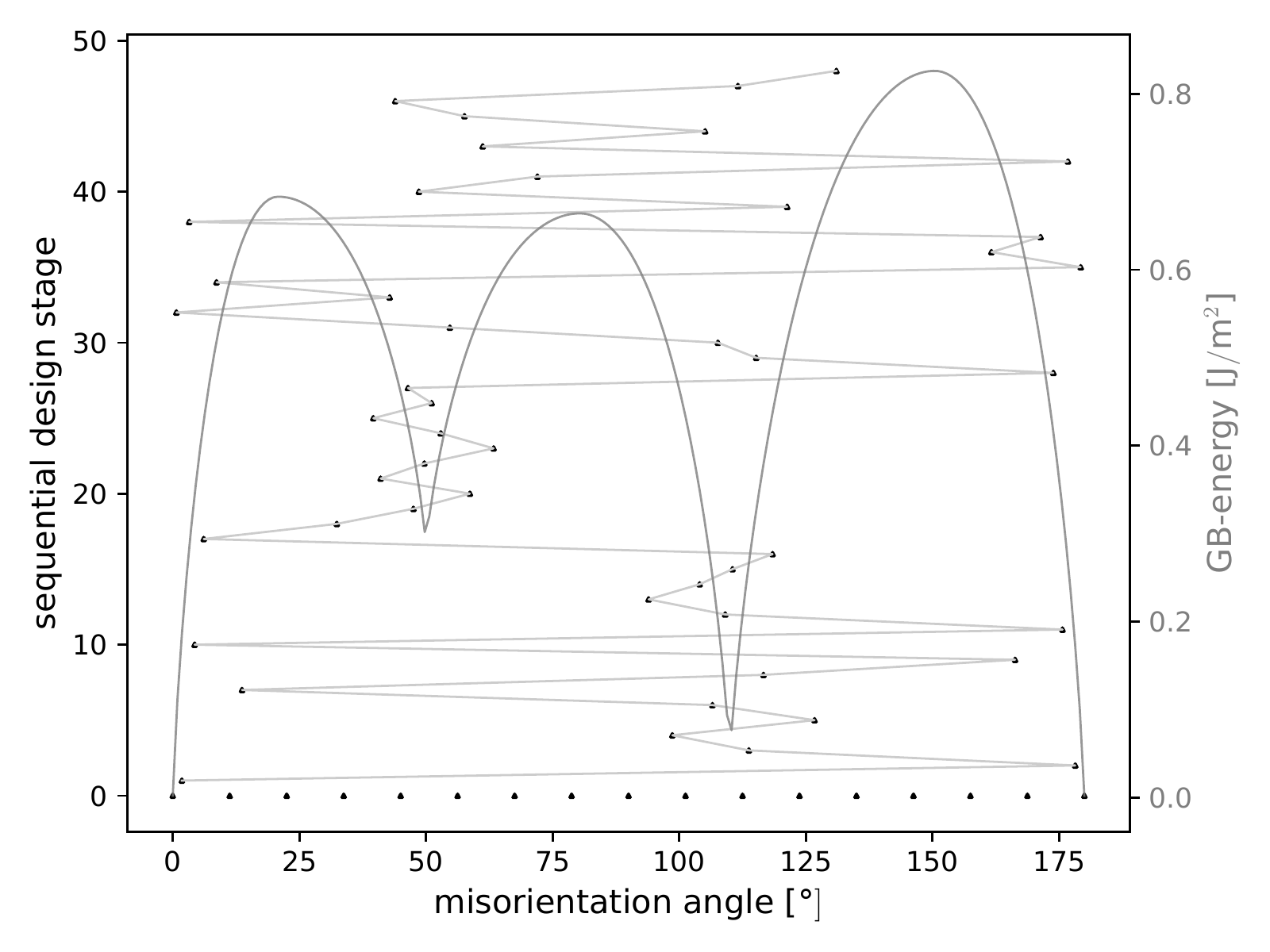}
     \end{subfigure}
     \begin{subfigure}[t]{0.49\textwidth}
         \centering
         \caption{}\label{fig:RSW:n_init_vs_n_total}
         \includegraphics[width=\textwidth]{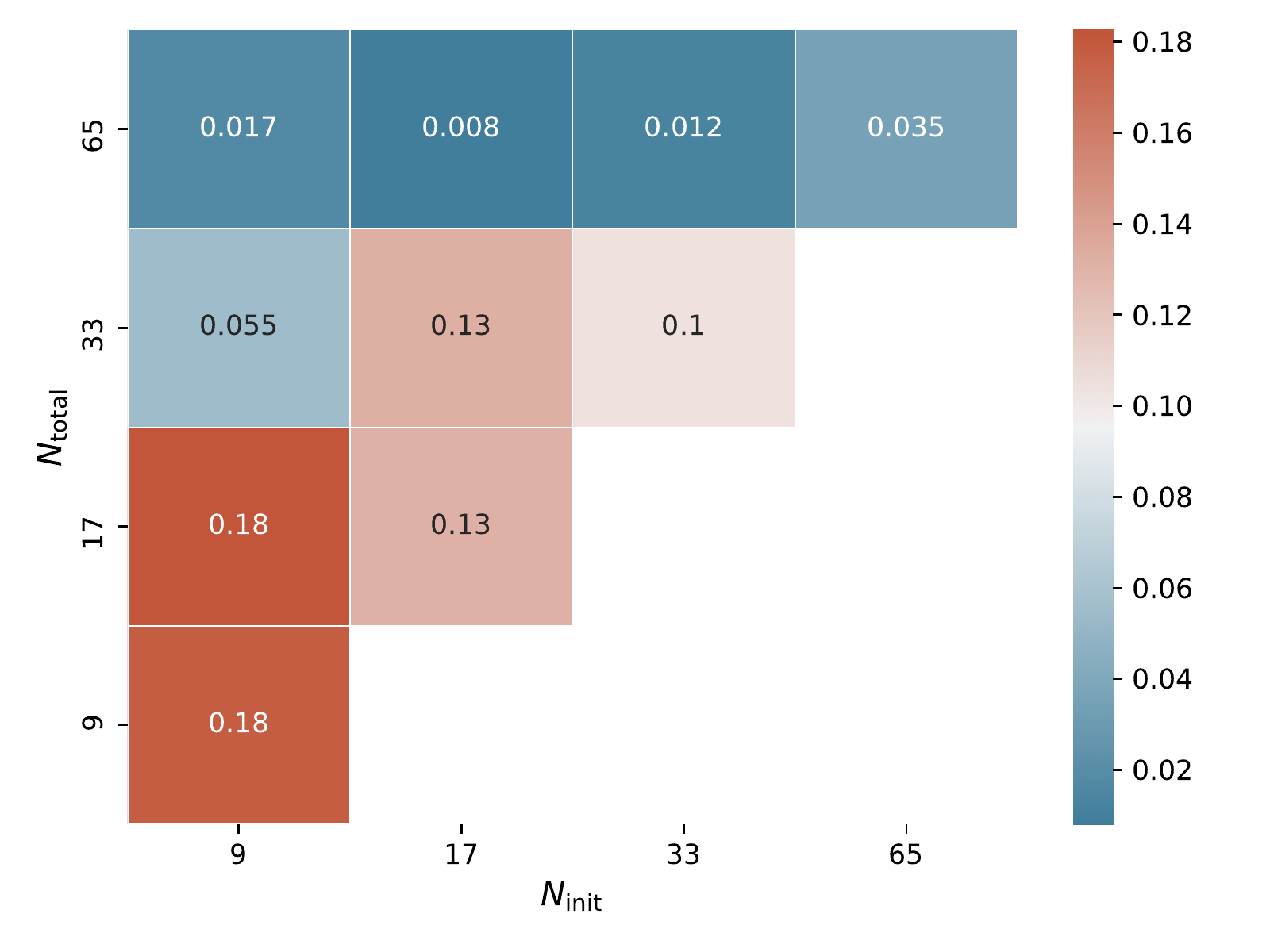}
     \end{subfigure}
    \caption{Left plot: dynamics of the sequential design algorithm: the $\theta$-value (angle) of the design points $\blacktriangle$ indicates the location in the interval $[0,\thetamax]$, the $y$-value the stage in the sequential design algorithm (stage $0$ corresponds to the initial design). The  $\FRSW$-function from which data are evaluated is plotted in gray. Right plot: maximum error of the Kriging interpolator in combination with the sequential design algorithm in dependence of the initial design size $\ninit$ and the total design size $\ntotal$.}\label{fig:plots:RSW:seq:2}
\end{figure}

Figure~\eqref{fig:RSW:dynamics} illustrates the dynamics of the sequential design algorithm.
It shows that the sequential design approach selects future sampling points alternately at different parts of the sampling domain $[0,\thetamax]$ with a high proportion of points close to the a priori unknown cusps.
A comparison of the prediction error of the Kriging interpolator of a non-sequential and sequential approach is summarized in Figure~\eqref{fig:RSW:n_init_vs_n_total}.
This figure shows  the  maximum error for different combinations of initial design sizes $\ninit$ and final design sizes $\ntotal$ after application of the sequential design algorithm (consequently, $\nseq = \ntotal - \ninit$ points were chosen by the sequential design strategy). For instance, the first line corresponds to
a total budget of $\ntotal=65$ experiments, with four entries 
 corresponding to the sample size $\ninit=9,$ $17$, $33$, and $65$ for the initial design (in the last case the are no sequential experiments). In this case the smallest error (namely $0.008$) is obtained for a sequential design with $\ninit=17$ equally spaced design points in the starting design and $\nseq = 48$ sequential observations. We observe that the sequential strategy can significantly outperform the non-sequential approach,
for which the maximum error is $4$ times larger.

\begin{figure}
     \centering
     \begin{subfigure}[b]{0.49\textwidth}
         \centering
\includegraphics[width=\textwidth]{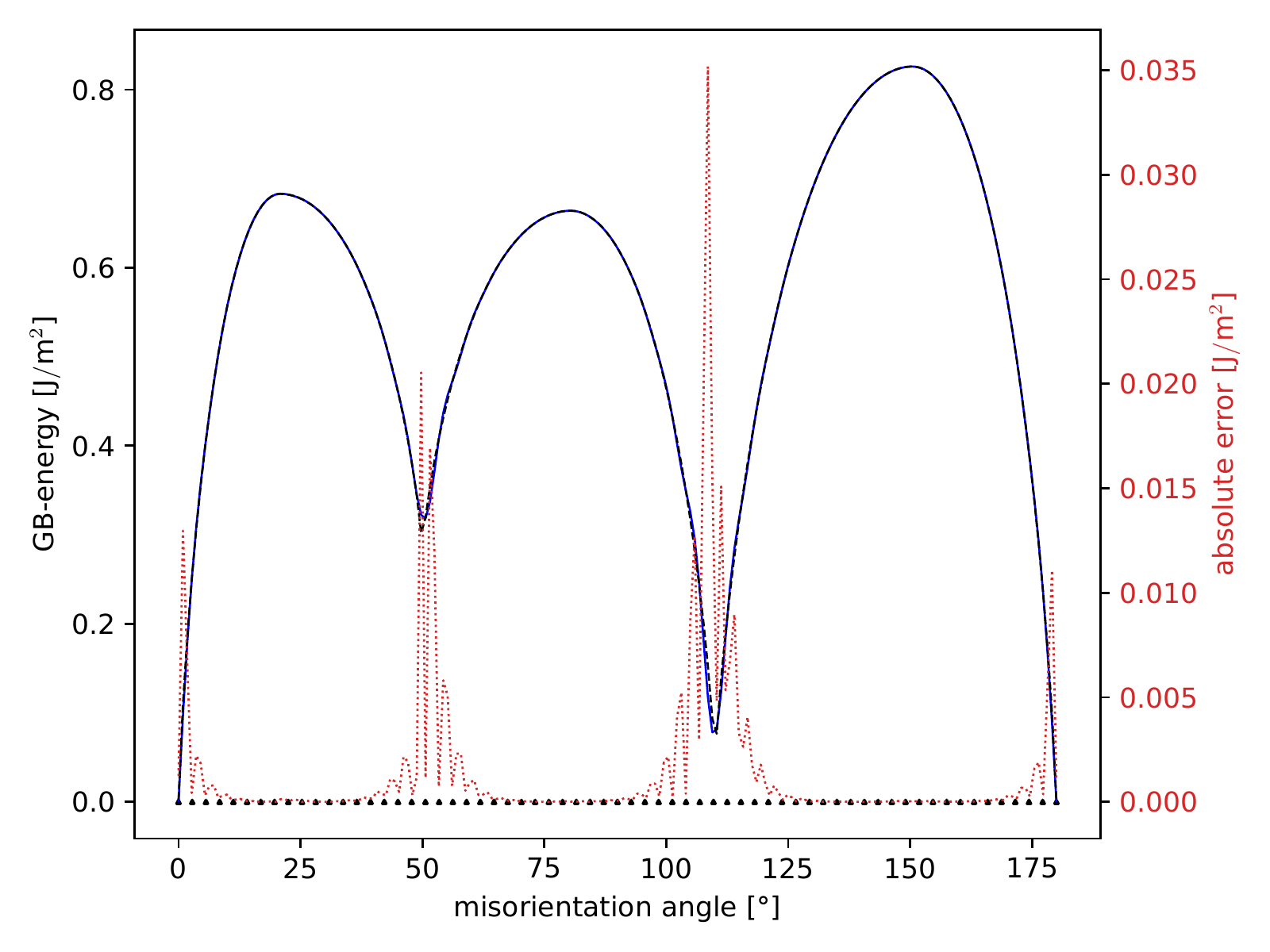}
     \end{subfigure}
     \begin{subfigure}[b]{0.49\textwidth}
         \centering
\includegraphics[width=\textwidth]{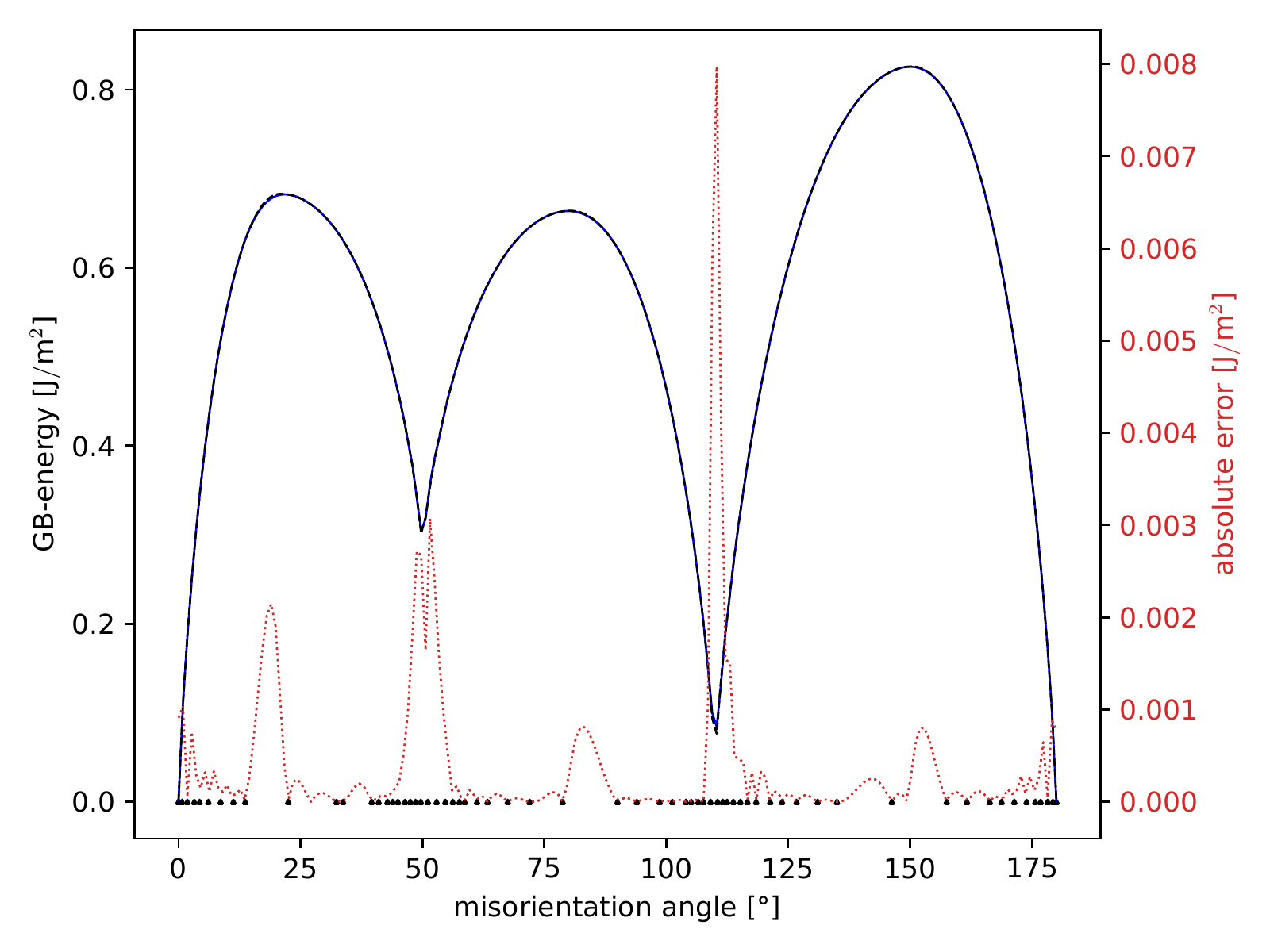}
     \end{subfigure}
    \caption{
    Kriging without and with sequential design strategy: the left plot shows the Kriging interpolator for a non-sequential design of $\ntotal = 65$ equidistant sampling points. The right plot shows the Kriging interpolator with a start design of $\ninit = 17$ sampling locations and $\nseq = 48$ sampling locations chosen by the the proposed sequential design algorithm. In both plots the interpolator is the blue solid line, the true energy function the black dashed line, and the red pointed line shows the absolute deviation between the two curves. Design points are indicated as $\blacktriangle$ on the $x$-axis.}
    \label{fig:plots:RSW:seq}
\end{figure}

In Figure~\ref{fig:plots:RSW:seq}
we display the non-sequential and sequential design, the resulting Kriging interpolator (blue line) together with the absolute error as a function of the misorientation angle.
We observe that the sequential algorithm fulfills our goal: It puts a significant number of sampling points (here indicated as $\blacktriangle$ on the $\theta$-axis) at regions of interest, given by the two cusps in this specific example. Note that the maximum error is attained at the right cusp (located at $\theta = 110\si{\degree}$) for both designs strategies, but the sequential design yields a substantially smaller error than the non-sequential design. The maximum error has decreased to $25\% $  of the original value by an application of the sequential design strategy.

We have also  compared the two design strategies for other subspaces of the DOF.
These results are presented in Appendix~\ref{app:subs:seq} and confirm the superiority of the sequential design approach. The sequential design strategy chooses more observations at locations where the function rises or descends steeply, or has a cusp-like structure.
Thus, for data simulated from the RSW model the proposed sequential design strategy yields substantial improvements in the performance of the Kriging interpolator in contrast to a regular design grid consisting only of equally spaced sampling locations. 
In the following section we will continue these investigations and study the methodology for real data from atomistic simulations.

\subsection{Sequential designs in atomistic simulations}
\label{sec23}

So far we showed that Kriging outperforms series interpolation and additionally that  Kriging can be  further improved by sequential sampling techniques. These  observations were made on the basis of  data which  was  generated from  the 
RSW model.
In this section we apply this methodology towards the sampling of grain boundary energy subspaces with atomistic simulations and demonstrate that, again, the sequential sampling technique outperforms a regular sampling technique with respect  the accuracy of prediction and the ability of 
sampling at regions close to the (unknown) cusps.

\label{sec:Atomoistic_results}
\begin{figure}
     \centering
     \begin{subfigure}{0.49\textwidth}
         \caption{}\label{fig:atom:data:100}
         \centering
         \includegraphics[width=\textwidth]{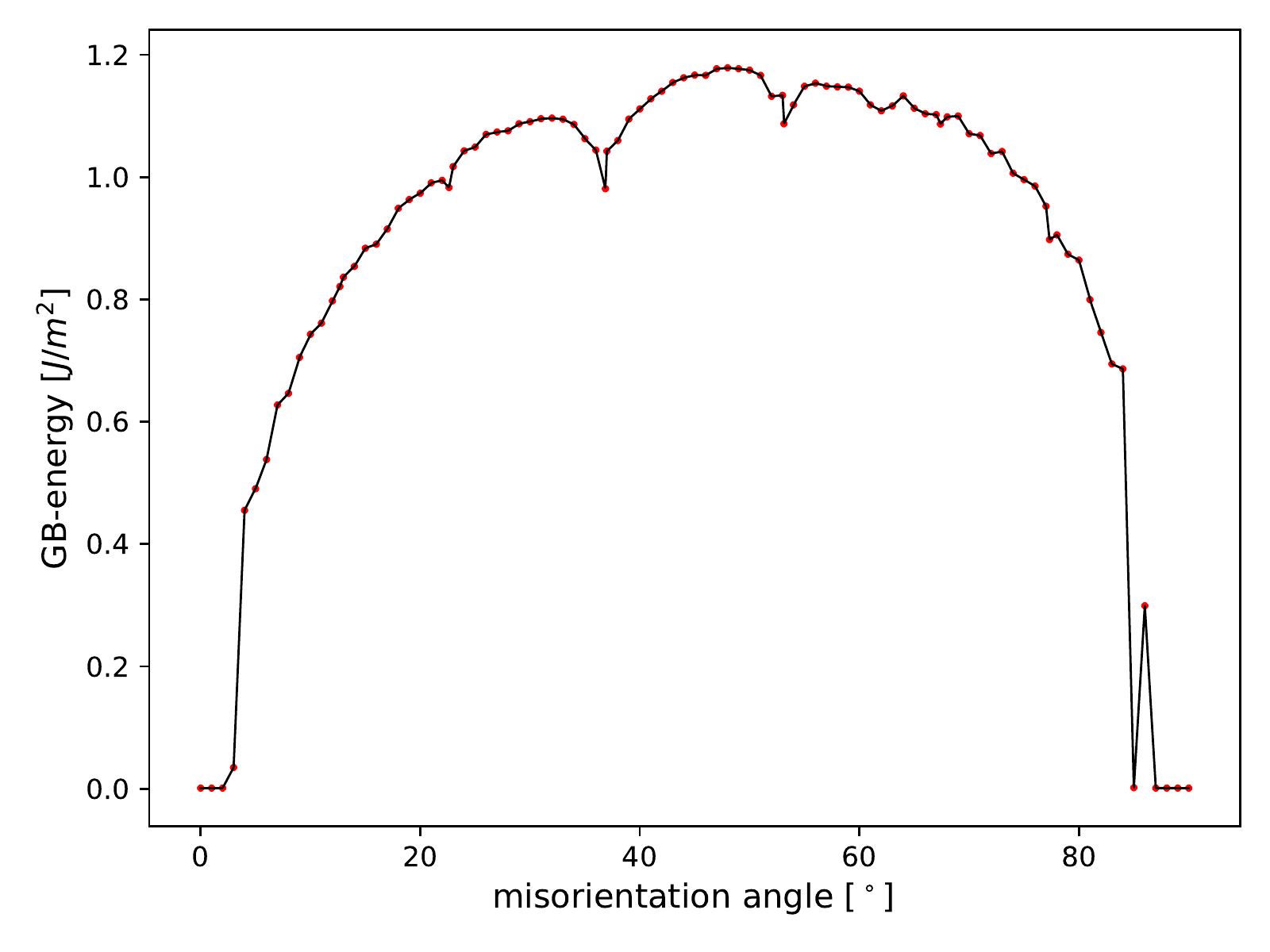}
\end{subfigure}
     \begin{subfigure}{0.49\textwidth}
         \caption{}
         \centering
         \includegraphics[width=\textwidth]{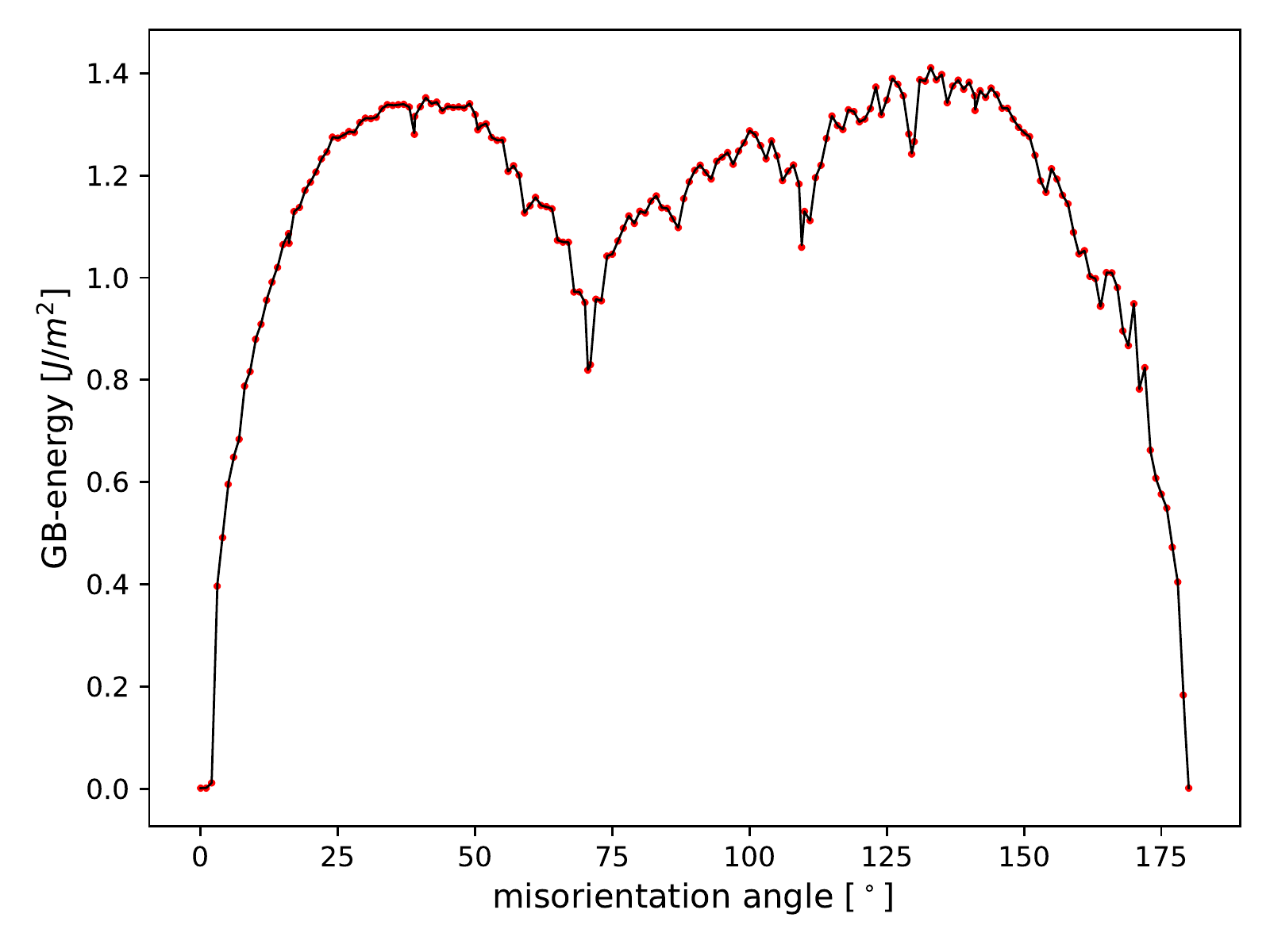}
\end{subfigure}
          \begin{subfigure}{0.49\textwidth}
         \caption{}\label{fig:atom:data:111}
         \centering
         \includegraphics[width=\textwidth]{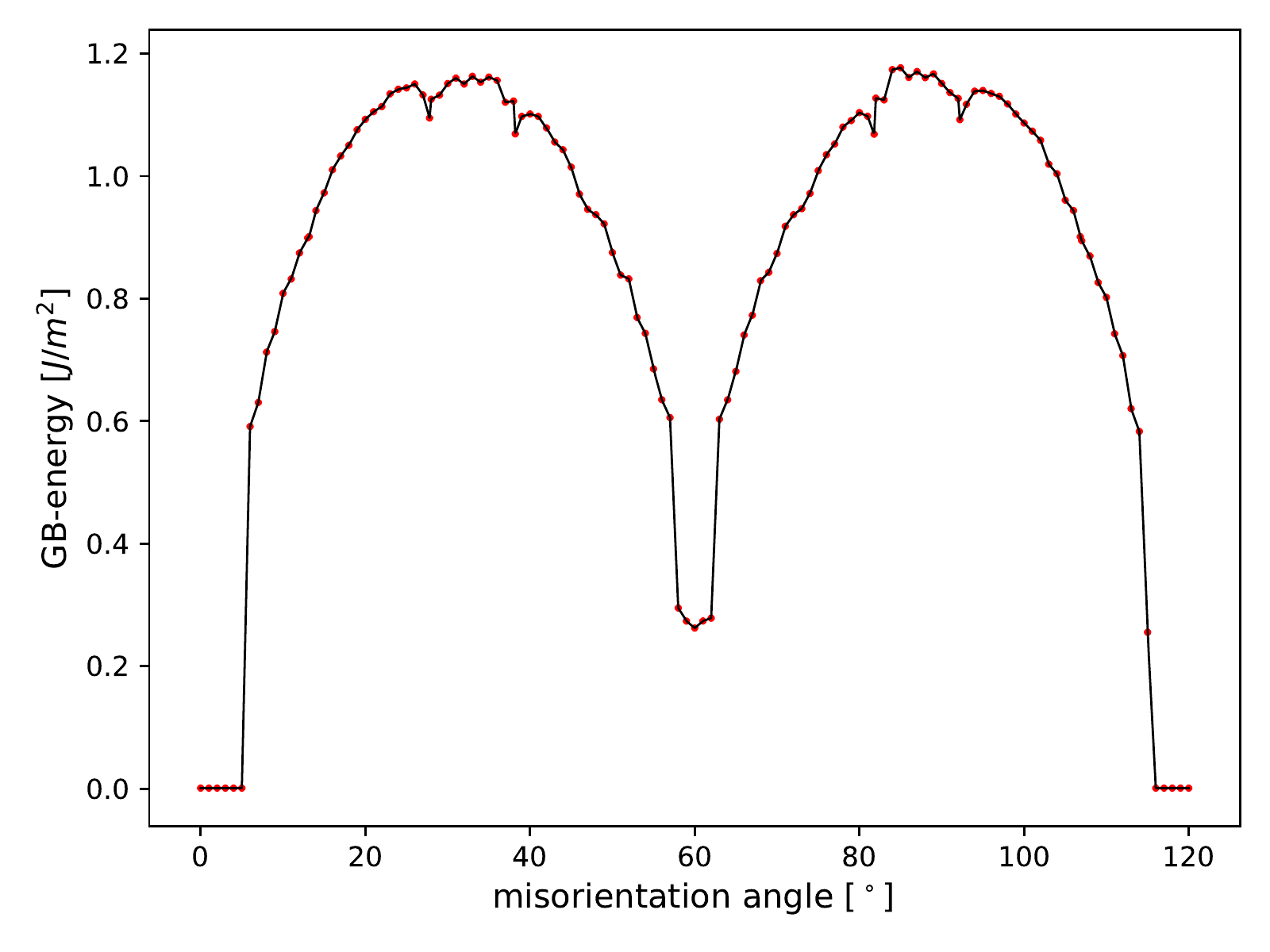}
\end{subfigure}
    \caption{Atomistic data (\textcolor{red}{\textbullet}) with $N_{\mathrm{ref}}$ simulated grain boundaries (including both the equidistant grain boundaries and the additional special $\Sigma$ grain boundaries) and linear interpolation 
(\textbf{\textcolor{black}{solid lines}}) 
  of the reference database 
for different subspaces: (a) $[100]$  ($N_{\mathrm{ref}} = 97$), (b) $[110]$ ($N_{\mathrm{ref}}= 187$),  (c) $[111]$ ($N_{\mathrm{ref}}= 125$).}
    \label{fig:plot:reference_database}
\end{figure}

\subsubsection{Reference data}\label{sec:reference-data}
As in the previous section, we want to evaluate the error which occurs due to an incomplete sampling of the parameter space. Instead of using the approximate, analytical RSW function for reference, we now generate a dense data set of energies from atomistic simulations, consisting of $N_{\mathrm{regular}}$ grain boundaries plus additional special $\Sigma$ grain boundaries (see Section \ref{sec:Data_generation}), and interpolate between them.
The atomistic reference data for the one-dimensional subspaces of $[100]$, $[110]$, and $[111]$ symmetric tilt grain boundaries (STGBs) is displayed in Figure \ref{fig:plot:reference_database}, together with the linear interpolation as black lines. It shows 
 that the interpolated function is much rougher than the idealized RSW function, which has been considered in Section \ref{sec22}. The rougher energy landscape makes the identification of the energy cusps more challenging and requires a careful selection of the design points and interpolation parameters.
 
Note that Figures~\eqref{fig:atom:data:100} and \eqref{fig:atom:data:111} seemingly show some outliers at angles close to $0\si{\degree}$ and $\thetamax$, which are equal to 0 $\si{\joule}/\si{\metre}^2$. These are data points for which the very small deviation from low-energy, stable structures leads to a relaxation of the atomic positions to these structures at the nearby angles.
In a real application, this problem might be avoided by increasing the size of the atomistic model, but for our proof of principle we simply ignore these data points.

In the following, we first compare non-sequential and sequential design strategies for a fixed smoothing parameter $\nu=0.5$. Next we study the impact of the choice of this parameter on the quality of the Kriging estimator and its ability to detect cusps.

\subsubsection{Effect of sequential design}
\begin{figure}
     \centering
     \begin{subfigure}{0.49\textwidth}
         \caption{}
         \centering
         \includegraphics[width=\textwidth]{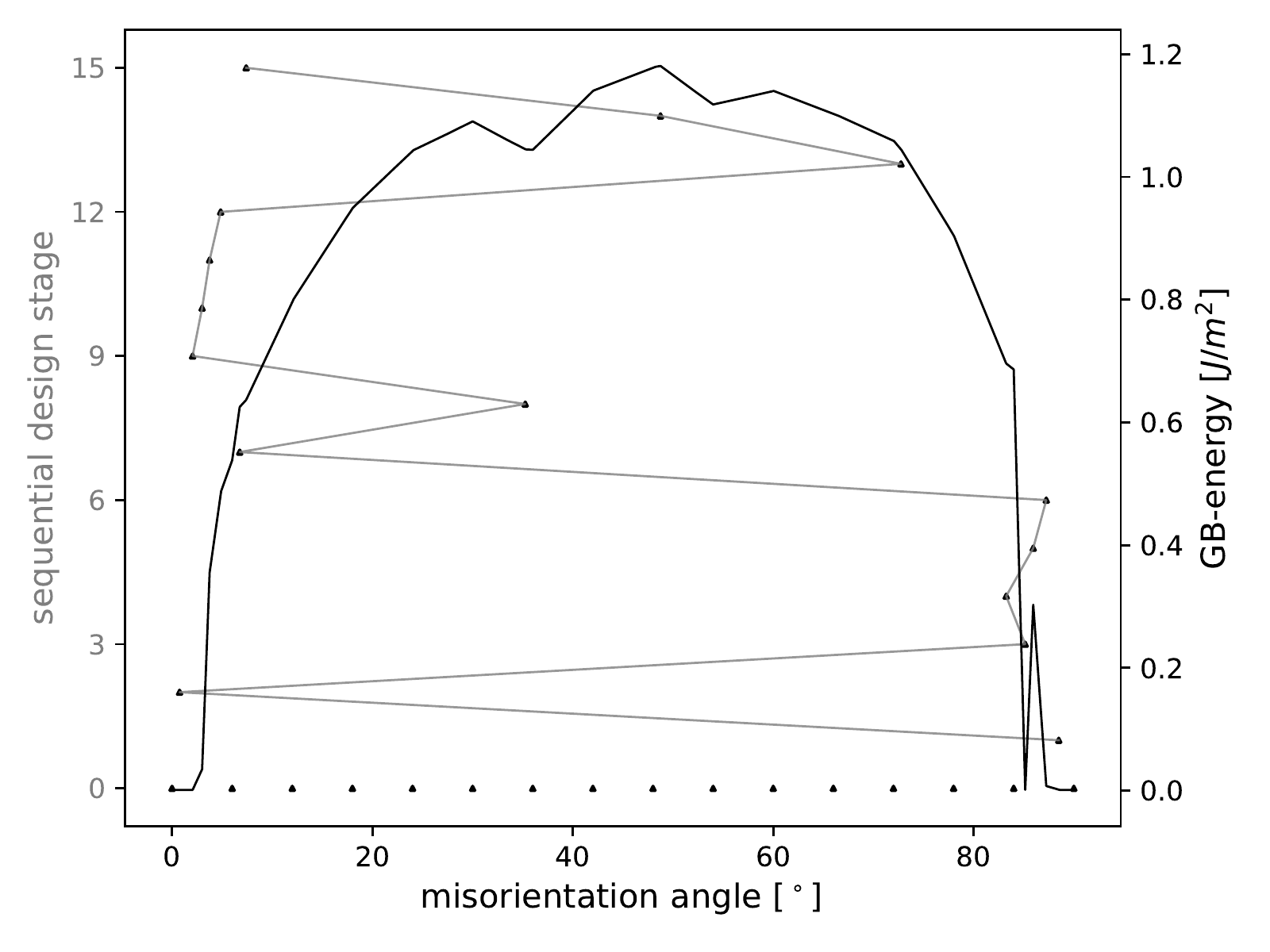}
\end{subfigure}
     \begin{subfigure}{0.49\textwidth}
         \caption{}
         \centering
         \includegraphics[width=\textwidth]{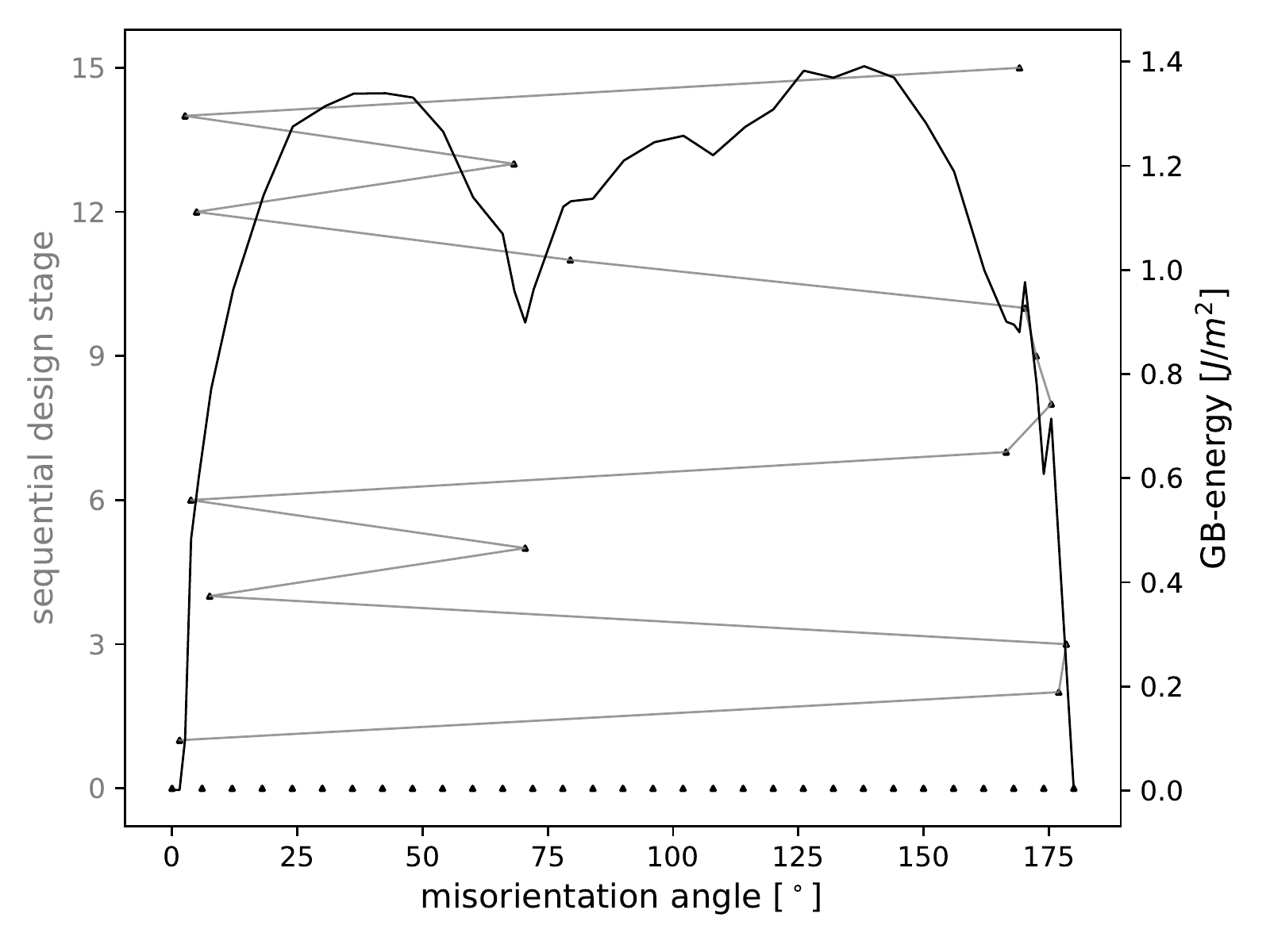}
\end{subfigure}
          \begin{subfigure}{0.49\textwidth}
         \caption{}
         \centering
         \includegraphics[width=\textwidth]{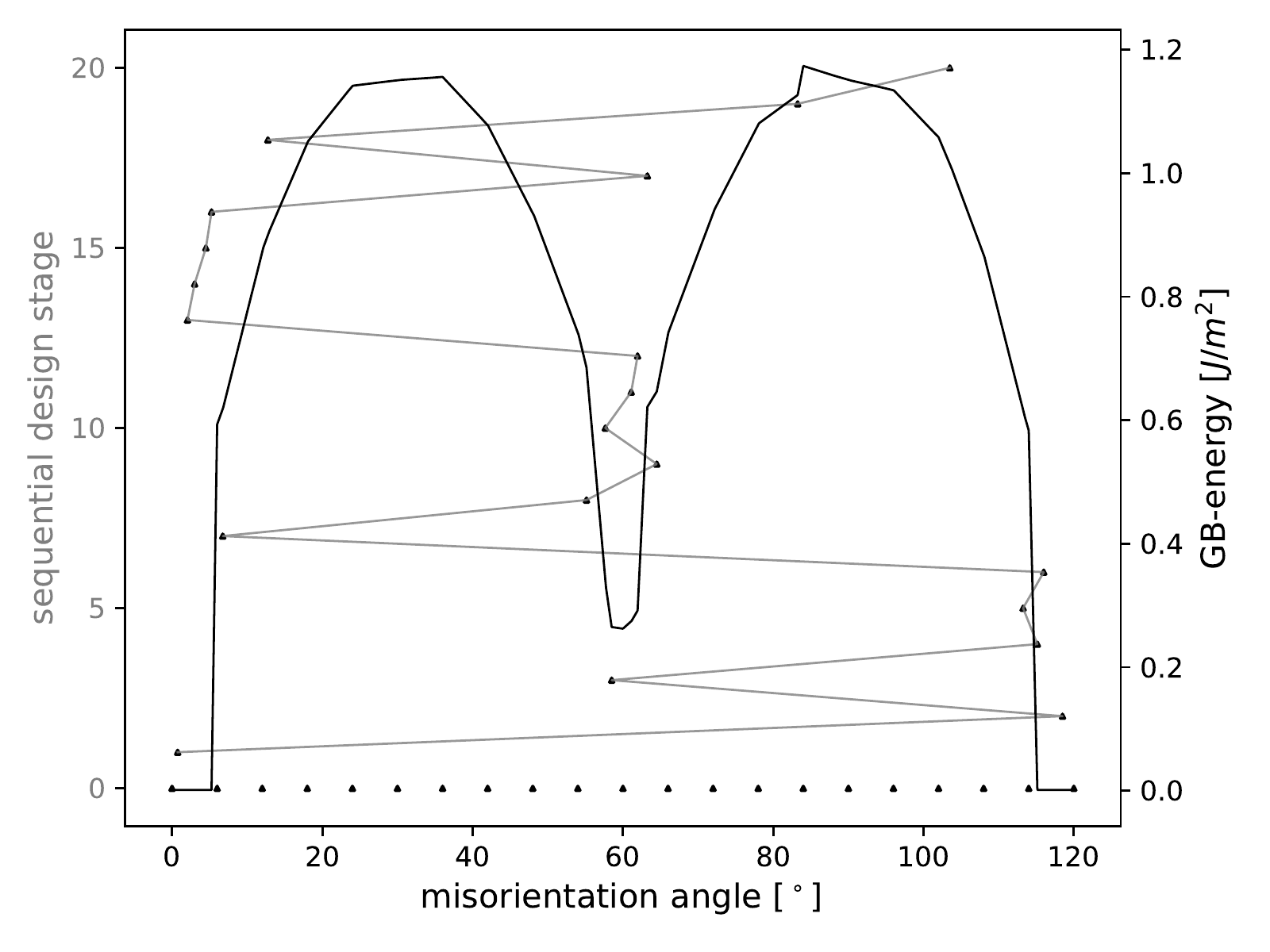}
\end{subfigure}
    \caption{Dynamics of the sequential design algorithm for exploring  different  STGB  subspaces with a
    Kriging interpolator ($\nu = 0.5$). 
    (a): $[100]$ ($15$ sequential points),
 (b): $[110]$ ($15$ sequential points),
 (c): $[111]$ ($20$ sequential points).
The positions of the design points (misorientation angle $\theta$) in the subspace are indicated by a $\blacktriangle$. The left $y$-value represents the stage in the sequential design algorithm (stage $0$ corresponds to the initial design), the right $y$-value the grain boundary energy in $\si{\joule}/\si{\metre}^2$ calculated by the Kriging interpolator.}
    \label{fig:plots:dynamic_plots_constant_nu_hat_0_5}
\end{figure}
In this subsection we focus on exploring  the $[100]$, $[110]$, and $[111]$ STGB subspaces by a Kriging interpolator with a fixed 
smoothing parameter $\nu=0.5$. Corresponding results for
$\nu = 1.5$, $2.5$, and the maximum likelihood estimator
 $\nuhat$  updated in every step of the sequential algorithm can be found in Appendix \ref{sec:appendix_progress_sequential_point_calc}. 
  We considered the three misorientation subspaces from $0\si{\degree}$ to $\thetamax$, with $\thetamax=90\si{\degree}$ for the $[100]$ subspace, $\thetamax=180\si{\degree}$ for the $[110]$ subspace, and $\thetamax=120\si{\degree}$ for the $[111]$ subspace.
  An initial design with $\ninit$ grain boundaries (16 for the $[100]$, 31 for the $[110]$, and 21 for the $[111]$ subspace) and a sequential design with $\nseq$ grain boundaries (15 for the $[100]$ and $[110]$ subspaces and 20 for the $[111]$ subspace) were used for the exploration of the subspaces. The resulting energies predicted by the Kriging interpolator
  are displayed in Figure \ref{fig:plots:dynamic_plots_constant_nu_hat_0_5} as black lines, and the values are given on the right $y$-axis. The left $y$-axis indicates the  stage of the sequential design, which places  the design points  at specific misorientation angles.
 
We observe that most of the experiments in the sequential designs 
are conducted in neighborhoods of the cusps. This is remarkable, because the starting designs contain no misorientation angles corresponding to cusps.
In particular, regions with large local energy gradients of the subspace are sampled in more detail by the sequential design. 
Compared to regular sampling, which does not guarantee the discovery of an unknown cusp of the subspace,  this is an  important advantage 
of the sequential approach.
Note also that the sequential design does not access the cusps one by one in full detail, but visits them all repeatedly, step by step. It thus  reduces the maximum error, which is usually located in the vicinity of the cusps. In this way, nearly every cusp of the STGB subspaces can be found within only a few iterations.

\begin{table}
\begin{center}
\begin{tabular}{c|c|c|c|c}
subspace & $N_{\mathrm{total}}$ & $N_{\mathrm{init}}$ & $N_{\mathrm{seq}}$ & max. abs. error $[\si{\joule}/\si{\metre}^2]$\\
 \hline
 \multirow{ 2}{*}{$[100]$} & \multirow{ 2}{*}{31} & 31 & 0 & 0.4560\\
  &  & 16 & 15 & \bfseries 0.0888\\
\hline
\multirow{ 2}{*}{$[110]$} & \multirow{ 2}{*}{46} & 46 & 0 & 0.2349\\
 & & 31 & 15 & \bf 0.1699\\
\hline
\multirow{2}{*}{$[111]$} & \multirow{ 2}{*}{41} & 41 & 0 & 0.3934\\
 &  & 21 & 20 & \bf 0.2161\\
\end{tabular}
\caption{Comparison of the maximum absolute error \eqref{h1} of the Kriging interpolator for purely regular sampling ($\ntotal = \ninit$, $\nseq=0$) and for sequential sampling ($\ntotal = \ninit + \nseq$ with $\nseq >  0$) with an initial design of size 
$N_{\mathrm{init}}$ and $N_{\mathrm{seq}}$ further grain boundaries chosen by the sequential design.}\label{table:maxerror}
\end{center}
\end{table}

\begin{figure}[H]
     \centering
     \begin{subfigure}{0.95\textwidth}
         \centering
         \includegraphics[width=\textwidth]{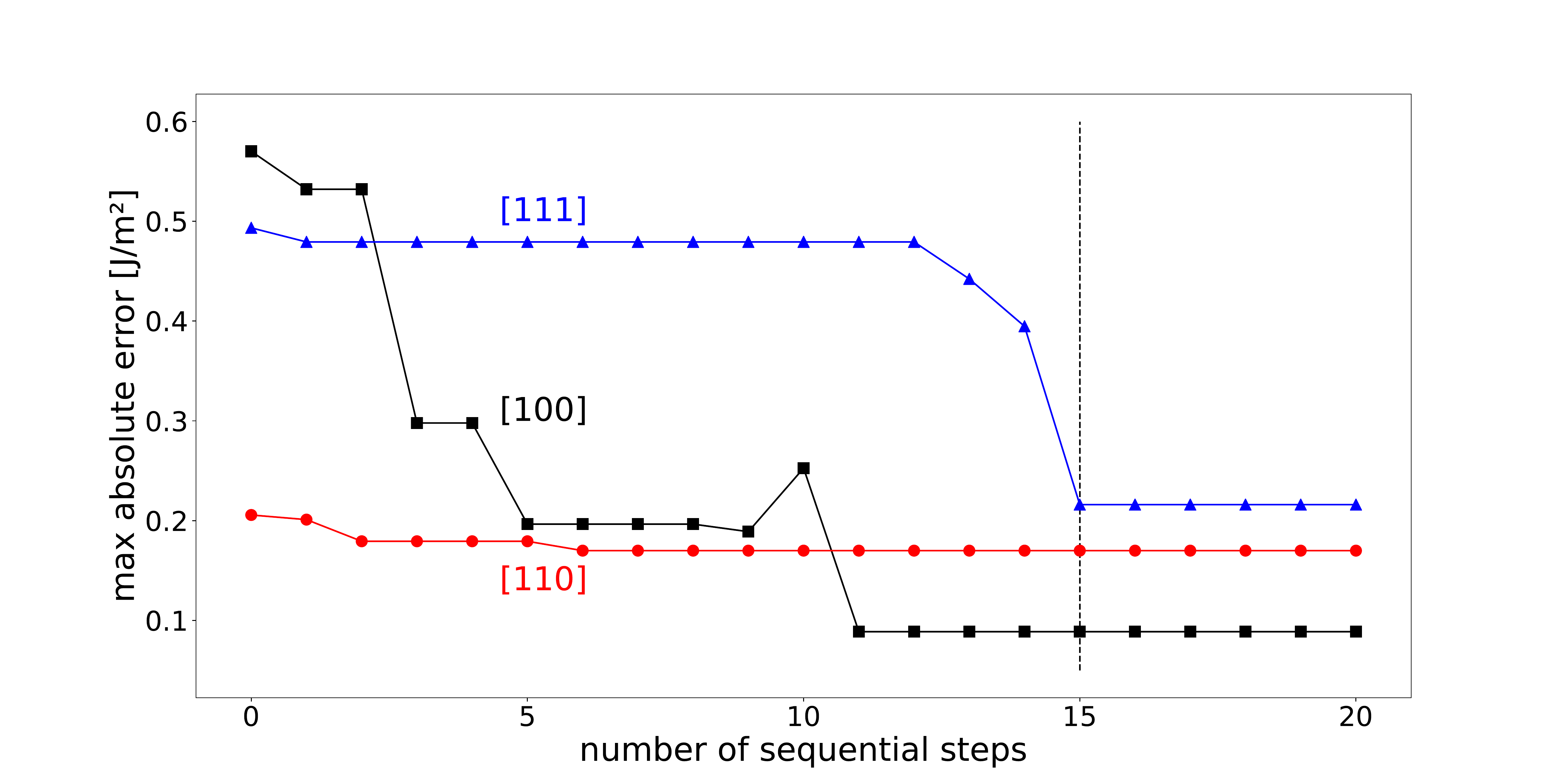}
\end{subfigure}
    \caption{Maximum absolute prediction error of the Kriging interpolator (with $\nu=0.5$)
    for $20$ steps of the sequential algorithm.
    Subspaces: 
    $[100]$ ($\blacksquare$), 
    $[110]$ (\textcolor{red}{\textbullet}) and
    $[111]$ (\textcolor{blue}{$\blacktriangle$}).}
    \label{fig:plots:max_error_nu_hat_0_5}
\end{figure}

In Table \ref{table:maxerror} we display the maximum error of the Kriging interpolator based on the two sampling schemes. Compared to the regular sampling technique
the sequential approach leads to a decrease of the maximum error from $0.4560$ $\si{\joule}/\si{\metre}^2$ to $0.0888$ $\si{\joule}/\si{\metre}^2$ for the $[100]$ subspace, from 0.2349 $\si{\joule}/\si{\metre}^2$ to 0.2161 $\si{\joule}/\si{\metre}^2$ for the $[110]$ subspace, and from 0.2349 $\si{\joule}/\si{\metre}^2$ to 0.1699 $\si{\joule}/\si{\metre}^2$ for the $[111]$ subspace.

An interesting question in this context is whether the error of the sequential design strategy always converges to a minimum value with increasing sample size. For large smoothness parameters $\nu$, an overfitting of the data, and hence no convergence, can be expected. For small $\nu$, however, adding data should always  improve the fit. This is demonstrated in Figure \ref{fig:plots:max_error_nu_hat_0_5}, which 
shows the maximum error of the Kriging interpolator as a function of the sample size of the sequential design. For example, the black  curve
corresponds to the $[100]$ subspace, where the sample size of the initial design is $N_{\mathrm{init}} = 16$. The line shows the maximum error of prediction by the Kriging interpolator for the next $20$ sequentially chosen points. We observe that the error decreases rapidly, because the sequential design allocates design points in  neighborhoods of the cusps. In  particular, no further improvement is obtained,
after $11$ sequential experiments have been conducted, and as consequence atomistic simulations could be stopped at this point in a real experiment. 
For the two other subspaces we observe a similar qualitative behavior. 
\subsubsection{Effect of the smoothing parameter \texorpdfstring{$\nu$}{nu} in atomistic simulations}

\begin{figure}
     \centering
     \begin{subfigure}{0.49\textwidth}
         \centering
         \caption{}
         \includegraphics[width=\textwidth]{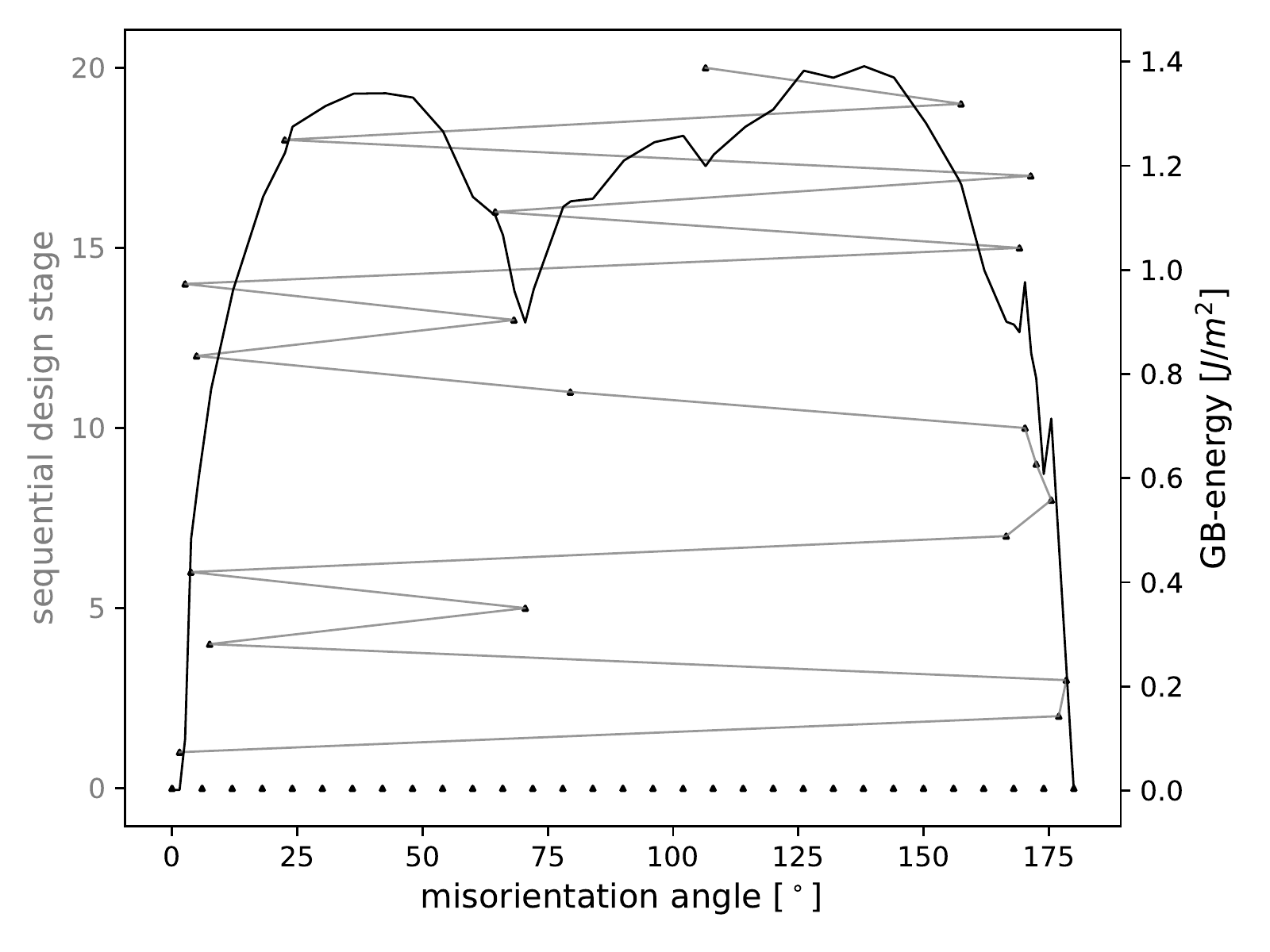}
\end{subfigure}
     \begin{subfigure}{0.49\textwidth}
         \centering
         \caption{}
         \includegraphics[width=\textwidth]{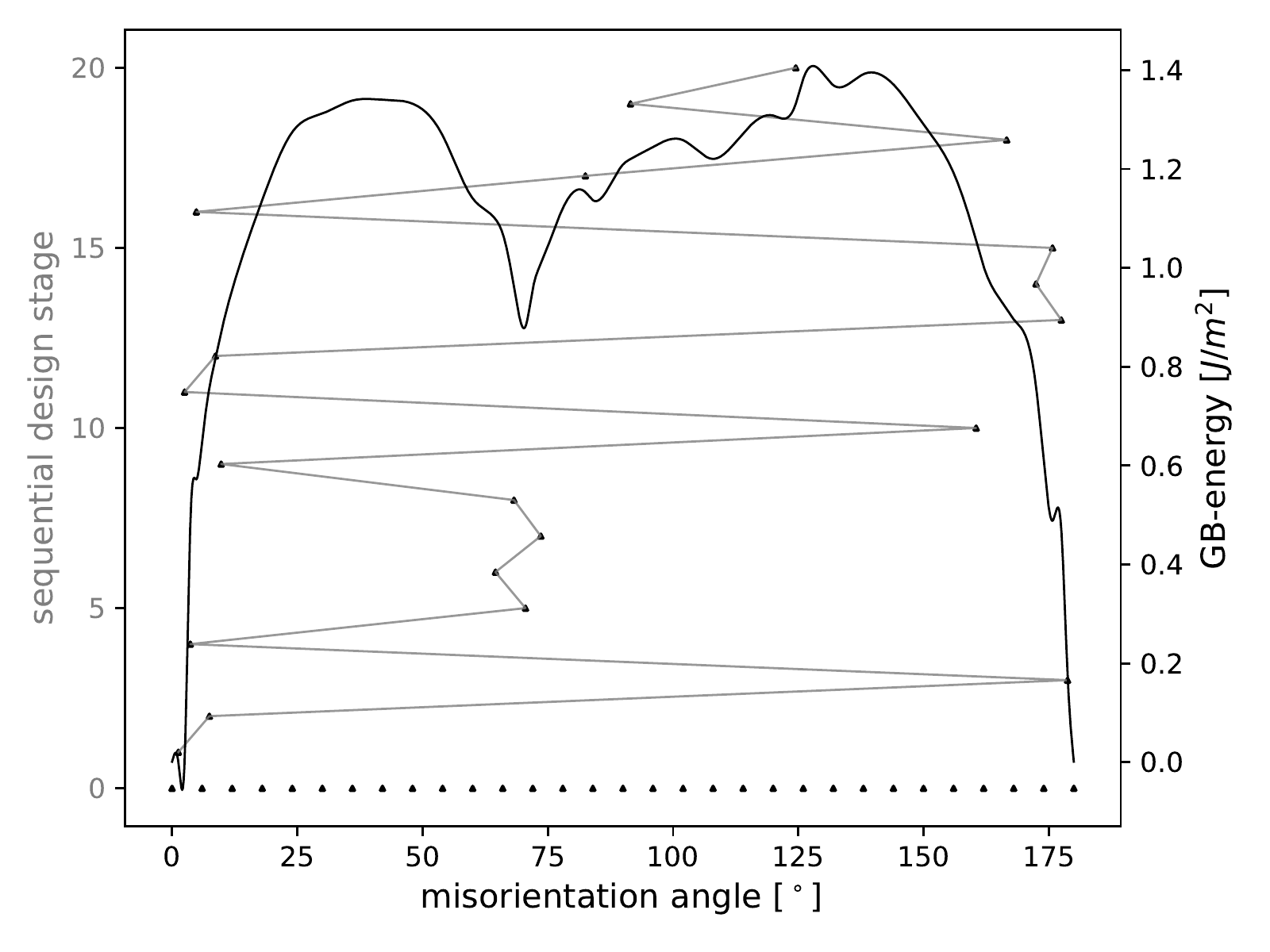}
\end{subfigure}
     \begin{subfigure}{0.49\textwidth}
         \centering
         \caption{}
         \includegraphics[width=\textwidth]{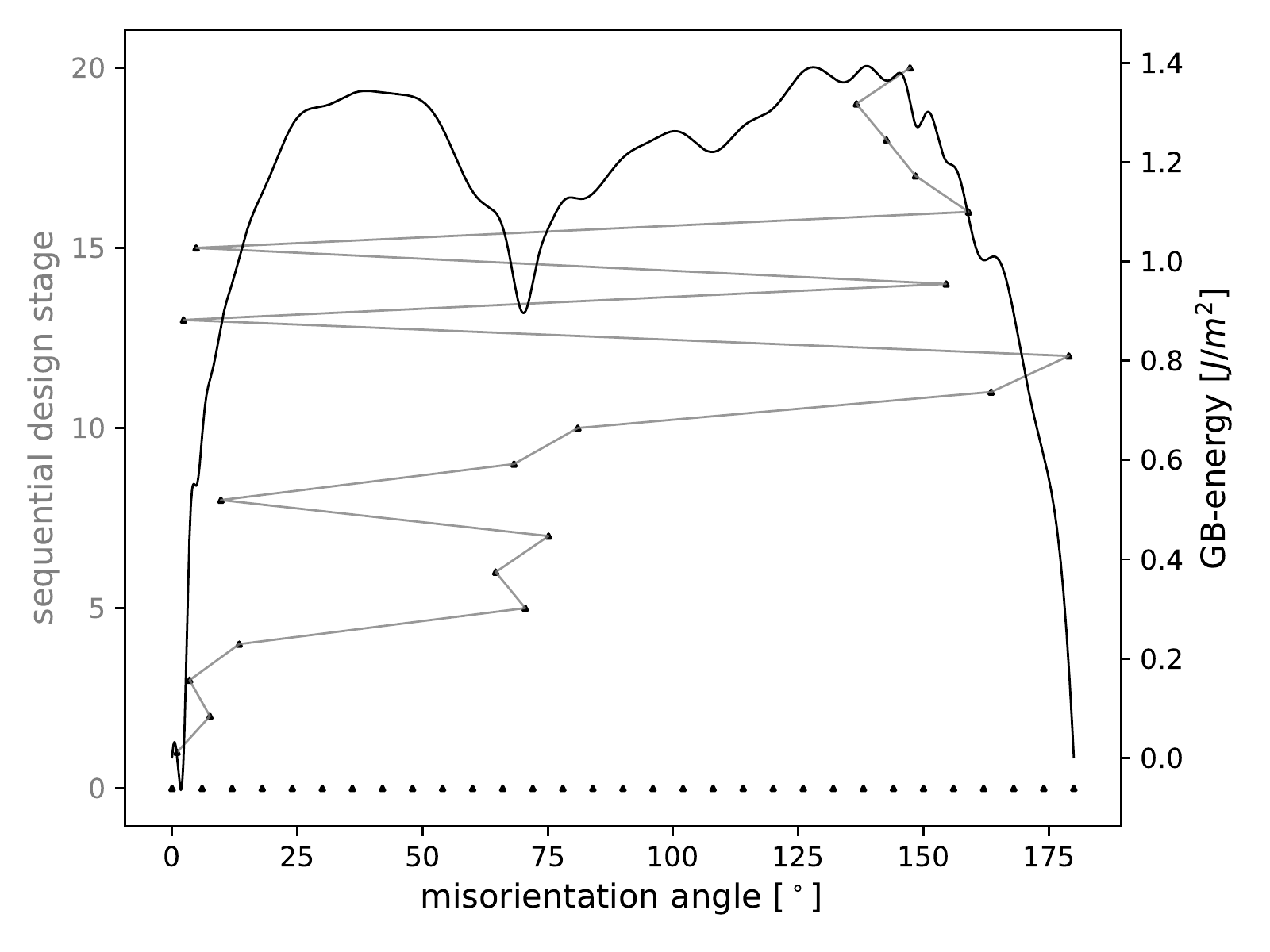}
\end{subfigure}
          \begin{subfigure}{0.49\textwidth}
         \centering
         \caption{}
         \includegraphics[width=\textwidth]{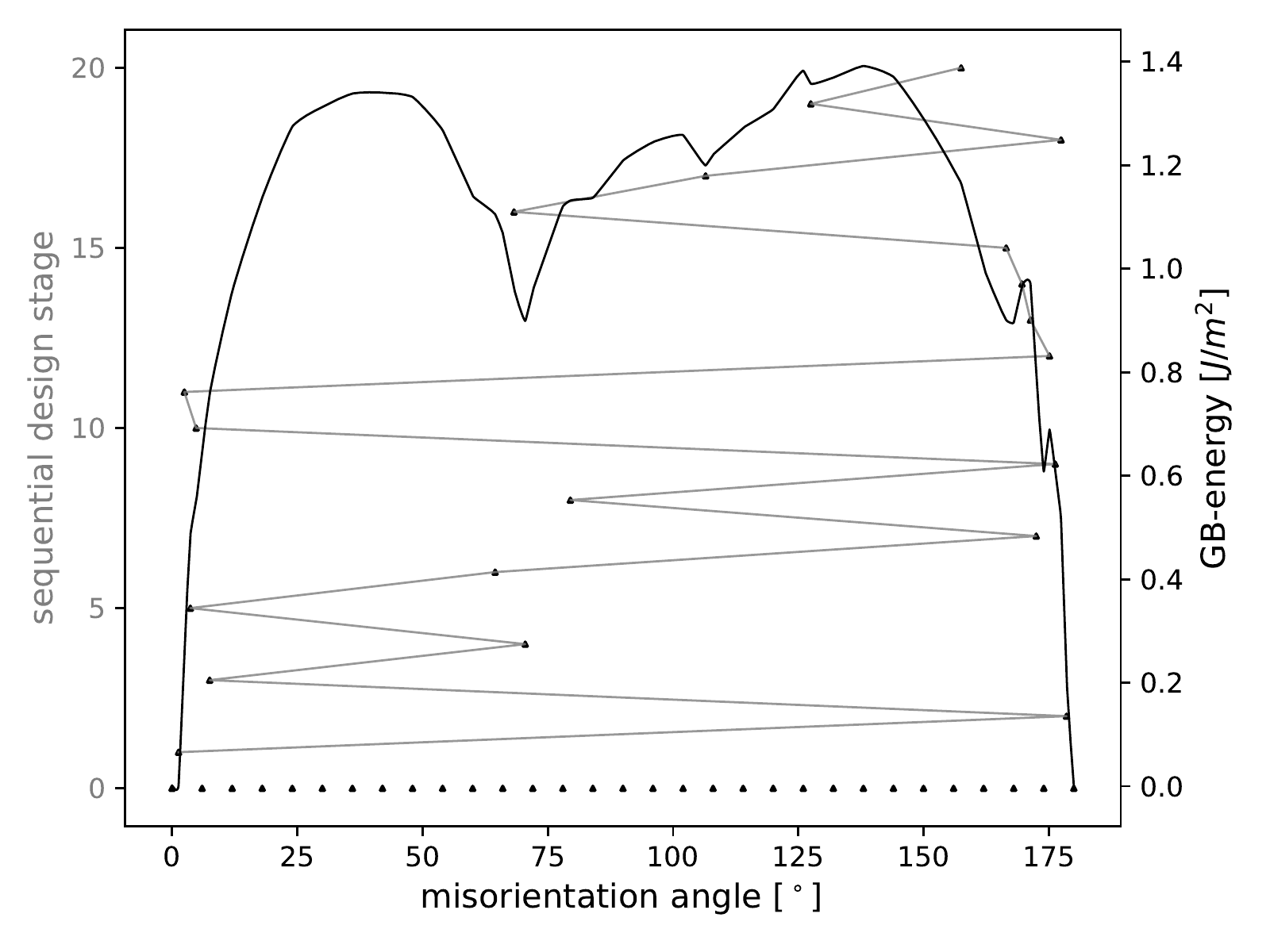}
\end{subfigure}
    \caption{Dynamics of the sequential design 
algorithm 
($20$ sequential design points)
for exploring the $[110]$ STGB subspace by a Kriging interpolator with different smoothing parameters.
    (a): $\nu=0.5$, (b): $\nu=1.5$, (c): $\nu=2.5$ and (d) $\nuhat$ calculated by MLE. The position of the design points (misorientation angle $\theta$) in the interval $[0\si{\degree},180\si{\degree}]$ is given by a $\blacktriangle$. The left $y$-value indicates the stage in the sequential design algorithm (stage $0$ corresponds to the initial design), the right $y$-value the grain boundary energy in $\si{\joule}/\si{\metre}^2$ calculated by the Kriging interpolator. 
    }
\label{fig:plots:dynamic_plots_110}
\end{figure}

In this subsection we briefly investigate the sensitivity of the method with respect to the choice of the smoothing  parameter $\nu$ in the Kriging interpolator, 
where  
we focus on the $[110]$ subspace.  Corresponding results for  the two other subspaces
show a similar picture and
can be found in 
Appendix~\ref{sec:appendix_progress_sequential_point_calc}.
In Figure~\ref{fig:plots:dynamic_plots_110} we display the dynamics of 
the sequential design and the resulting Kriging interpolator for different values of
$\nu = 0.5, 1.5, 2.5$, and $\nuhat$ calculated in an adaptive way (maximum likelihood estimation).
We observe  that  the  sequential design  strategy chooses more points in neighborhoods of the cusps, independently of the choice of $\nu$, but the resolution of the sampling depends on the choice of $\nu$. 
While for $\nu$ = 0.5 the  sequential design allocates at least one point to  every cusp of the subspace, the design points of the sequential design for $\nu$ = 2.5 remain close to the three deepest cusps  at $0\si{\degree}$, $70.53\si{\degree}$ ($\Sigma$3) and $180\si{\degree}$.
Sequential sampling of the $[110]$ STGB subspace with a data adaptive choice of smoothing parameter 
(in the interval $[0.5, 2.5]$), which is 
 updated in each  iteration, 
showed the following behaviour: the interpolation started with $\nuhat\approx 1.5$ for the initial design. Adding sequential points then caused an adjustment of $\nuhat$ towards smaller values, finally varying between $0.6$ and $0.65$. This means that an increasing number of 
sequentially chosen data points leads to a Kriging interpolator that is close to linear interpolation.
In contrast to the data generated from the RSW model, where  the choice  $\nu = 2.5$ for the smoothing parameter
yields the best performance of the Kriging interpolator, the results in this section demonstrate that for atomistic data
small values of $\nu$ should be preferred. 
This is up to what should be expected as
atomistic simulations produce much rougher curves
than the idealized RSW model and as a consequence the
parameter $\nu$ should be chosen much smaller for this type of data.
\section{Discussion } 
The numerical results for the RSW model demonstrate  that the Kriging interpolator (with Mat\'ern kernel) has the capacity to outperform the interpolator based on   series estimation  in a wide range of scenarios.
The suggested sequential design algorithm can even further improve the performance of the Kriging interpolator due to its tendency to prefer observations at locations where the target energy function has higher fluctuations and is more difficult to estimate.
Our  results for atomistic simulations of one dimensional STGB subspaces
demonstrate  that sequential sampling techniques can lead to substantial improvements in the prediction of grain boundary energies.
 In particular,  grain boundaries chosen by the sequential design are  located close to cusps of the unknown energy functions, which are 
 the most important areas to sample due to their high energy gradient.
 Moreover, the proposed sequential design strategy chooses grain boundaries for the atomistic 
 simulation  by  jumping between different cusps 
 and is therefore able to sample grain boundaries at multiple cusps simultaneously. 
Consequently, the sequential sampling technique can discover unknown cusps, which opens the opportunity to sample even more complex subspaces.
The accuracy of the Kriging interpolator  
and the number of iterations required to find the cusps depend on the choice of a smoothness parameter 
but estimating this parameter via a maximum likelihood approach yields very satisfying results. Moreover,
for atomistic simulations smaller values of the smoothness parameter  close  to $0.5$ can be used as well. 
In this case the maximum error  decreases with an increasing 
number of  sequential steps and  converges after a few  sequential iterations.

Due to the different dimensions of the subspaces and datasets, our new sequential design strategy and Kriging interpolator cannot be directly compared with previous sampling methods \cite{kim2011identification,restrepo2014artificial}.
However, the obvious advantage is the ability of our method to identify the energy cusps. In \cite{kim2011identification} a priori knowledge of their location is needed for a meaningful interpolation between the data, and also a satisfying training of an ANN \cite{restrepo2014artificial} requires the inclusion of at least some of the special grain boundaries in the training data. In contrast to these approaches we demonstrated in this paper that a Kriging interpolator combined with a sequential sampling technique is able to identify cusps of the subspaces automatically without any prior information concerning their location. 
Moreover,  one can keep the sample size to a minimum
by  making use of the convergence properties mentioned in the previous paragraph.
This strategy will be advantageous for any application with strong, localised fluctuations in the values of the unknown function.
 
\section{Methods}

\subsection{The Read-Shockley-Wolf model} \label{sec42}
The original Read-Shockley model \cite{read1950dislocation} relies on the fact that GB structures with small misorientation angles can be viewed as two-dimensional arrays of lattice dislocations.
In this case, elasticity theory predicts a logarithmic dependency of GB energy on the misorientation angle. Wolf \cite{wolf1989read} extended the Read-Shockley model into a piecewise logarithmic function which spans the whole range of misorientation angles in the one-dimensional subspace of STGBs, and can capture all cusps and maxima, by connecting the appropriate number of logarithmic segments. This fully empirical extension can be motivated by the existence of grain boundary dislocations with continuous and much smaller Burgers vectors than lattice dislocations \citep{sutton1995interfaces}.

Given specific misorientation angles $\boldsymbol \beta = (\beta_0,\ldots,\beta_k)$ with $0 = \beta_0 < \beta_1 < \ldots < \beta_k = \thetamax$, which mark the beginning and ending of the individual logarithmic segments, and energy levels $\boldsymbol \gamma = (\gamma_0,\ldots,\gamma_{k})$ the RSW function at the misorientation angle $\theta$ is given by
\begin{align*}
       F_{\textrm{RSW}}&(\theta,\boldsymbol \beta,\boldsymbol \gamma) = \sum_{i=1}^{k} \1_{[\beta_{i-1},\beta_i]}(\theta)\\
        & \times \left[  \1_{(-\infty,0)}(\gamma_{i} - \gamma_{i-1})\cdot \left( \gamma_i + (\gamma_{i-1} - \gamma_{i}) \cdot f_{\textrm{RSW}}\left( \frac{\beta_i - \theta}{\beta_i - \beta_{i-1}}\right) \right) \right.\\
        &\hspace{2em}+ \left. \1_{[0,\infty]}(\gamma_i - \gamma_{i-1}) \cdot \left( \gamma_{i-1} + (\gamma_i - \gamma_{i-1}) \cdot f_{\textrm{RSW}}\left( \frac{\theta-\beta_{i-1}}{\beta_i - \beta_{i-1}}\right) \right) \right],
    \end{align*}
    where the function $f_{\textrm{RSW}}$ is defined by 
    \begin{equation*}
        f_{\textrm{RSW}}(\theta) = \sin(\pi \theta/2) (1-a\log(\sin (\pi \theta/2)))
    \end{equation*}
    for some real valued parameter $a$ (in our experiments, we exclusively consider $a=1$), and $ \1_A$ is the indicator function of the set $A$, that is 
    \begin{equation*}
        \1_A(\theta) = \begin{cases}1, & \text{ if } \theta \in A,\\ 0, & \text{ if } \theta \notin A. \end{cases}
    \end{equation*}
  The indicator function determines the orientation of the logarithmic segment, i.e.~if it describes a non-descending ($\gamma_i - \gamma_{i-1} \geq 0$)
  or descending ($\gamma_i - \gamma_{i-1} < 0)$ branch of the energy.
    
In our  study we define $F_{\mathrm{RSW}}$ for STGBs with misorientation axis [110]  via $k=6$ logarithmic segments between the following misorientation angles and energy offset values, 
\begin{align*}
    \boldsymbol \beta = (\beta_0,\beta_1,\beta_2,\beta_3,\beta_4,\beta_5,\beta_6) &= (0\si{\degree}, 20.775\si{\degree}, 50\si{\degree}, 80.4199\si{\degree}, 110\si{\degree}, 150.49\si{\degree}, 180\si{\degree}),\\
    \boldsymbol \gamma = (\gamma_0,\gamma_1,\gamma_2,\gamma_3,\gamma_4,\gamma_5,\gamma_6) &= (0, 0.683, 0.285, 0.664, 0.043, 0.826, 0).
\end{align*}
\subsection{Fourier series and  Kriging interpolation}\label{s:comparison}
\label{sec43}
We are interested in the prediction of one-dimensional sections of the whole grain boundary energy landscape from a data set of observations that should be as small as possible. This landscape can be parametrised by one single parameter, the misorientation angle.
More precisely, the target quantity is the energy function
\begin{equation}\label{eq:def:f}
y= f(\theta) 
\end{equation}
that maps an angle $\theta \in [0,\thetamax]$ to the corresponding grain boundary energy $y$.
Based on data $\Dc_N = \{ (\theta_1,y_1),\ldots,(\theta_N,y_N) \}$ 
satisfying the functional relationship \eqref{eq:def:f}, our aim is to predict the energy also at angles $\theta$ where the corresponding energy has not been observed.
Such a predictor is desirable since the direct computation of the energy both by physical or computer experiments is often exceptionally expensive.

In this section we propose two approaches to tackle the prediction problem. Both methods  have the desirable property that the resulting curves interpolate exactly through the observed data.

\subsubsection{Truncated Fourier series expansion}\label{subs:Fourier}
    The idea of the first approach is to model the function $f$ in \eqref{eq:def:f} as a truncated Fourier series. This is inspired by the use of spherical harmonics in materials science. For instance, the solid-liquid interface free energy has been expanded in a series of hyperspherical harmonics for application in a phase field model of dendritic growth \cite{Hoyt2001,Haxhimali2006}. In this context the function is used to reflect the surface energy anisotropy, but it can also be used to model grain boundary energies. This was demonstrated 
    by Banadaki and Patala~\citep{Banadaki2016}, who fitted a  series of symmetrized hyperspherical harmonics to investigate the dependence of interfacial energies on the boundary-plane inclination of grain boundaries. For the one-dimensional energy functions of STGBs, the analogue of this representation of a 2D, spherical function by spherical harmonics, is the representation in terms of trigonometric basic functions.
    \begin{equation*}
        f_{\cb}(\theta) = \sum_{j=0}^T c_j \phi_j(\theta),
    \end{equation*}
    where the basis functions $\phi_j$
are defined by 
    $\phi_0(\theta) = \frac{1}{\sqrt{\thetamax}} \1_{[0,\thetamax]}(\theta)$,
    \begin{align*}
\phi_{2j-1}(\theta) &= \sqrt{\frac{2}{\thetamax}} \cos \left( \frac{2\pi j \theta}{\thetamax} \right) ~,~~ \phi_{2j}(\theta) = \sqrt{\frac{2}{\thetamax}} \sin \left( \frac{2\pi j \theta}{\thetamax} \right) 
        ~~, ~j = 1,2,\ldots 
    \end{align*}
 The coefficients can be estimated from the data $\Dc_N$ by means of a least squares approach, that is, $\cbhat = (\chat_0,\ldots,\chat_T)$ is determined as the minimizer of the 
 sum of squares
    \begin{equation}\label{eq:c:LS}
\sum_{i=1}^N (y_i - f_{\cb}(\theta_i))^2 =    \sum_{i=1}^N  \Big (y_i - \sum_{j=0}^T c_j \phi_j(\theta_i) \Big )^2 .
    \end{equation}
    The resulting function $f_{\cbhat} (\theta) =\sum_{j=0}^T \chat_j \phi_j(\theta)$ is then used to predict energies at unobserved angles.
    If $T=N-1$ (that is, the number of basis functions equals the number of observations), minimizing \eqref{eq:c:LS}  with respect
    $c_1,\ldots,c_n$ leads to a system of 
$n$ linear equations for the parameters $\chat_1,\ldots,\chat_n$ with a unique solution (at least if the sampling points form an equidistant grid of the interval $[0,\thetamax]$).
    Then, the resulting function $f_{\chat}$ interpolates exactly through the given data as has been illustrated in Figure~\eqref{fig:trig} above.

\subsubsection{Simple Kriging interpolator}\label{subs:Kriging}
  
The second approach that we consider is called the (simple) Kriging interpolator which is a popular technique in applied statistics for interpolation of scattered data: Originally proposed in the context of geostatistical applications \cite{cressie1993statistics,stein1999interpolation} it has also become a standard method in the area of computer experiments \cite{santner2018design}.
In the case of noisy observations (that is, the observations in \eqref{eq:def:f} are perturbed by some random error) Kriging is also referred to as Gaussian process regression.
The Kriging interpolator depends on the choice of a positive definite \emph{kernel function}
$$
k(\theta,\theta')~,~~~ \theta,\theta' \in [0,\thetamax]
$$
(we refer to \citep{paulsen2016introduction}, Section~3.2 for a rigorous definition of kernel functions). Using the data $\Dc_N = \{ (\theta_1,y_1),\ldots,(\theta_N,y_N) \}$ one next calculates the matrix $\Kb _{\Dc_N}= (k(\theta_i,\theta_j))_{i,j=1}^N \in \R^{N \times N}$ and the vector $\kb_{\Dc_N} = (k(\theta,\theta_1),\ldots,k(\theta,\theta_N)) \in \R^{1 \times N}$. Here $\theta$ is the point of interest where one wants to predict the energy. The matrix $\Kb _{\Dc_N}$ is called Gramian matrix and 
is invertible for all combinations of kernel and sampling locations considered in this paper. 
The Kriging interpolator $\fhat_{\Dc_N}$ based on the set $\Dc_N$ of observations is finally given by the formula
\begin{equation*}
    \fhat_{\Dc_N}(\theta) = \kb_{\Dc_N} \Kb_{\Dc_N}^{-1} \yb^\top,
\end{equation*}
where $\yb = (y_1,\ldots,y_N) \in \R^{1 \times N}$ for $y_i = f(\theta_i)$.
An important advantage of the Kriging
 interpolator consists in the fact that it allows to quantify the uncertainty of the prediction
using  a Bayesian interpretation.
More precisely, modeling a random function as a centred Gaussian process with covariance kernel $k$, the Kriging interpolator is the posterior mean after observing the data $\Dc_N$ (\citep{rasmussen2006gaussian}, Equation~(2.19)).
In this case, the posterior variance (at location $\theta$) is given by
\begin{equation} 
\label{h2}
    \tau_{\Dc_N}(\theta) = k(\theta,\theta) - \kb_{\Dc_N} \Kb_{\Dc_N}^{-1} \kb_{\Dc_N}^\top,
    \end{equation}
which does not depend on the observed responses $y_1,\ldots,y_N$.
Roughly speaking, large values of $\tau_{\Dc_N}$ indicate locations with a rather large uncertainty concerning the target function $f$ whereas small values are attained at locations where one has already a better understanding of $f$.
In particular,  the posterior variance $\tau_{\Dc_N}$ vanishes at the points 
$\theta_1,\ldots,\theta_N$, where simulations have already been performed.
    
To specify the Kriging estimator, one has to choose an appropriate kernel function.
In this work, we consider the \emph{Mat\'ern class} of kernels given through
\begin{equation*}
    k(\theta,\theta') = \sigma^2 ~ \frac{2^{1-\nu}}{\Gamma(\nu)} \Big ( \frac{\sqrt{2\nu} \lvert \theta - \theta' \rvert}{\thetamax} \Big  )^\nu K_\nu \Big ( \frac{\sqrt{2\nu} \lvert \theta - \theta' \rvert}{\thetamax} \Big ),
    \end{equation*}
    where $\nu > 0$ denotes a smoothness parameter, $\sigma > 0$ a variance parameter, and $K_\nu$ denotes a modified Bessel function.
    The \emph{hyperparameters} $\nu$ and $\sigma$ have to be estimated (if we do not fix them a priori).
    The Matérn class contains a large class of interesting kernels as special cases (see Table~1 in \cite{guttorp2006studies}).
    For instance, in the case $\nu = 0.5$, the Kriging interpolator is nearly a piecewise linear function (provided that the scale parameter is not too small) and the corresponding kernel is referred to as the exponential kernel.
    Larger values of $\nu$ lead to smoother interpolators.
    A typical Kriging estimator  for $\nu = 2.5$ is depicted  in Figure~\eqref{fig:Kriging}.
    In some machine learning applications, the (in terms of smoothness interpretable) parameter $\nu$ is fixed in advance (see, for instance,  \cite{cornford2002modelling} for an application in wind field modelling with fixed choice $\nu=2.5$), and only the remaining hyperparameter has to be estimated.
    In our study, we have considered both scenarios where one either fixes the parameter $\nu$ in advance or one estimates $\nu$ from the given data.
    As the method of choice for the estimation of hyperparameters we choose the maximum likelihood approach that was also used in \cite{kleijnen2004application}.

\subsection{Sequential designs for Kriging}\label{subs:sequential}
\label{sec44}

Given a limited budget concerning the number of possible experiments, the question of
an appropriate design is of particular importance in material science simulations.
For this purpose several strategies such as space-filling or Latin hypercube designs have been proposed in the statistical literature
\citep{mckay1979comparison,morris1995exploratory,stinstra2003constrained,joseph2016space}.
Roughly speaking, one distinguishes between non-sequential and sequential designs.
In the non-sequential case one fixes
a design at the beginning and conducts the  experiments according to this design.
In the sequential design only a part of the budget is used as an initial sample (as in the non-sequential case), and
for the remaining  observations
the experimental conditions are sequentially
updated depending  on the output 
of the previous experiments. Both, sequential and non-sequential design strategies have their pros and cons (depending on the particular application), and 
the advantages of the Kriging interpolator based on
a sequential design strategy for the prediction of grain boundary energies 
have been demonstrated in Sections \ref{sec22} and  \ref{sec23}.

We will now develop the sequential design algorithm  used  in this paper which is  an adaption of an algorithm proposed in Kleijnen and van Beers~
\cite{kleijnen2004application} 
to GB energy prediction.
Given $N$ observations $\Dc_N = \{ (\theta_1,y_1),\ldots,(\theta_N,y_N) \}$ where $N \geq \ninit$ we describe how to find the ${(N+1)}$-th sampling location $\theta_{N+1}$.
This procedure consists of the following steps:
\begin{enumerate}
\item \textbf{Selection of candidate points}

In a first step, $\ncand$ candidate points $\xcand_1,\ldots,\xcand_{\ncand}$ (among which $\theta_{N+1}$ will be selected) are determined via the following pseudo-code:

\begin{itemize}[label=]
    \item Set $\Dctilde := \Dc_N$
    \item For $i=1,\ldots,\ncand$:
    \begin{itemize}[label=]
        \item[-] 
        Choose $\xcand_i $  as the point in the experimental region $ [0,\thetamax] $, which maximizes the posterior variance  $\tau_{\Dctilde}(\theta)$ in \eqref{h2}
        \item[-]  Set $\Dctilde := \Dctilde \cup \{ \xcand_i \}$
    \end{itemize}
\end{itemize}
Note that the definition of the posterior variance $\tau_{\Dctilde}(\theta)$ indeed depends only on the sampling locations but not on the response values. Therefore, it can  be computed without performing any simulation-based measurements at all.
The idea behind the above pseudo-code is to add successively sampling locations where the uncertainty (measured in the Bayesian interpretation by the value of the posterior variance) is maximal.

\item \textbf{Cross-validation} 

\nopagebreak

In order to select the next sampling point $\theta_{N+1}$, we aim at estimating the variance of the predicted output at each candidate input based only on Kriging computations which are computationally cheap in comparison with the expensive computations of the computer experiment of interest.
For this purpose, we consider Kriging estimators based on cross-validation.
Given the data $\Dc_N$ we denote with $\Dc_N^{(-j)}$ the data with the $j$-th observation deleted, that is,
\begin{equation*}
    \Dc_N^{(-j)} = \{ (\theta_1,y_1),\ldots,(\theta_{j-1},y_{j-1}),(\theta_{j+1},y_{j+1}),\ldots,(\theta_N,y_N) \}.
\end{equation*}
Based on the sets $\Dc_N$ and $\Dc_N^{(-i)}$ for $i=1,\ldots,N$ we can compute the Kriging estimates
\begin{align*}
    &\fhat_{\Dc_N}(\xcand_i) \qquad \text{for } i = 1,\ldots,\ncand,\\
    &\fhat_{\Dc_N^{(-j)}}(\xcand_i) \qquad \text{for } j=1,\ldots,N \text{ and } i = 1,\ldots,\ncand,
\end{align*}
at all the candidate points $\xcand_i$, either based on the whole datasets $\Dc_N$ or on the reduced datasets $\Dc_N^{(-j)}$.

\item \textbf{Computation of jackknifing variance}

\nopagebreak

Using the Kriging estimates, define the \emph{jackknife's pseudo-value} for candidate $i$ as
\begin{equation*}
    \ytilde_{i,j} = N \fhat_{\Dc_N}(\xcand_i) - (N-1) \fhat_{\Dc_N^{(-j)}}(\xcand_i),
\end{equation*}
for $j=1,\ldots,N$ and $i=1,\ldots,\ncand$.
From the $\ytilde_{i,j}$ one can compute the \emph{jackknife variance} for the candidate location $\xcand_i$ as
\begin{equation*}
    \stilde^2(\xcand_i) = \frac{1}{N(N-1)} \sum_{j=1}^N (\ytilde_{i,j} - \overline{\ytilde_i})^2,
\end{equation*}
where $\overline{\ytilde_i} = \frac{1}{N} \sum_{j=1}^N \ytilde_{i,j}$. 

\item \textbf{Selection of the next sampling location $\theta_{N+1}$}

Finally, the winning candidate $\theta_{N+1}$ among $\xcand_1,\ldots,\xcand_{\ncand}$ is determined as
\begin{equation*}
    \theta_{N+1} \in \argmax_{\xcand \in \{ \xcand_1,\ldots,\xcand_{\ncand} \} } \stilde^2 (\xcand).
\end{equation*}
\end{enumerate}
Having determined $\theta_{N+1}$, one will determine the corresponding response $y_{N+1}$ in a (simulation) experiment.
Afterwards, the sequential design algorithm can be used again to determine the next sampling location $\theta_{N+2}$ and so on.
The choice of $\ncand$ permits the following heuristic: Taking $\ncand = 1$ leads to the, in the computer experiments literature  well-known, {\it algorithm of selecting the next design point among the maximizers of the posterior variance.}
In this special case, the next sampling point does not depend on the response values anymore and accumulation of sampling at region of interests (in our case, cusps) cannot be achieved.
For increasing $\ncand$ more points close to already existing observations are allowed and the desired kind of accumulation becomes more likely.
In all our simulations we put $\ncand = 75$.

\subsection{Atomistic simulation method}

\label{sec:Atomistic_simulations}

To determine the grain boundary energy for a given set of DOF, a molecular statics simulation has been performed using LAMMPS (Large-scale Atom\-ic/Mo\-lecu\-lar Massively Parallel Simulator, \citep{Plimpton1995LAMMPS}). We chose body centred cubic iron as an example, using the second embedded atom model (EAM) potential from \cite{mendelev2003potential} which predicts a lattice constant of Fe of $a_{\mathrm{Fe}} = 2.85 \si{\angstrom}$. The grain boundary structures were generated by adapting the method introduced by Lee and Choi \citep{lee2004computation}, which allows simulating non-periodic grain boundaries by using a spherical model. Firstly, two spheres of atoms with a radius of $r_{\mathrm{o}}$ are created (in this work, $r_{\mathrm{o}} = 35 a_{\mathrm{Fe}}$) , and one of the spheres is rotated by the desired misorientation angle $\theta$ around the rotation axis $\vec{a}$. Secondly, both grains are split in half-spheres and combined to create again a full sphere with a grain boundary with the desired normal vector $\vec{n}$. To obtain the minimum grain boundary energy for this set of macroscopic degrees of freedom, the microscopic DOF have to be probed. 
For each set of macroscopic DOF, initial translations of 0 and 1.5 $\si{\angstrom}$ perpendicular to the grain boundary plane were applied, to probe the excess volume of the GB. Similarly, initial in-plane translations parallel to the grain boundary plane were carried out in steps of 0 to 5$\si{\angstrom}$ for the direction perpendicular to the tilt axis, and in steps of $d_{hkl}/5$ along the tilt axis, with $d_{hkl}$ being the interplanar spacing along this direction, $d_{hkl} =a/\sqrt{h^2+k^2+l^2}$ ($[hkl]$ are the Miller indices of the rotation axis). In total, we used $144$ different starting configurations of the microscopic DOF to relax the atomic positions and obtain the optimized structure for one set of macroscopic DOF of a grain boundary.
After relaxation of all structures, the one with the minimum energy is kept in the database. Note that there is no guarantee that our minimum grain boundary energies are the absolute minimum energies for the respective set of macroscopic DOF, but an analysis of the relaxed trial structures shows that on average more than $25\%$ of all initial sets of microscopic DOF relax into the same state, which has the lowest energy of the set, or deviates by $\leq1\%$ from it.
After relaxation of the atomic positions the grain boundary energy is calculated using only the atoms of an inner sphere with a radius of $r_{\mathrm{i}}$ ($r_{\mathrm{i}} = 20 a_{\mathrm{Fe}}$), to avoid surface effects. Note that the difference between $r_{\mathrm{o}}$ and $r_{\mathrm{i}}$ has to be larger than the cut-off radius of the used interatomic potential plus the maximum applied parallel shift. The grain boundary energy is calculated according to the following equation:
\begin{equation*}
    E_{\mathrm{GB}} = \frac{\sum_{n=1}^{N}E_{\mathrm{pot},n}}{\pi r_{\mathrm{i}}^{2}},
\end{equation*}
where $N$ is the number of atoms in the inner sphere and $E_{\mathrm{pot},n}$ the energy of the $n$-th atom in the inner sphere. Of all the initial microscopic DOF, those which result in the minimum grain boundary energy for the given macroscopic DOF, are kept, and represent one datapoint in the energy landscape.

\subsection{Atomistic data generation}
\label{sec:Data_generation}

The macroscopic DOF, in this case the misorientation angle, was varied between 0 and a maximum angle $\thetamax$ for each subspace, which is defined by the rotation axis: $\thetamax=90\si{\degree}$ for $[100]$,  $180\si{\degree}$ for $[110]$ and $120\si{\degree}$ for $[111]$. \\
A reference database with $N_{\mathrm{regular}}$ grain boundaries was generated, with a regular high throughput sampling technique and a spacing of $\theta=1{\si{\degree}}$, plus additional special $\Sigma$ grain boundaries, resulting in $N_{\mathrm{ref}}= 91+6$, $181+8$, and $121+6$ data points, respectively. 
Information about the additional special $\Sigma$ grain boundaries can be found in Tables~\ref{table:SigmaSTGBs_100}--\ref{table:SigmaSTGBs_111} in Appendix~\ref{sub:Appendix_reference_datasets}.
The runs using the sequential sampling technique have $N_{\mathrm{init}} $ initial design points taken as equidistant subsets of the reference database.
The points in the sequential design 
proposed in Section \ref{sec44}  
are chosen  from an equidistant grid 
with 
\begin{equation*}
    N_{\mathrm{grid}} = 20\thetamax+1
\end{equation*}
points. Here the corresponding GB-energies are obtained 
from the reference database 
by linear interpolation.

Figure~\eqref{fig:plots:max_error_nu_hat_0_5} 
shows that the maximum absolute error generally decreases with increasing number of sequential iterations, and  we decided to limit $N_{\mathrm{seq}}$ to $20$, 
to prove that our technique is able to sample a subspace with as few grain boundaries as possible with a high accuracy. 
The information regarding the different sampling parameters is summarized in Table~\ref{table:atomistic_simulation data}.

\begin{table}[H]
\begin{center}
 \begin{tabular}{c|c|c|c|c|c|c}
 subspace & $\thetamax [\si{\degree}]$& $N_{\mathrm{ref}}$ & $N_{\mathrm{total}}$ & $N_{\mathrm{init}}$ & $N_{\mathrm{seq}}$ & $N_{\mathrm{grid}}$\\
 \hline
 $[100]$ & 90 & 97 & 46 & 16 & 20 & 1801\\
 $[110]$ & 180 & 187 & 91 & 31 & 20 & 3601\\
 $[111]$ & 120 & 125 & 61 & 21 & 20 & 2401\\
\end{tabular}
\caption{Sampling parameters for each subspace, defined by rotation axis $\vec{\rho}$: Number of points in the reference database, $N_{\mathrm{ref}}$, for the regular sampling, $N_{\mathrm{total}}$, number $N_{\mathrm{init}}$ of initial design and of $N_{\mathrm{seq}}$ sequential points for the sampling with the sequential sampling technique, number of grid points, $N_{\mathrm{grid}}$, for the Kriging interpolation.}
\label{table:atomistic_simulation data}
\end{center}
\end{table}
 To analyse the influence of $\nu$ towards the sampling of the subspace, first of all every subspace has been sampled with the sequential sampling technique three times with different fixed $\nu$ values each. The different $\nu$ values chosen for the sampling were $0.5, 1.5$, and $2.5$. After that the sampling was repeated with $\nu$ limited between $0.5$ and $2.5$ but estimated from the atomistic data using the MLE technique.

\section{Data availability}

The code for the RSW model, which can be used to reproduce all the results obtained for this model, is available at
\begin{center}\scriptsize
    \url{https://gitlab.com/kroll.martin/rsw-model}.
\end{center}
Data from atomistic simulations can be downloaded from
\begin{center}\scriptsize
     \url{https://git.noc.ruhr-uni-bochum.de/Schmalofski.Timo/atomistic-data-STGBs-1D}.
\end{center} 
\bigskip

\noindent

\bigskip

{\bf Funding.} This research has been supported by the German Research Foundation (DFG), project number 414750139.\\
\bibliographystyle{plain}
 
\newpage

\pagenumbering{arabic}\renewcommand*{\thepage}{A\arabic{page}}
\appendix

\section{Further results for the RSW model}

\begin{figure}[H]
     \centering
     \begin{subfigure}[t]{0.49\textwidth}
         \centering
         \caption{}
         \includegraphics[width=\textwidth]{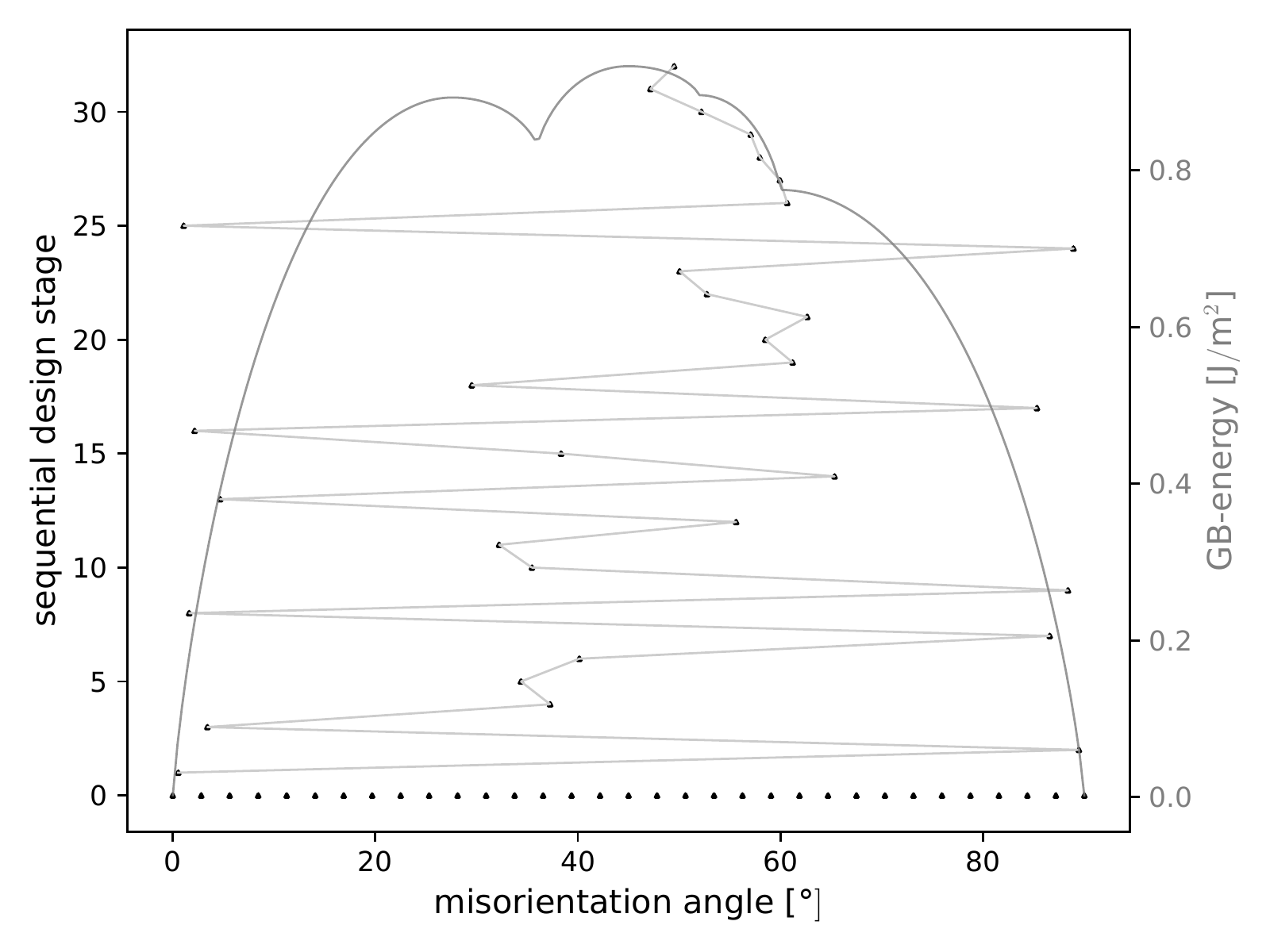}
\end{subfigure}
     \begin{subfigure}[t]{0.49\textwidth}
         \centering
         \caption{}
         \includegraphics[width=\textwidth]{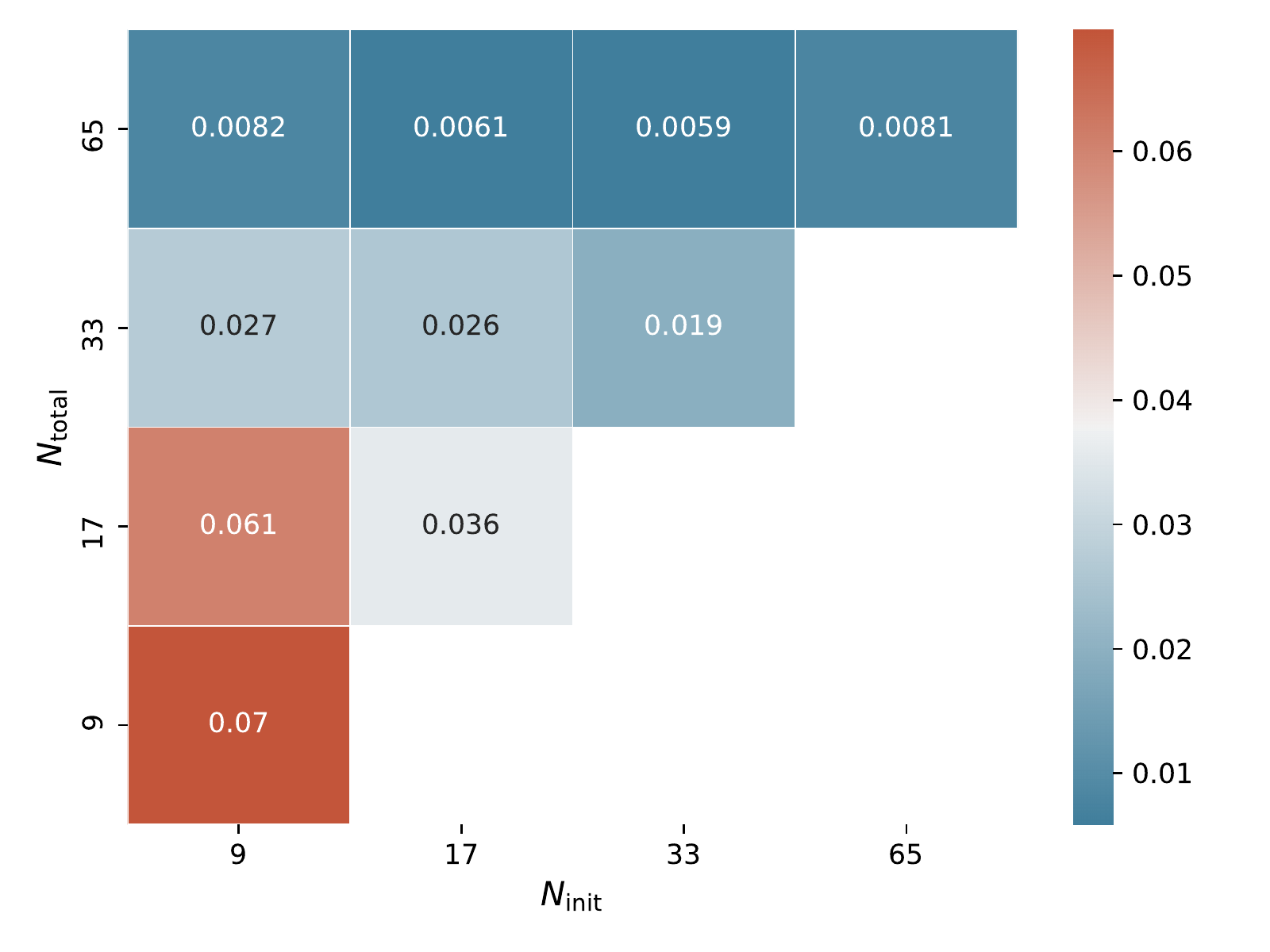}
\end{subfigure}
    \caption{
    $F_1$: $[100]$ STGB subspace.
    }
    \label{fig:plots:RSW:seq:1}
\end{figure}

\begin{figure}[H]
     \centering
     \begin{subfigure}[t]{0.49\textwidth}
         \centering
         \caption{}
         \includegraphics[width=\textwidth]{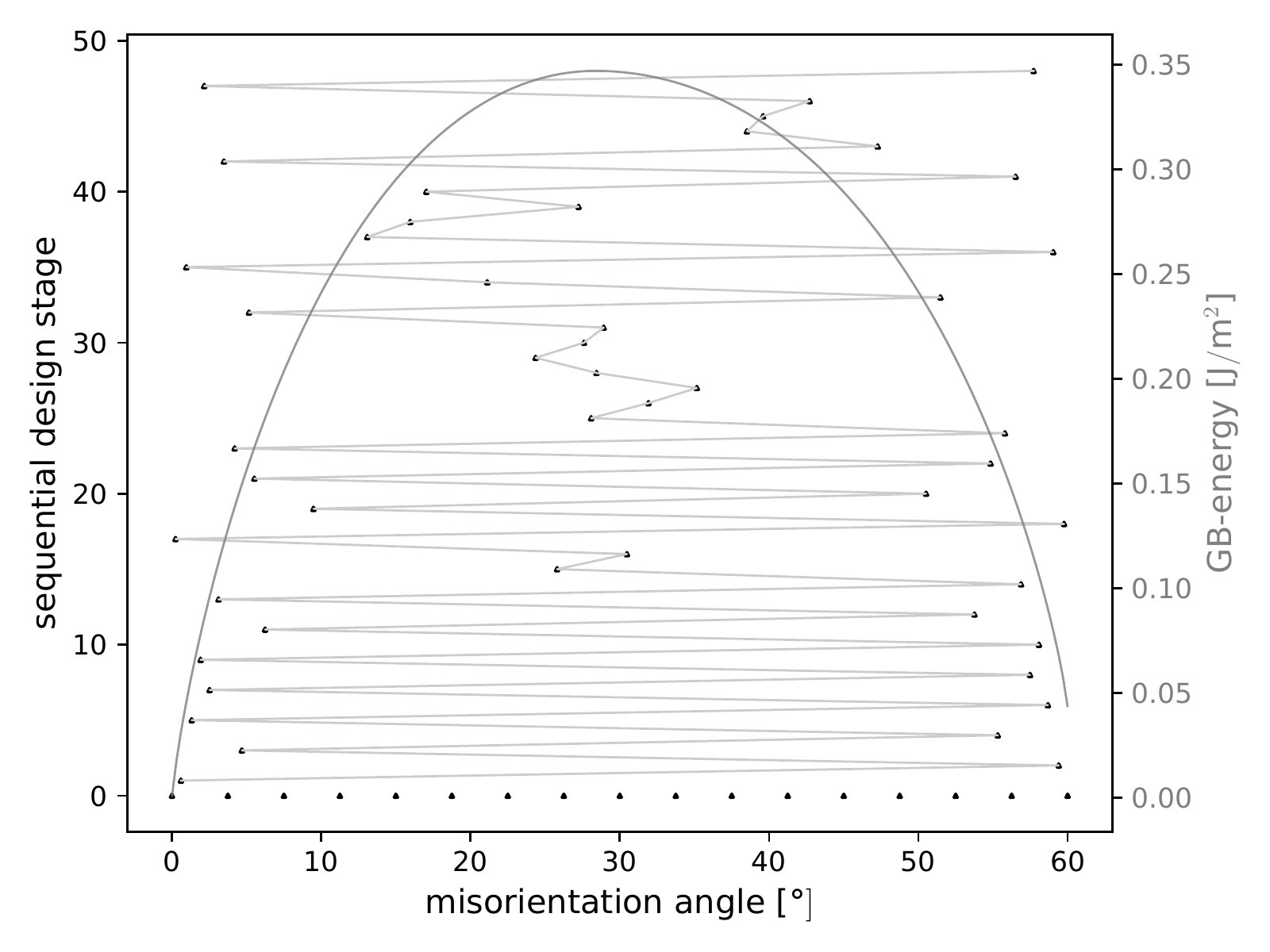}
\end{subfigure}
     \begin{subfigure}[t]{0.49\textwidth}
         \centering
         \caption{}
         \includegraphics[width=\textwidth]{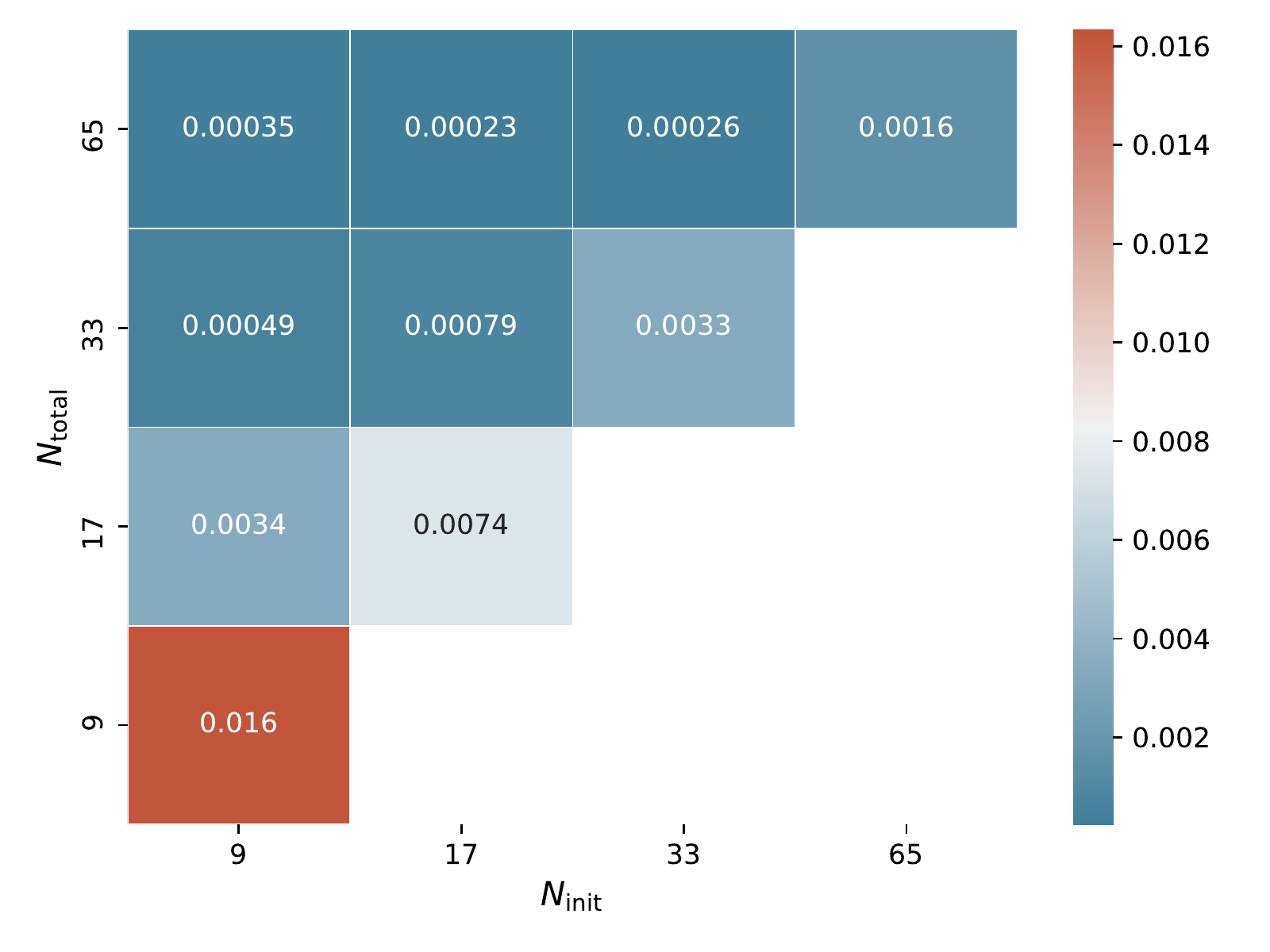}
\end{subfigure}
    \caption{$F_6$: $[111]$ twist grain boundaries subspace.}
    \label{fig:plots:RSW:seq:6}
\end{figure}

In this part of the supplementary material we continue the investigation of Section \ref{sec22} and
provide results for five more test functions, which were considered in \citep{dette2017efficient} and are depicted in the left plots of Figures~\ref{fig:plots:RSW:seq:1}--\ref{fig:plots:RSW:seq:6}, respectively.
The functions $F_1$ and $F_3$ are examples of $[100]$ and $[111]$ symmetric tilt grain boundaries. The
RSW function $\FRSW$ considered in Section \ref{sec22} of the main part of the paper also belongs to this class, representing the $[110]$ symmetric tilt grain boundary.
On the other hand the functions $F_4$--$F_6$ represent the $[100]$, $[110]$, and $[111]$ twist grain boundaries.
Note that not all functions satisfy the boundary condition $f(0)=f(\thetamax)$ which is automatically satisfied by the trigonometric interpolator.
For this reason we first transformed the data via $y \mapsto \widetilde y = y - (a\theta + b)$ where the numbers $a,b$ are chosen such that $\widetilde y(0) = 0 = \widetilde y (\thetamax)$.
Prediction was performed for the data transformed in this way, and afterwards the predicted data were transformed back.

\subsection{Comparison of trigonometric series and Kriging interpolator}\label{app:subs:comparison}

In analogy to Table~\ref{table:RSW} we state in Tables \ref{tab:RSW1}--\ref{tab:RSW6} the maximum absolute error \eqref{h1} for the remaining F$_{\mathrm{RSW}}$ functions for symmetrical tilt and twist grain boundaries with rotation axis [100], [110], and [111], 
and the interpolators under consideration.
In all cases, different values for the parameter $\nu$ including a data-adaptive choice $\nuhat$ by means of maximum likelihood estimation (MLE) were considered for the Kriging interpolator.
These results confirm the findings from Section \ref{sec22}: 
For simulated data from the RSW model Kriging outperforms series interpolation for grain boudary energy prediction 
in all cases under consideration, where the value $\nu = 2.5$ and MLE result in the best results for the Kriging estimator.

\begin{table}[H]
\centering
\begin{tabular}{c|c|c|c|c|c}
$N$ & trig. series & $\nu = 0.5$ & $\nu = 1.5$ & $\nu = 2.5$ & $\widehat \nu$ via MLE \\ 
\hline 
9 & 0.2163 & 0.1136 & 0.0949 & \bfseries 0.0698 & 0.0925 \footnotesize($\widehat \nu$ = 1.5648) \\ 
17 & 0.1206 & 0.0540 & 0.0434 & \bfseries 0.0358 & 0.0365 \footnotesize($\widehat \nu$ = 1.9147) \\ 
33 & 0.0690 & 0.0263 & 0.0206 & \bfseries 0.0194 & 0.0198 \footnotesize($\widehat \nu$ = 1.8653) \\ 
65 & 0.0384 & 0.0131 & 0.0100 & \bfseries 0.0082 & 0.0086 \footnotesize($\widehat \nu$ = 1.7820) \\ 
\end{tabular}
\caption{$F_1$: $[100]$ STGB subspace.}\label{tab:RSW1}
\end{table}

\begin{table}[H]
\centering
\begin{tabular}{c|c|c|c|c|c}
$N$ & trig. series & $\nu = 0.5$ & $\nu = 1.5$ & $\nu = 2.5$ & $\widehat \nu$ via MLE \\ 
\hline 
9 & 0.1020 & 0.0600 & 0.0519 & \bfseries 0.0356 & 0.0357 \footnotesize($\widehat \nu$ = 2.4938) \\ 
17 & 0.0573 & 0.0291 & 0.0240 & \bfseries 0.0160 & \textbf{0.0160} \footnotesize($\widehat \nu$ = 2.4938) \\ 
33 & 0.0318 & 0.0143 & 0.0114 & \bfseries 0.0073 & \textbf{0.0073} \footnotesize($\widehat \nu$ = 2.4938) \\ 
65 & 0.0175 & 0.0071 & 0.0055 & \bfseries 0.0035 & \textbf{0.0035} \footnotesize($\widehat \nu$ = 2.4938) \\ 
\end{tabular}
\caption{$F_3$: $[111]$ STGB subspace.}
\end{table}

\begin{table}[H]
\centering
\begin{tabular}{c|c|c|c|c|c}
$N$ & trig. series & $\nu = 0.5$ & $\nu = 1.5$ & $\nu = 2.5$ & $\widehat \nu$ via MLE \\ 
\hline 
9 & 0.0414 & 0.0252 & 0.0229 & \bfseries 0.0155 & 0.0156 \footnotesize($\widehat \nu$ = 2.4938) \\ 
17 & 0.0228 & 0.0124 & 0.0105 & \bfseries 0.0069 & \textbf{0.0069} \footnotesize($\widehat \nu$ = 2.4938) \\ 
33 & 0.0125 & 0.0061 & 0.0049 & \bfseries 0.0031 & \textbf{0.0031} \footnotesize($\widehat \nu$ = 2.4938) \\ 
65 & 0.0068 & 0.0031 & 0.0024 & \bfseries 0.0015 & \textbf{0.0015} \footnotesize($\widehat \nu$ = 2.4955) \\ 
\end{tabular}
\caption{$F_4$: $[100]$ twist grain boundary subspace.}
\end{table}

\begin{table}[H]
\centering
\begin{tabular}{c|c|c|c|c|c}
$N$ & trig. series & $\nu = 0.5$ & $\nu = 1.5$ & $\nu = 2.5$ & $\widehat \nu$ via MLE \\ 
\hline 
9 & 0.1461 & 0.1046 & 0.0871 & \bfseries 0.0608 & 0.0655 \footnotesize($\widehat \nu$ = 2.2559) \\ 
17 & 0.0812 & 0.0499 & 0.0403 & \bfseries 0.0269 & 0.0270 \footnotesize($\widehat \nu$ = 2.4938) \\ 
33 & 0.0451 & 0.0244 & 0.0191 & \bfseries 0.0123 & 0.0154 \footnotesize($\widehat \nu$ = 1.9192) \\ 
65 & 0.0252 & 0.0121 & 0.0093 & \bfseries 0.0060 & 0.0074 \footnotesize($\widehat \nu$ = 1.9423) \\ 
\end{tabular}
\caption{$F_5$: $[110]$ twist grain boundary subspace.}
\end{table}

\begin{table}[H]
\centering
\begin{tabular}{cc|c|c|c|c}
$n$ & trig. series & $\nu = 0.5$ & $\nu = 1.5$ & $\nu = 2.5$ & $\widehat \nu$ via MLE \\ 
\hline 
9 & 0.0581 & 0.0278 & 0.0238 & \bfseries 0.0163 & 0.0164 \footnotesize($\widehat \nu$ = 2.4938) \\ 
17 & 0.0333 & 0.0134 & 0.0110 & \bfseries 0.0074 & \textbf{0.0074} \footnotesize($\widehat \nu$ = 2.4938) \\ 
33 & 0.0186 & 0.0066 & 0.0052 & \bfseries 0.0034 & \textbf{0.0034} \footnotesize($\widehat \nu$ = 2.4938) \\ 
65 & 0.0103 & 0.0033 & 0.0025 & \bfseries 0.0016 & \textbf{0.0016} \footnotesize($\widehat \nu$ = 2.4938) \\ 
\end{tabular}
\caption{$F_6$: $[111]$ twist grain boundary subspace.}\label{tab:RSW6}
\end{table}

\subsection{Performance of the sequential design algorithm}\label{app:subs:seq}

The following figures summarize the results concerning the sequential design algorithm for the $[100]$ and $[111]$ STGB subspaces ($F_1$ and $F_3$) and the $[100]$, $[110]$ and the $[111]$ twist grain boundaries subspaces ($F_4$--$F_6$).
The illustration of our results is in complete analogy to the one used in Figure~\ref{fig:plots:RSW:seq:2}.
The left plots show the respective dynamics of the sequential design algorithm for the best initial design leading to a final design of $\ntotal=65$ design points.
As in Figure~\eqref{fig:RSW:dynamics}, the $\theta$-value (angle) of the design points $\blacktriangle$ indicates the location in the interval $[0,\thetamax]$, the $y$-value the stage in the sequential design algorithm (stage $0$ corresponds to the initial design). The  $\FRSW$-function from which data are evaluated is plotted in gray.
The right plots state the maximum absolute error of the Kriging interpolator in combination with the sequential design algorithm in dependence of the initial design size $\ninit$ and the total design size $\ntotal$. 

The results confirm our finding from Section \ref{sec22}. In all cases under consideration the sequential design leads to a substantial improvement of the Kriging interpolator and allocates more design points to regions close to cusps.

\begin{figure}[H]
     \centering
     \begin{subfigure}[t]{0.49\textwidth}
         \centering
         \caption{}
         \includegraphics[width=\textwidth]{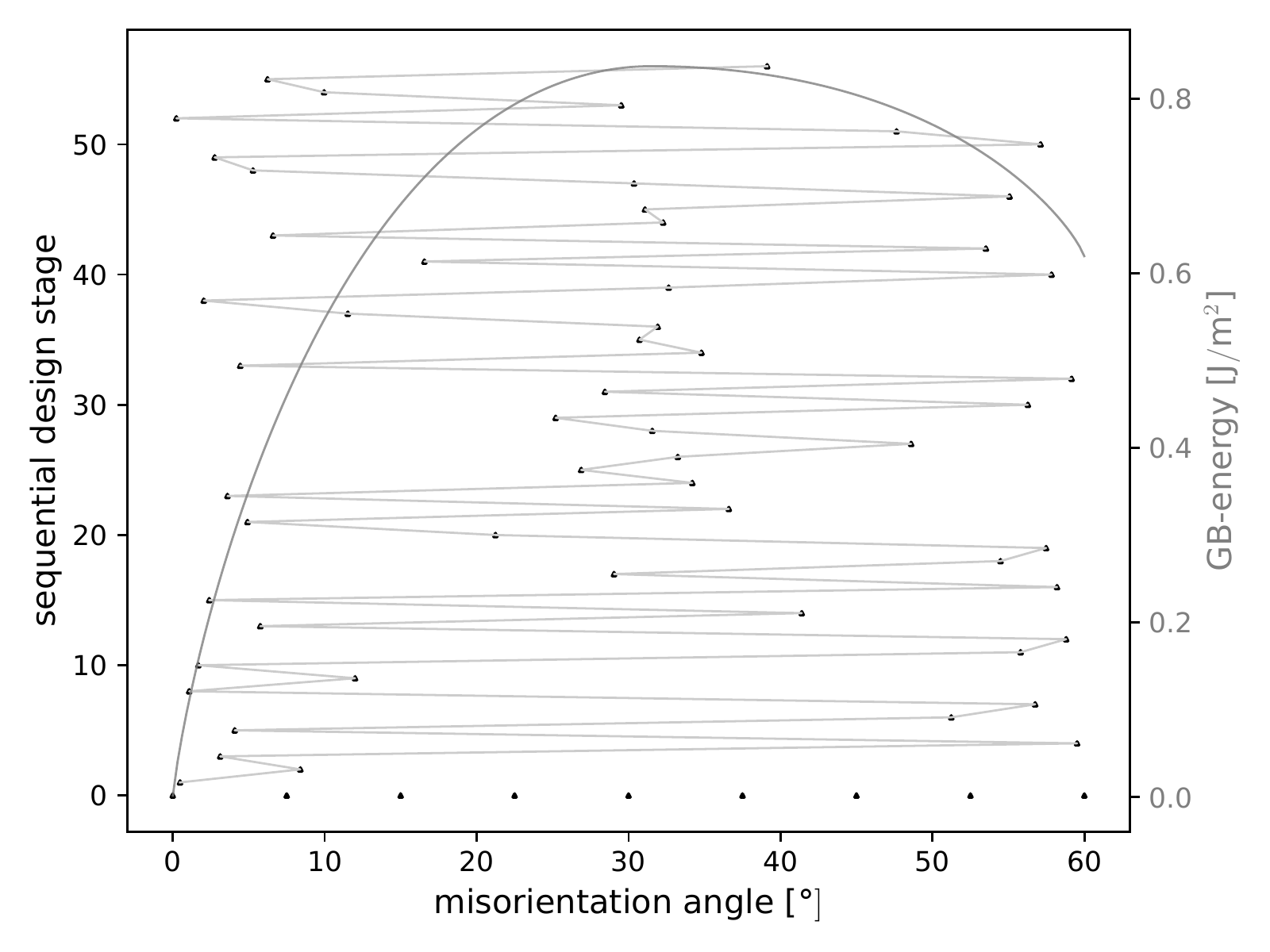}
\end{subfigure}
     \begin{subfigure}[t]{0.49\textwidth}
         \centering
         \caption{}
         \includegraphics[width=\textwidth]{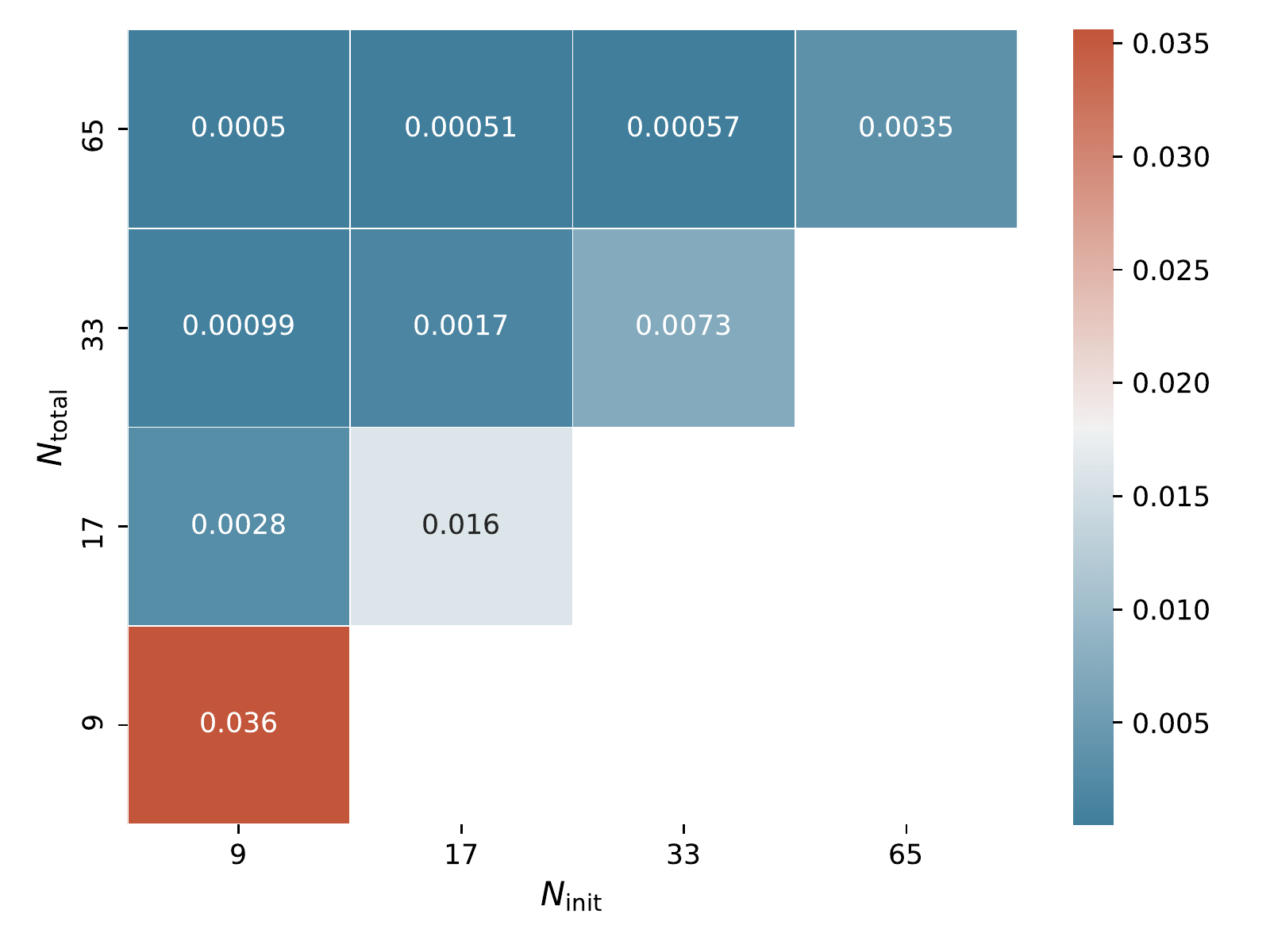}
\end{subfigure}
    \caption{$F_3$: $[111]$ STGB subspace.}
\end{figure}

\begin{figure}[H]
     \centering
     \begin{subfigure}[t]{0.49\textwidth}
         \centering
         \caption{}
         \includegraphics[width=\textwidth]{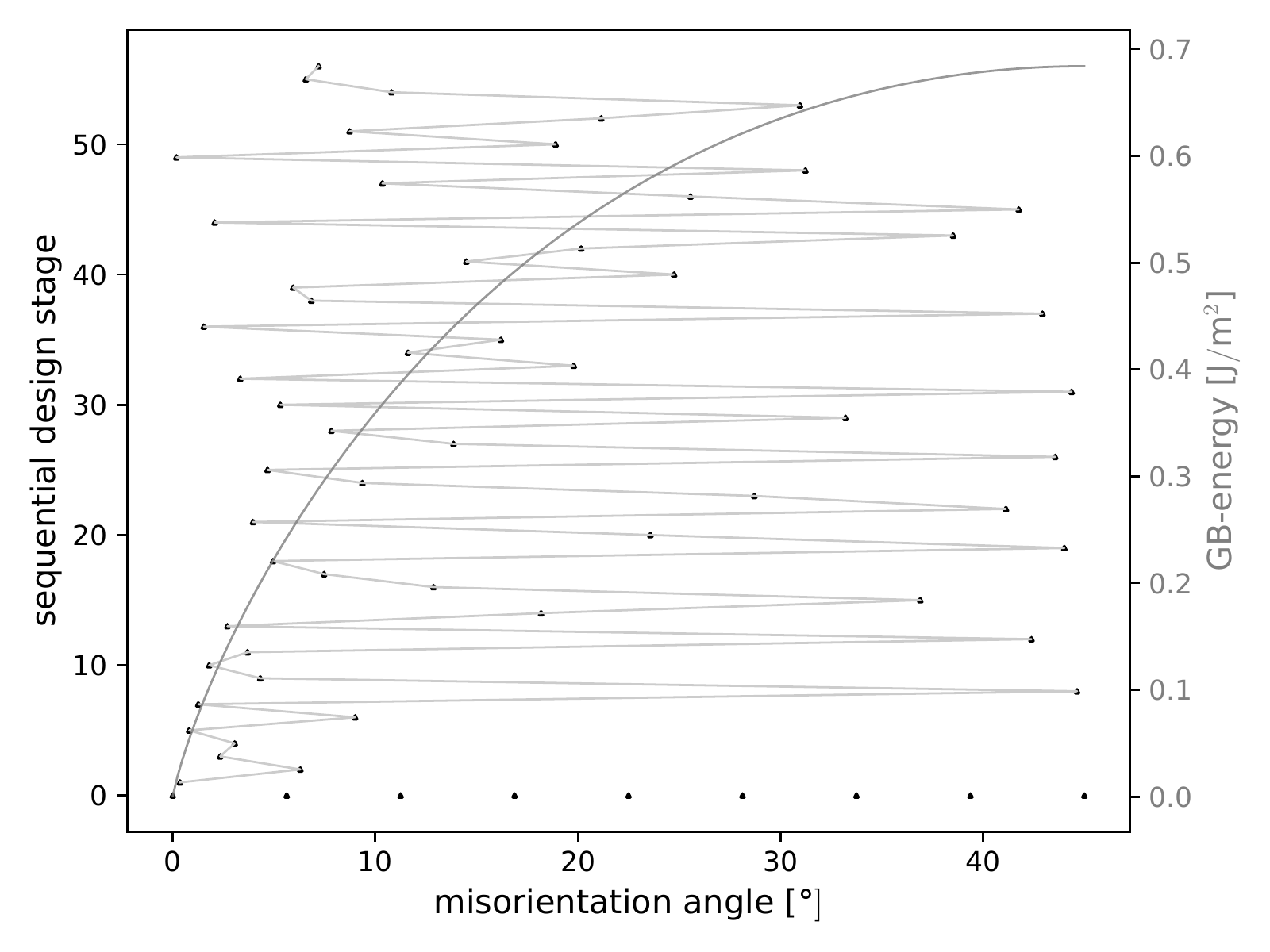}
\end{subfigure}
     \begin{subfigure}[t]{0.49\textwidth}
         \centering
         \caption{}
         \includegraphics[width=\textwidth]{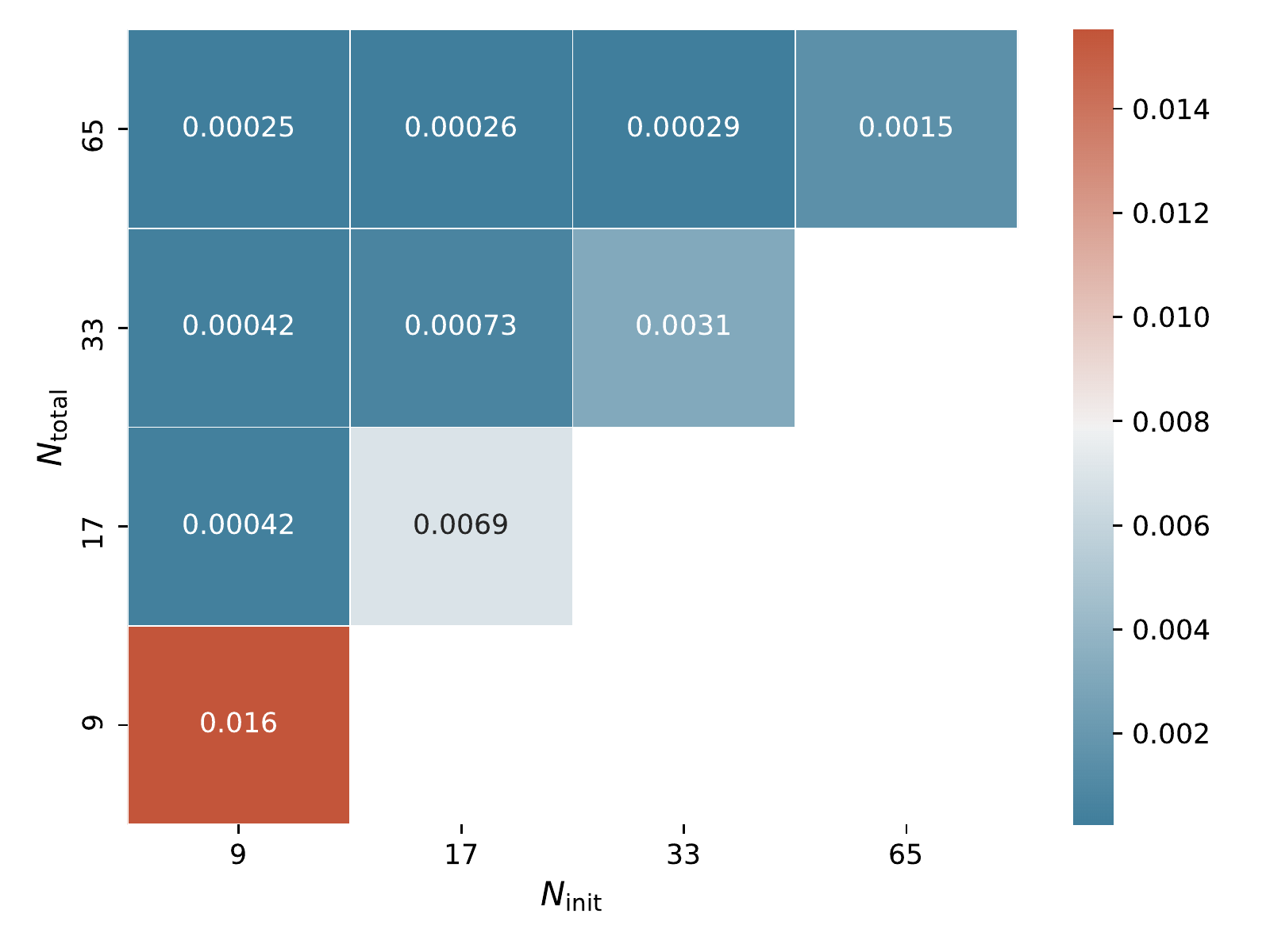}
\end{subfigure}
    \caption{$F_4$: $[100]$ twist grain boundaries subspace.}
\end{figure}

\begin{figure}[H]
     \centering
     \begin{subfigure}[t]{0.49\textwidth}
         \centering
         \caption{}
         \includegraphics[width=\textwidth]{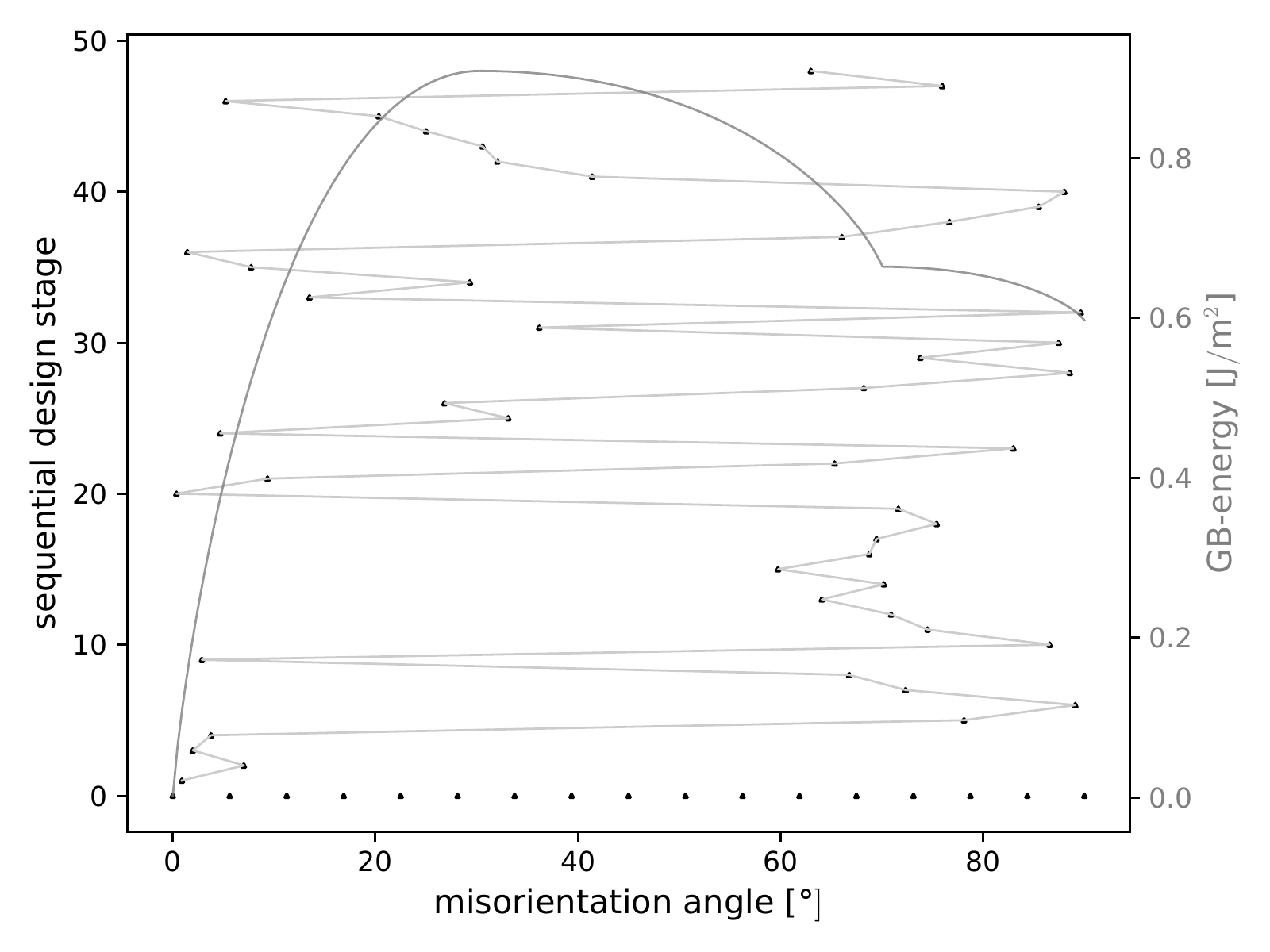}
\end{subfigure}
     \begin{subfigure}[t]{0.49\textwidth}
         \centering
         \caption{}
         \includegraphics[width=\textwidth]{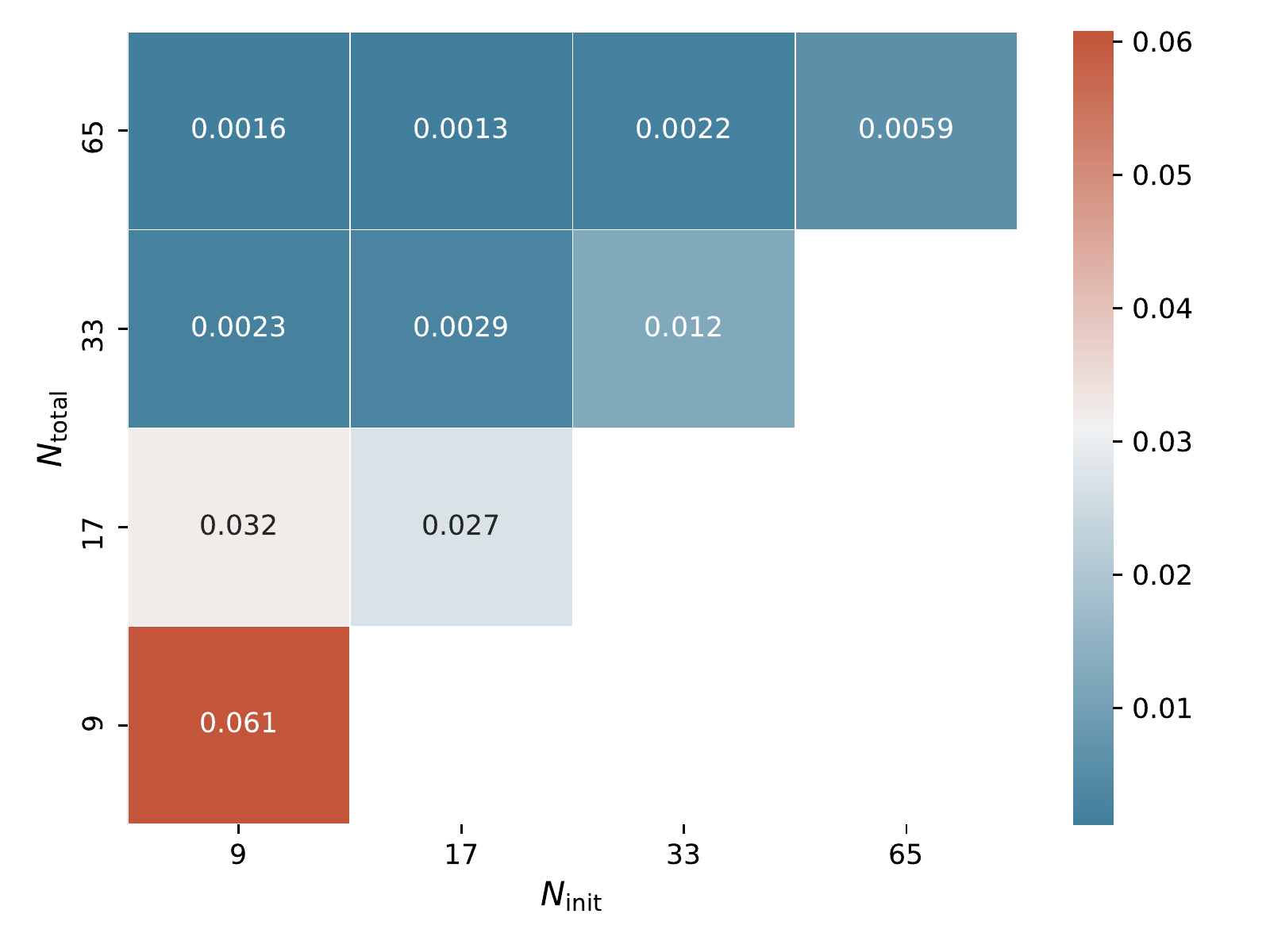}
\end{subfigure}
    \caption{$F_5$: $[110]$ twist grain boundaries subspace.}
\end{figure}

\section{Special STGBs of the reference data sets}
\label{sub:Appendix_reference_datasets}
The data generation of the reference data sets is described in Section \ref{sec:Data_generation}. In addition to the equally spaced design points of the regular sampling, the positions of the special, CSL based, grain boundaries were added. Their energies represent the deep cusps in the energy function. The orientation relationships of these grain boundaries are displayed in Table \eqref{table:SigmaSTGBs_100} for the $[100]$ subspace, in Table \eqref{table:SigmaSTGBs_110} for the $[110]$ subspace, and in Table \eqref{table:SigmaSTGBs_111} for the $[111]$ subspace together with their $\Sigma$ value.

\begin{table}[H]
\begin{center}
\begin{subtable}{\textwidth}
\caption{}\label{table:SigmaSTGBs_100}
\centering
\begin{tabular}{c|c|c|c|c|c|c}
$\Sigma$ & 5 & 5 & 13 & 13    & 41    & 41 \\
\hline
$\theta[\si{\degree}]$ & 36.87 & 53.13 & 22.62 & 67.38 & 12.68 & 77.32 \end{tabular}
\end{subtable}
\\[\bigskipamount]
\begin{subtable}{\textwidth}
\centering
\caption{}\label{table:SigmaSTGBs_110}
\begin{tabular}{c|c|c|c|c|c|c|c|c}
$\Sigma$               & 3     & 3      & 9     & 9      & 11    & 11     & 51    & 51  \\
\hline
$\theta[\si{\degree}]$ & 70.53 & 109.47 & 38.94 & 141.06 & 50.48 & 129.52 & 16.10 & 163.90 
\end{tabular}
\end{subtable}
\\[\bigskipamount]
\begin{subtable}{\textwidth}
\centering
\caption{}\label{table:SigmaSTGBs_111}
\begin{tabular}{c|c|c|c|c|c|c}
$\Sigma$ & 7     & 7     & 13    & 13    & 57    & 57      \\
\hline
$\theta[\si{\degree}]$ & 38.21 & 81.79 & 27.80 & 92.20 & 13.17 & 106.83 
\end{tabular}
\end{subtable}
\\[\bigskipamount]
\caption{$\Sigma$ STGBs added to the reference data sets for the (a) $[100]$, (b) $[110]$, and (c) $[111]$ STGB subspace, which is defined by the rotation axis~$\vec{\rho}$.}  \label{table:SigmaSTGBs}
\end{center}
\end{table}

\section{Further atomistic simulation results}
\label{sec:appendix_progress_sequential_point_calc}
\begin{figure}
     \centering
     \begin{subfigure}{0.49\textwidth}
         \centering
         \caption{}
         \includegraphics[width=\textwidth]{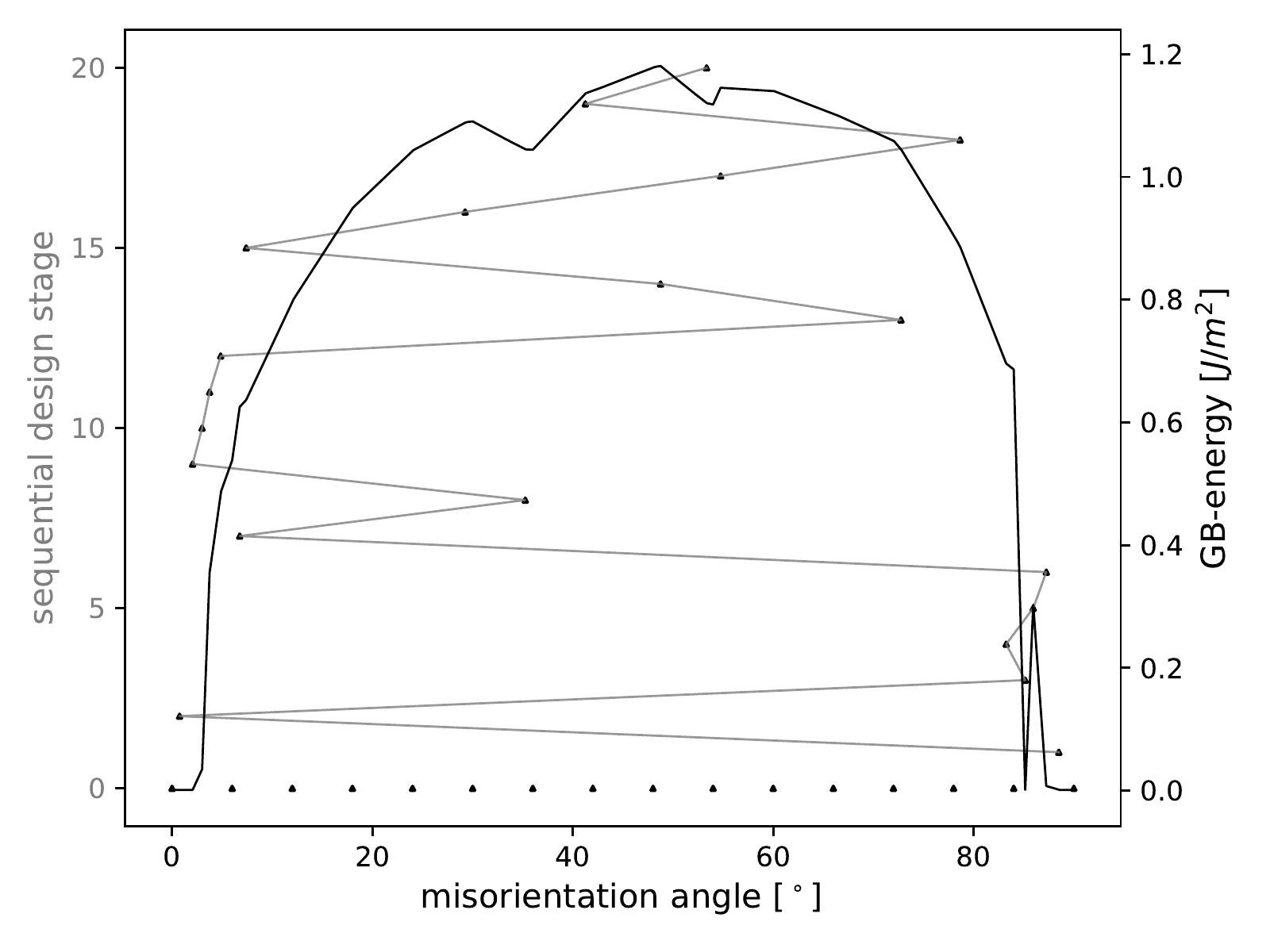}
\end{subfigure}
     \begin{subfigure}{0.49\textwidth}
         \centering
         \caption{}
         \includegraphics[width=\textwidth]{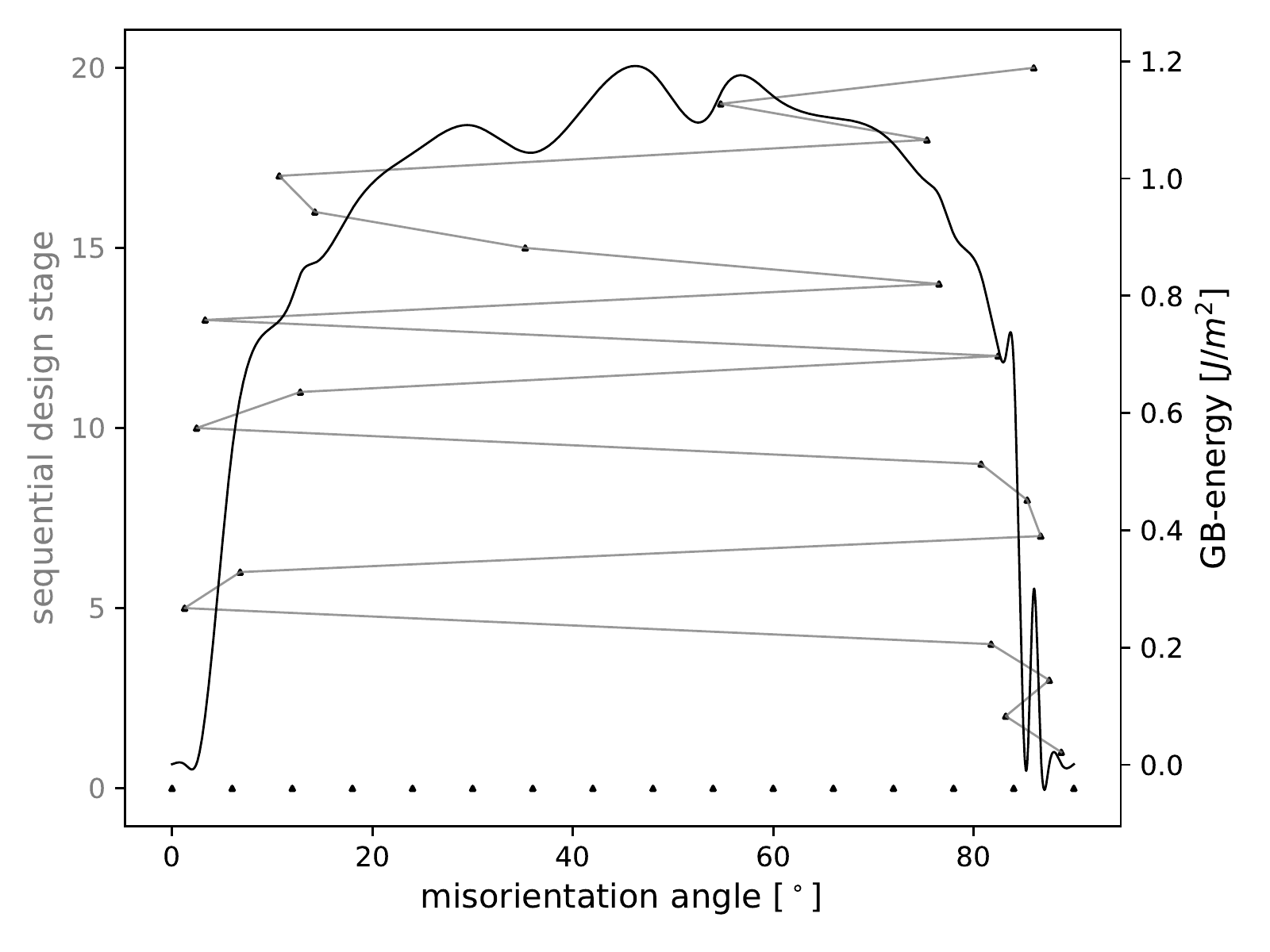}
\end{subfigure}
          \begin{subfigure}{0.49\textwidth}
         \centering
         \caption{}
         \includegraphics[width=\textwidth]{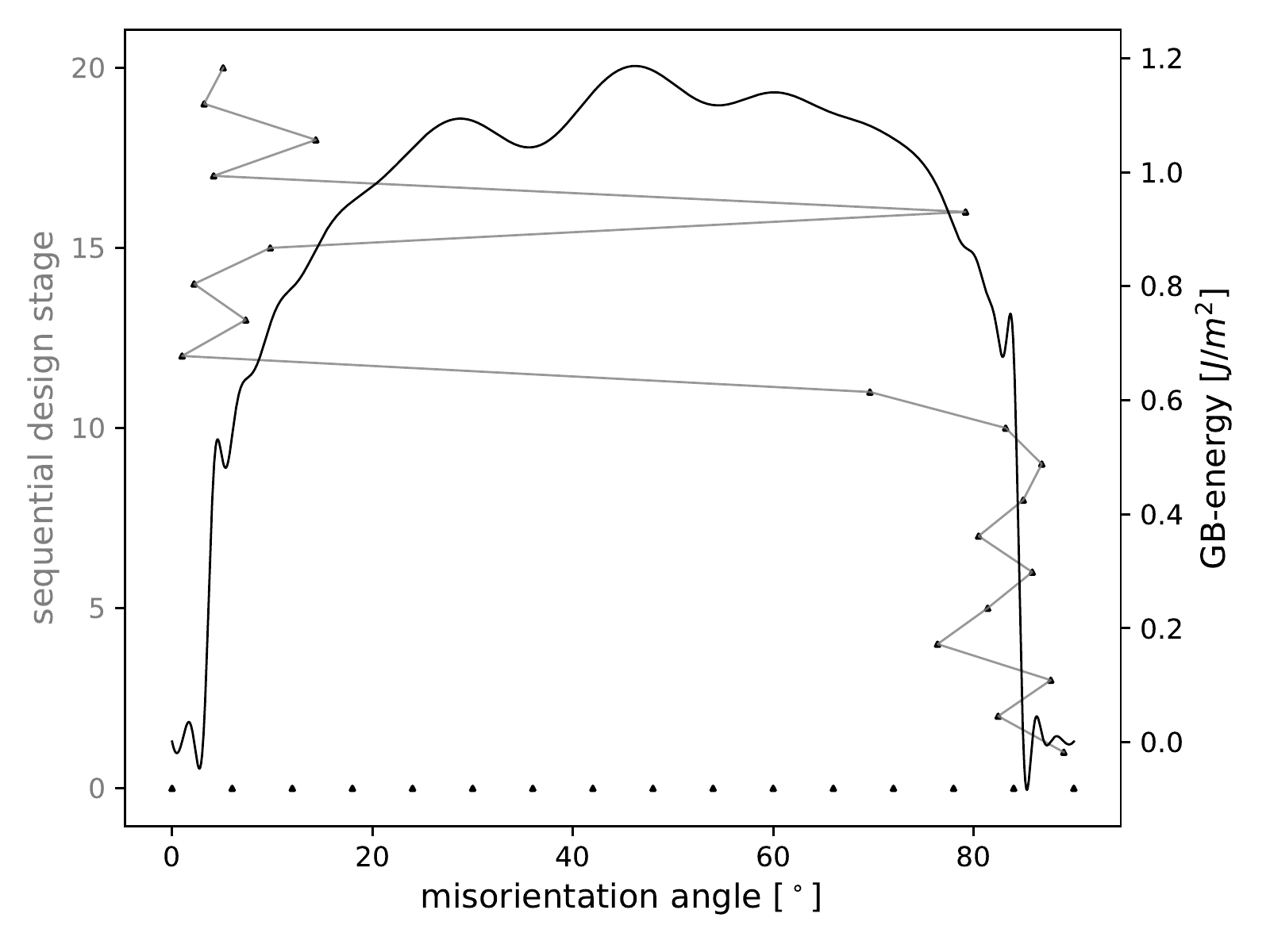}
\end{subfigure}
          \begin{subfigure}{0.49\textwidth}
         \centering
         \caption{}
         \includegraphics[width=\textwidth]{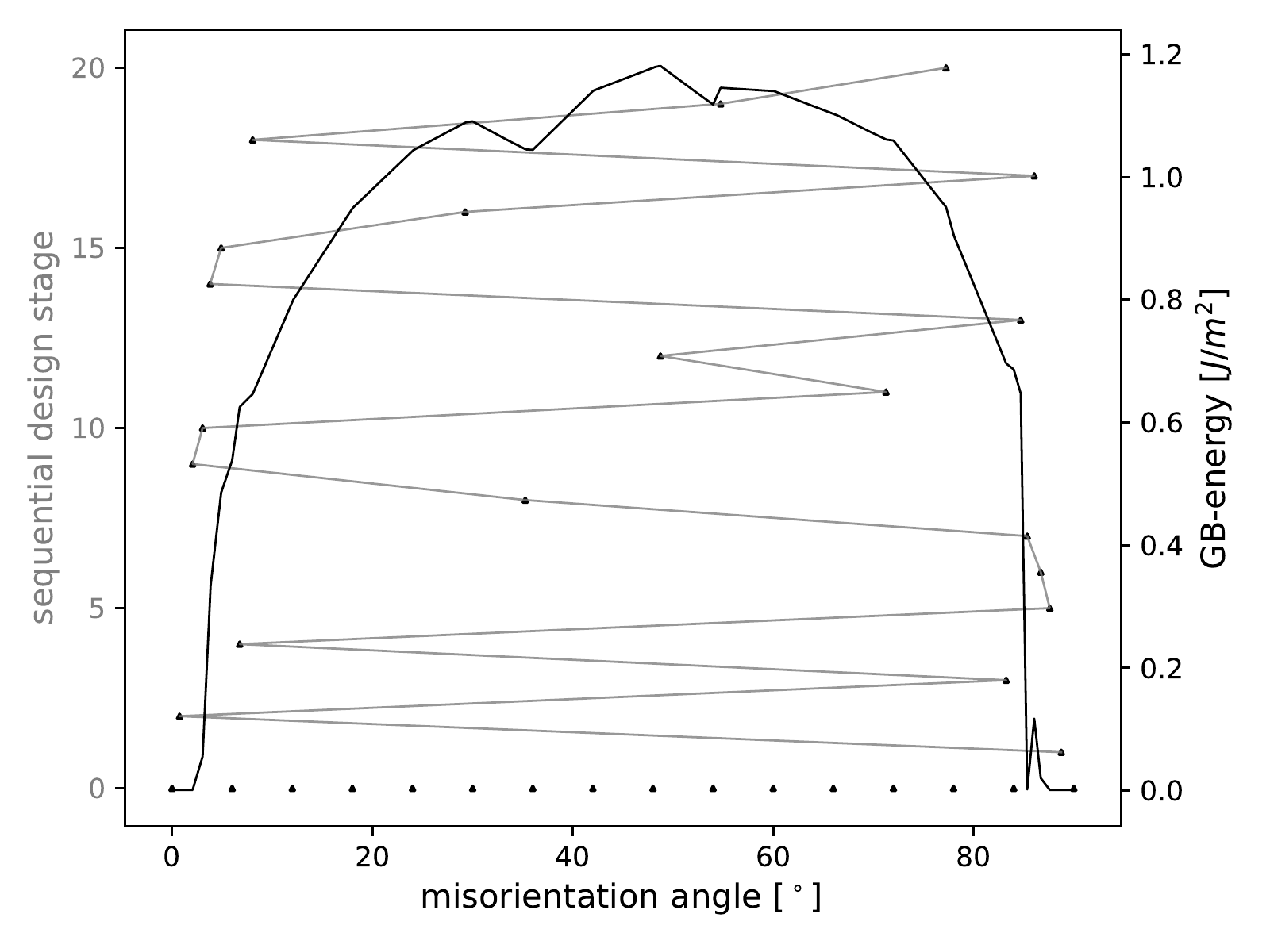}
\end{subfigure}
    \caption{Dynamics of the sequential design 
algorithm 
($\nseq = 20$ sequential design points)
for exploring the $[100]$ STGB subspace by a Kriging interpolator with different smoothing parameters.
    (a): $\nu=0.5$, (b): $\nu=1.5$, (c): $\nu=2.5$, and (d) $\nuhat$ calculated by MLE.}
    \label{fig:plots:dynamic_plots_100_appendix}
\end{figure}

\begin{figure}
     \centering
     \begin{subfigure}{0.49\textwidth}
         \centering
         \caption{}
         \includegraphics[width=\textwidth]{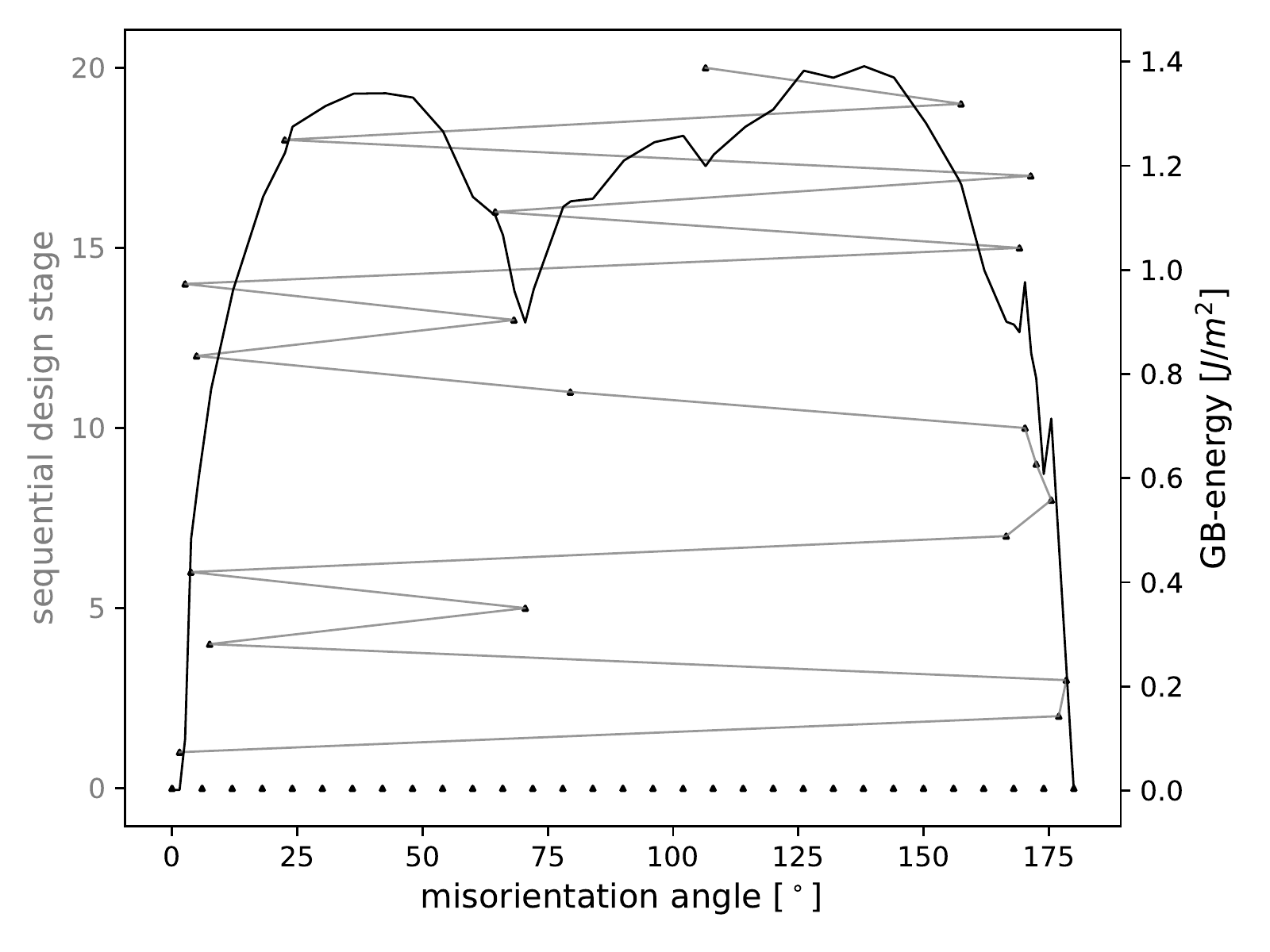}
\end{subfigure}
     \begin{subfigure}{0.49\textwidth}
         \centering
         \caption{}
         \includegraphics[width=\textwidth]{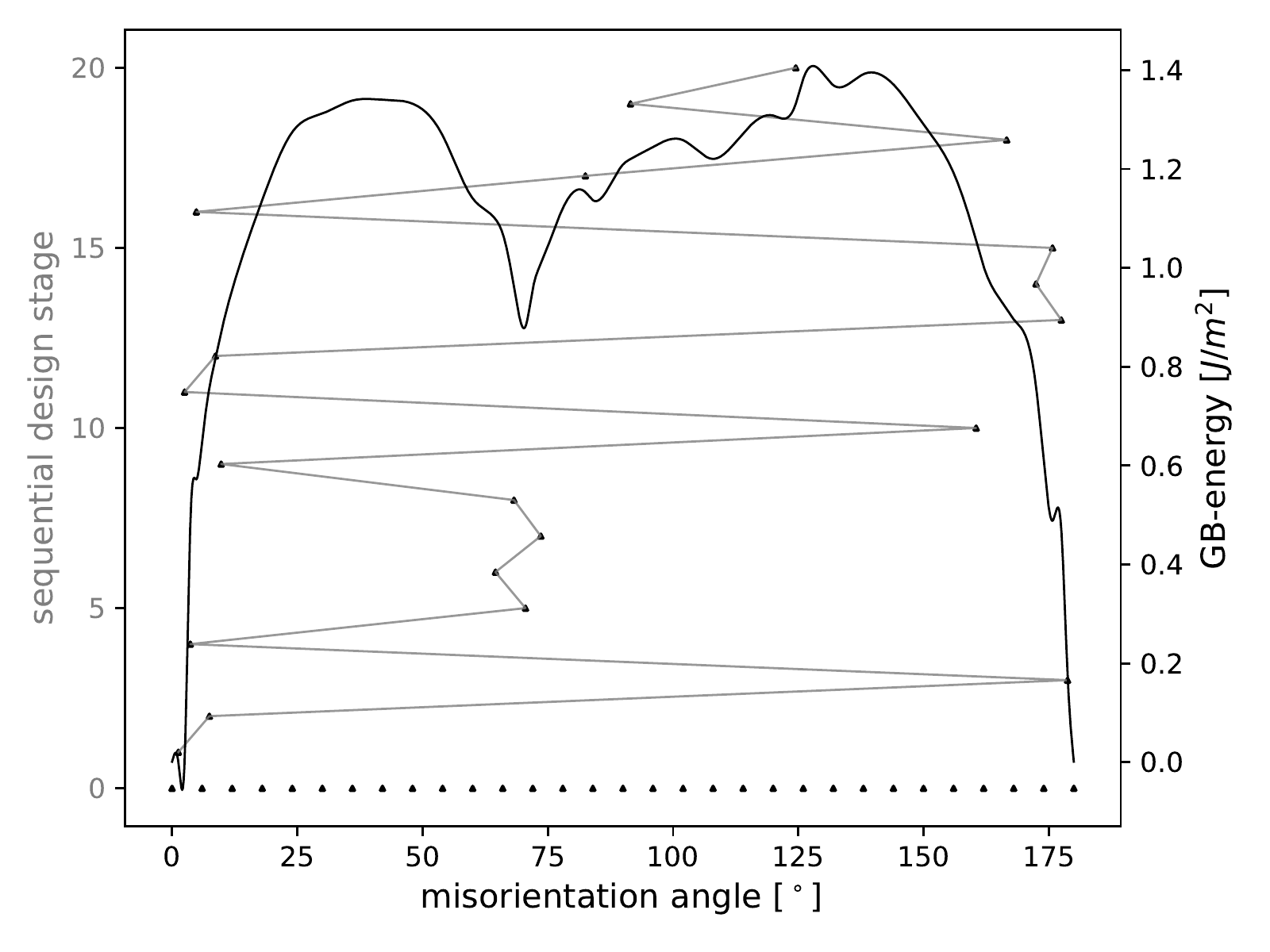}
\end{subfigure}
          \begin{subfigure}{0.49\textwidth}
         \centering
         \caption{}
         \includegraphics[width=\textwidth]{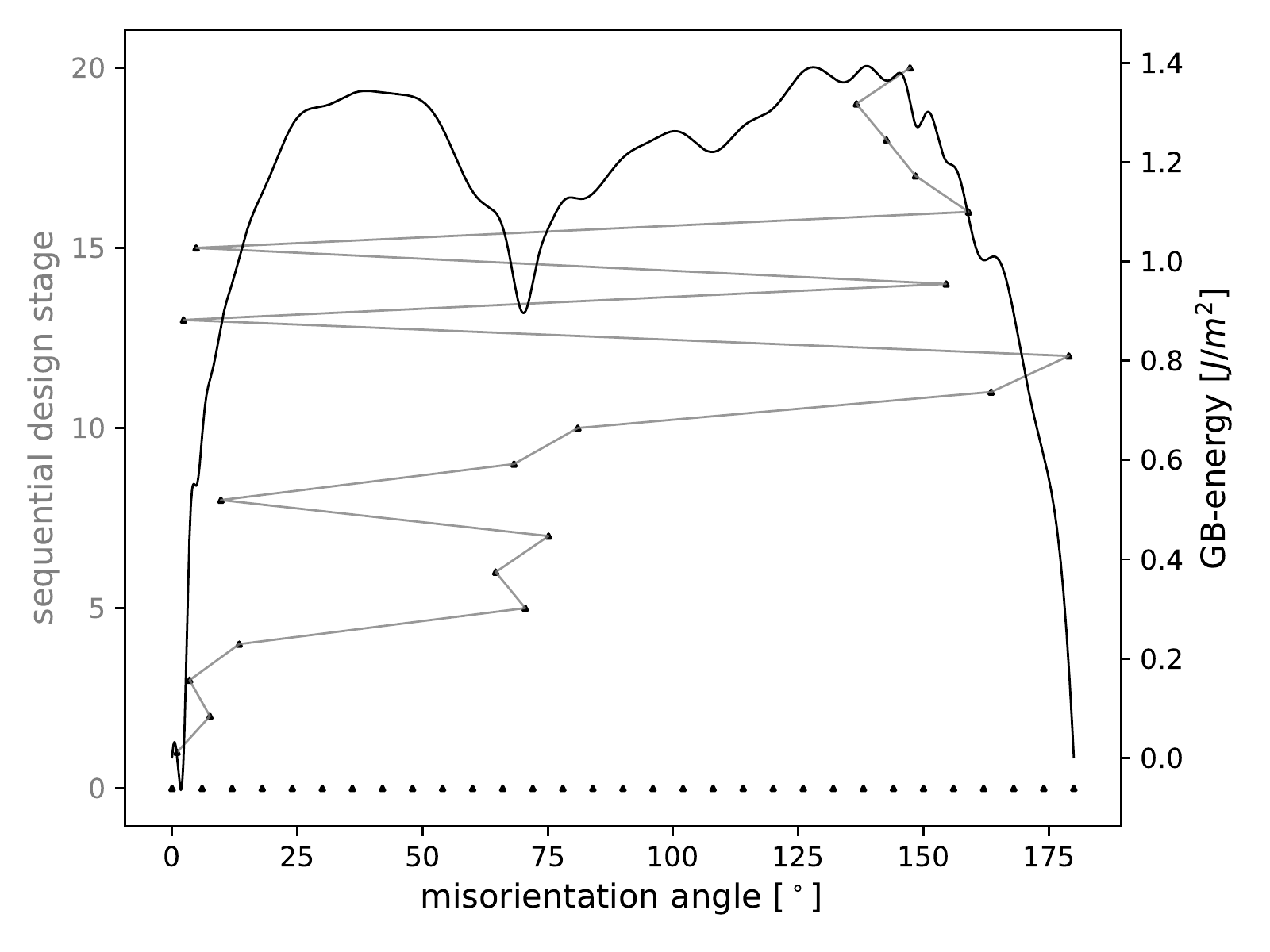}
\end{subfigure}
        \begin{subfigure}{0.49\textwidth}
         \centering
         \caption{}
         \includegraphics[width=\textwidth]{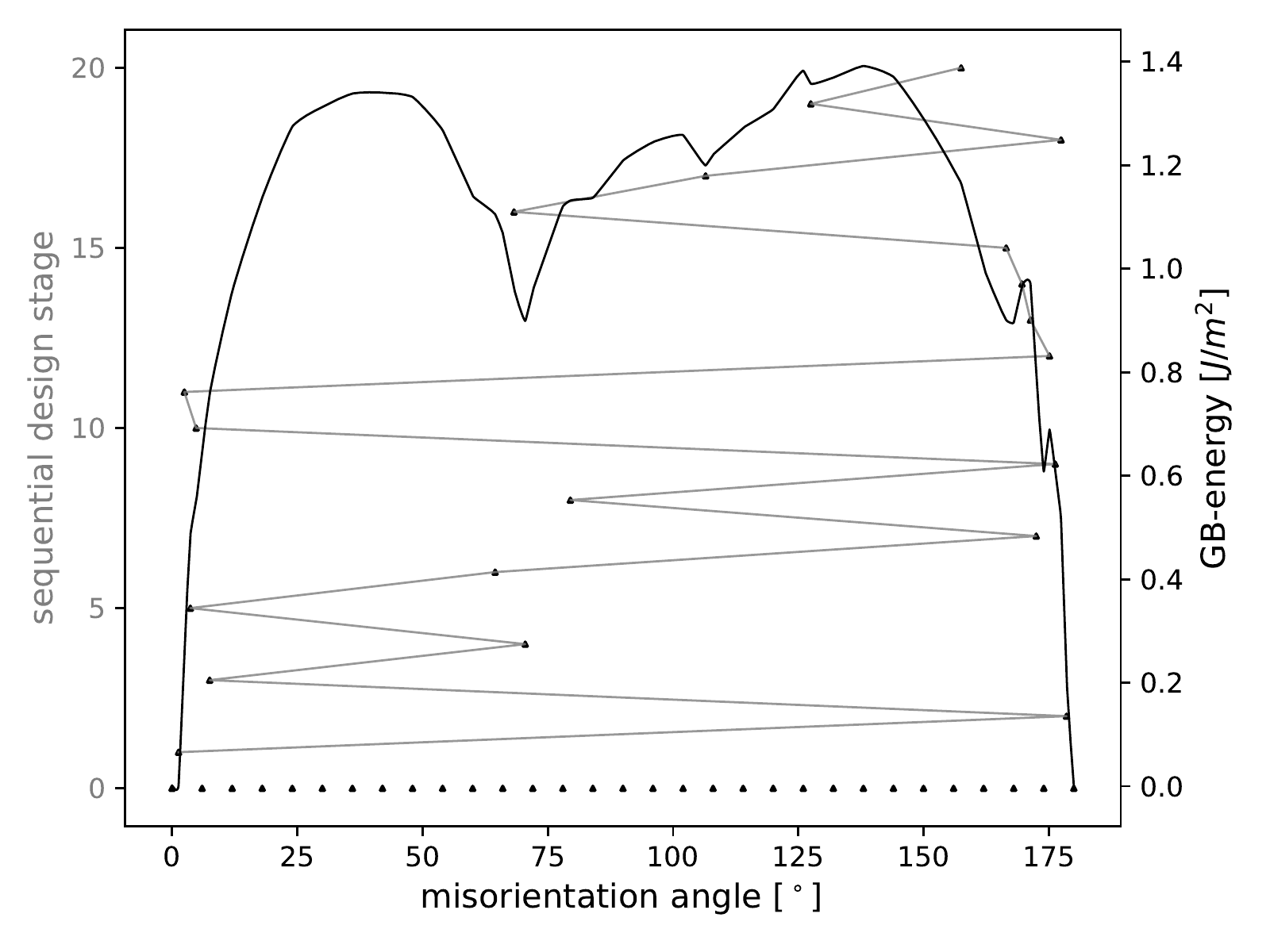}
\end{subfigure}
    \caption{Dynamics of the sequential design 
algorithm 
($\nseq=20$ sequential design points)
for exploring the $[110]$ STGB subspace by a Kriging interpolator with different smoothing parameters.
    (a): $\nu=0.5$, (b): $\nu=1.5$, (c): $\nu=2.5$, and (d) $\nuhat$ calculated by MLE.}
    \label{fig:plots:dynamic_plots_110_appendix}
\end{figure}

\begin{figure}
     \centering
     \begin{subfigure}{0.49\textwidth}
         \centering
         \caption{}
         \includegraphics[width=\textwidth]{plots/atomistic/n_start_21_n_total_61_41_ordinary_dynamics_Kleijnen_0_5.pdf}
\end{subfigure}
     \begin{subfigure}{0.49\textwidth}
         \centering
         \caption{}
         \includegraphics[width=\textwidth]{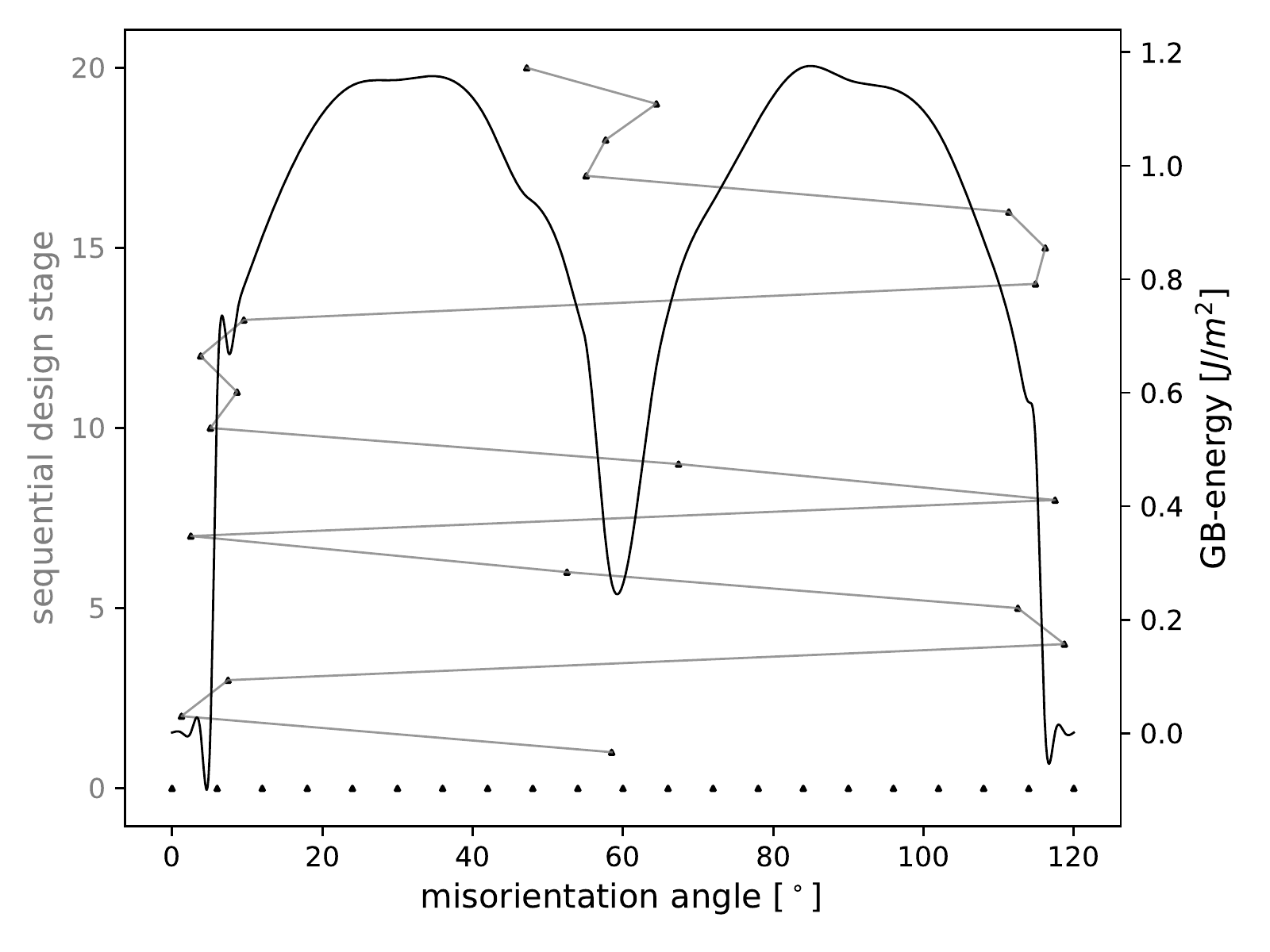}
\end{subfigure}
          \begin{subfigure}{0.49\textwidth}
         \centering
         \caption{}
         \includegraphics[width=\textwidth]{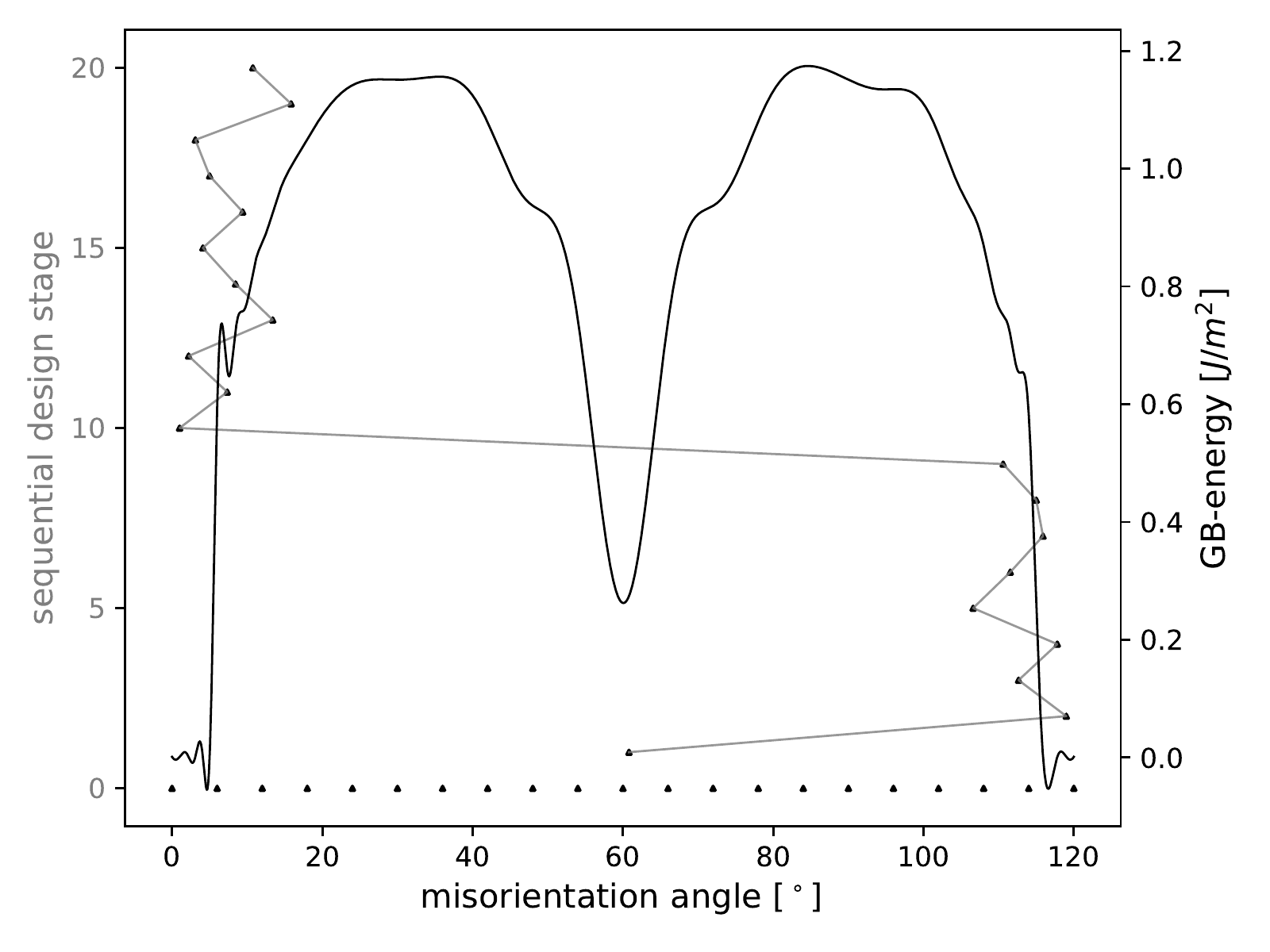}
\end{subfigure}
     \begin{subfigure}{0.49\textwidth}
         \centering
         \caption{}
         \includegraphics[width=\textwidth]{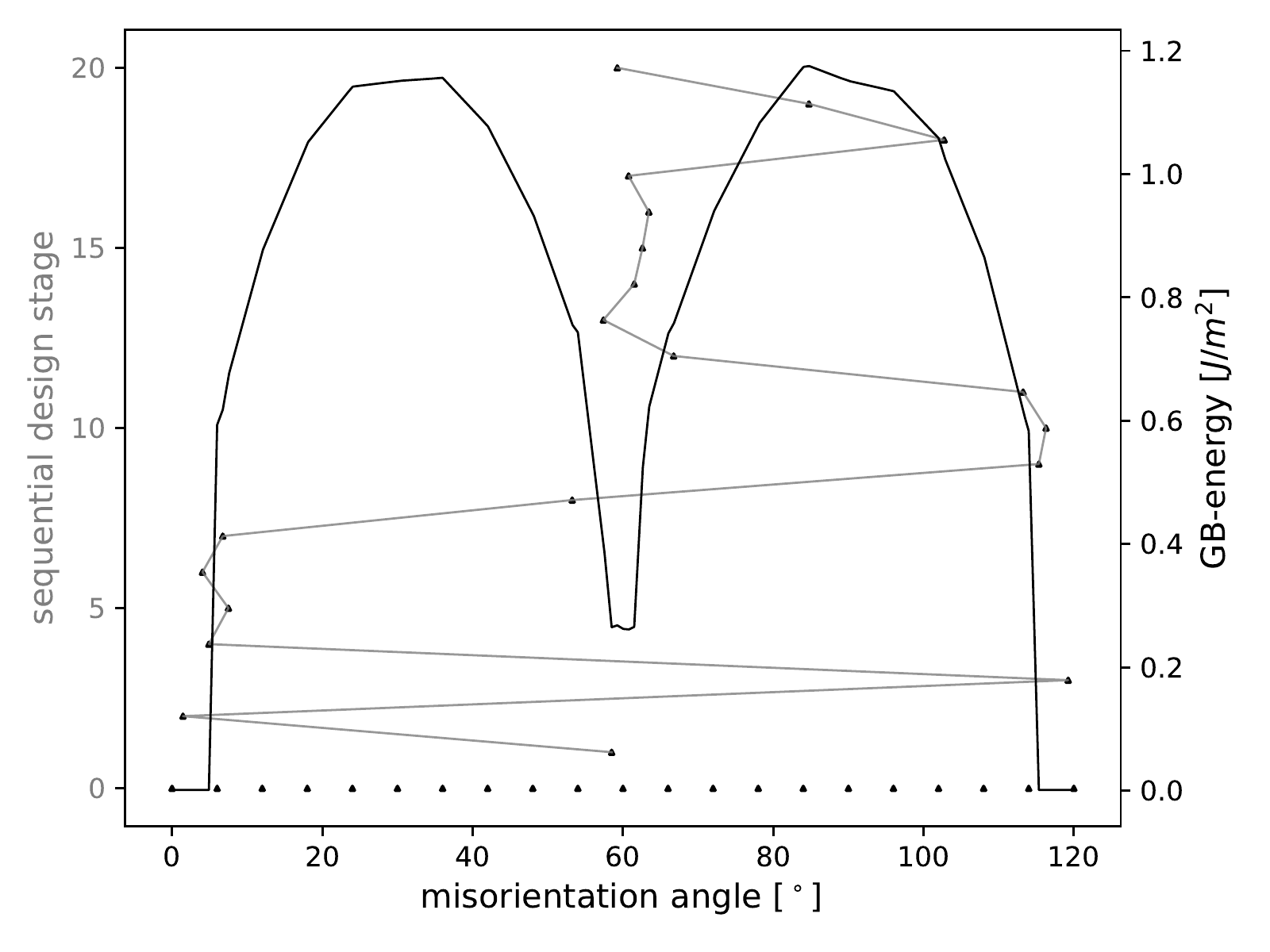}
\end{subfigure}
    \caption{Dynamics of the sequential design 
algorithm 
($\nseq = 20$ sequential design points)
for exploring the $[111]$ STGB subspace by a Kriging interpolator with different smoothing parameters.
    (a): $\nu=0.5$, (b): $\nu=1.5$, (c): $\nu=2.5$, and (d) $\nuhat$ calculated by MLE.}
    \label{fig:plots:dynamic_plots_111_appendix}
\end{figure}

In this section we will present additional results for the situation discussed in Section \ref{sec:Atomoistic_results}.
The dynamics of the sequential design algorithm and the Kriging predictions for different subspaces are shown in Figure~\ref{fig:plots:dynamic_plots_100_appendix} ($[100]$ STGB subspace),  Figure~\ref{fig:plots:dynamic_plots_110_appendix} ($[110]$ STGB subspace), and  Figure~\ref{fig:plots:dynamic_plots_111_appendix} ($[111]$ STGB subspace), respectively.
As in the main part of the paper, different choices of the smoothness parameter including a data-adaptive choice via maximum likelihood estimation were considered.
The positions of the design points (misorientation angle $\theta$) in the interval  $[0\si{\degree},\thetamax]$ ($[100]$: $\thetamax= 90\si{\degree}$, $[110]$: $\thetamax= 180\si{\degree}$, $[111]$: $\thetamax= 120\si{\degree}$) are indicated by a $\blacktriangle$. The left $y$-value indicates the stage in the sequential design algorithm (stage $0$ corresponds to the initial design), the right $y$-value the grain boundary energy in $\si{\joule}/\si{\metre}^2$ calculated by the Kriging interpolator.
As for the simulations described in Section~\ref{sec:Atomoistic_results} the sequential design places a significant number of points close to a priori unknown cusps.
Again, we observe that the sequential design  strategy chooses more points in neighborhoods of the cusps, independently of the choice of $\nu$, but the resolution of the sampling depends on the choice of $\nu$.
Finally, we briefly discuss the data adaptive choice of the smoothing parameter by the maximum likelihood
method. For
the $[100]$  and the $[111]$ subspace 
the data adaptive 
choice of the smoothing parameter
 starts at $\nuhat = 0.85$ and  $\nuhat =1.51$, respectively. For both subspaces $\nuhat$ decreases to $0.5$ 
 during the sequential  algorithm and stays close to this value after a few iterations. If the 
 Kriging interpolator and the sequential design
 are used for exploring the $[110]$ subspace, 
 the algorithm starts with  $\nuhat= 1.39$, decreases 
during the iterations, but does not converge $0.5$. Instead it decreases towards $0.6$ and varies between $0.6$ and $0.65$.

\end{document}